\documentclass[english]{article}
\usepackage[T1]{fontenc}
\usepackage[latin9]{inputenc}
\usepackage{geometry}
\usepackage{array}
\usepackage{float}
\usepackage{textcomp}
\usepackage{multirow}
\usepackage{amstext}
\usepackage{amssymb}
\usepackage{graphicx}
\usepackage{esint}
\usepackage{natbib}
\usepackage{hyperref}

\usepackage{geometry}
\makeatother

\usepackage{babel}
\begin{document}

\title{Surface reflectance of Mars observed by CRISM/MRO: 2. Estimation of surface photometric properties in Gusev Crater and Meridiani Planum}

\author{J. Fernando$^{1,2}$, F. Schmidt$^{1,2}$,
X. Ceamanos$^{3,4}$, P. Pinet$^{5,6}$, S. Dout\'e$^{3,4}$ and Y. Daydou$^{5,6}$}

\maketitle

\textbf{Abstract:} The present article proposes an approach to analyze the photometric
properties of the surface materials from multi-angle observations
acquired by the Compact Reconnaissance Imaging Spectrometer for Mars
(CRISM) on-board the Mars Reconnaissance Orbiter. We estimate photometric parameters using 
Hapke\textquoteright{}s model in a Bayesian inversion
framework. This work also represents a validation of the atmospheric
correction provided by the Multi-angle Approach for Retrieval of Surface Reflectance
from CRISM Observations (MARS-ReCO) proposed in the companion article.The latter algorithm retrieves photometric curves of surface materials in reflectance units after removing the aerosol
contribution. This validation is
done by comparing the estimated photometric parameters to those obtained
from in situ measurements by Panoramic Camera instrument at the Mars
Exploration Rover (MER)-Spirit and MER-Opportunity landing sites.
Consistent photometric parameters with those from in situ measurements
are found, demonstrating that MARS-ReCO gives
access to accurate surface reflectance. Moreover the assumption of
a non-Lambertian surface as included in MARS-ReCO is shown to be significantly
more precise to estimate surface photometric properties from space
in comparison to methods based on a Lambertian surface assumption.
In the future, the presented method will allow us to map from orbit the surface bidirectional
reflectance and the related photometric parameters in order to characterize the
Martian surface.

\section{Introduction}

Reflectance of planetary surfaces is tightly controlled by the composition
of the present materials but also by their granularity, the internal
heterogeneities, porosity, and roughness. The reflectance can be characterized
by measurements at different wavelengths, viewing geometries (emergence
direction) and solar illuminations (incidence direction). Such investigations
have been conducted for Mars using telescopes, instruments on-board
spacecrafts and rovers. A summary of these studies is available in the \citet{Johnson2008} review chapter. 
One can find studies related to: the Viking Landers \citep{guinness1997}, 
the Pathfinder Lander \citep{johnson1999}, the Hubble Space Telescope \citep{BellIII1999}, the PANoramic CAMera
(Pancam) instrument on-board Mars Exploration Rovers (MER) \citep{Johnson2006a,Johnson2006b},
the Observatoire pour la Min\'{e}ralogie, l'Eau, les Glaces et l'Activit\'{e}
(OMEGA) instrument on-board Mars Express (MEx) \citep{pinet2005}
and the High Resolution Stereo Camera (HRSC) instrument on-board MEx \citep{Jehl2008}. 
Recently, \citet{Shaw2012} derived maps of mm- to cm- scale surface roughness at MER-Opportunity 
landing site by using multi-angle hyperspectral imager spectrometer called Compact Reconnaissance 
Imaging Spectrometer for Mars (CRISM) on-board Mars Reconnaissance Orbiter (MRO). 

In order to derive compositional and structural information from reflectance
measurements, physical models describing the interaction of light
with natural media are needed. \citet{chandrasekhar1960} proposed
the radiative transfer equation describing the loss and gains of multidirectional streams
of radiative energy within media considered as continuously
absorbing and scattering where grains are separated by a distance greater
than the wavelength (e.g. atmospheres). In the case of a dense medium (e.g., surfaces),
two different solutions are developed. First, solutions based on Monte 
Carlo ray tracing methods handled the medium complexity \citep[e.g. ][]{Grynko2007}. Unfortunately,
this direct approach requires large computing times and large parameter
space, limiting the inversion. Second, solutions based on an empirical or semi-empirical approach
were proposed by adapting the radiative transfer equation to granular
media \citep[e.g., ][]{hapke1981,hapke11981,hapke1986,Hapke1993,hapke2002,Shkuratov1999a,doute1998}.
These techniques are more relevant for inversion. Several parameters
characterize natural surfaces such roughness and compaction, while other 
parameters characterize an average grain, such the single scattering albedo or the phase function.

Previous photometric studies suggest that variations
in scattering properties are controlled by local processes. For example, photometric variations
observed by Pancam at Columbia Hill and the cratered plains of Gusev
Crater are mainly caused by aeolian and impact cratering processes \citep{Johnson2006a}. These conclusions encourage
us to expand the estimation of photometric properties from in
situ observations to the entire planet using orbital data to go further
in the interpretations.

Observations acquired from space, however, require the correction
for atmospheric contribution (i.e., gases and aerosols) in
the remotely sensed signal prior to the estimation of the bidirectional
reflectance of the surface materials. Previous orbital photometric
studies of Martian surfaces were conducted without atmospheric correction
but using the lowest aerosols content observations \citep[e.g., ][]{Jehl2008,pinet2005}.
Using data acquired by the multi-angle hyperspectral imaging spectrometer
called CRISM on-board MRO \citep{murchie2007},
our objective is to estimate accurate (i) surface bidirectional reflectance
of the surface of Mars and (ii) photometric parameters associated with
the materials. \citet{Ceamanos2012} presents a method
referred to as Multi-angle Approach for Retrieval of Surface Reflectance
from CRISM Observations (MARS-ReCO). This original technique takes
advantage of the multi-angular capabilities of CRISM to determine
the bidirectional reflectance of the Martian surface. This is done
through the atmospheric correction of the signal sensed at the top
of atmosphere (TOA). We propose an approach to 
analyze the photometric parameters of the surface materials in terms of structural information by inverting Hapke's photometric model in a Bayesian framework, as discussed below. The validation
of the methods proposed in this work and in the companion article \citep{Ceamanos2012} is performed 
by comparing the estimated photometric parameters to those obtained
from in situ measurements by Pancam instrument at the MER-Spirit and
MER-Opportunity landing sites (respectively at Gusev Crater and Meridiani
Planum) \citep{Johnson2006a,Johnson2006b}.

This article is organized as follows. First, the methodology to obtain
photometric surface parameters is described in Section \ref{sec:Methodology}.
Second, the estimated photometric parameters are presented in Section
\ref{sec:Results-of-retrieved}. Third, results are compared to experimental
studies, independent orbital measurements and in situ measurements
in Section \ref{sec:Validation}. The significance of the photometric
results shall be discussed in Section \ref{sec:Analysis-of-the}.

\section{Methodology \label{sec:Methodology}}

\begin{figure}
\begin{raggedright}
\centering\includegraphics[scale=0.45]{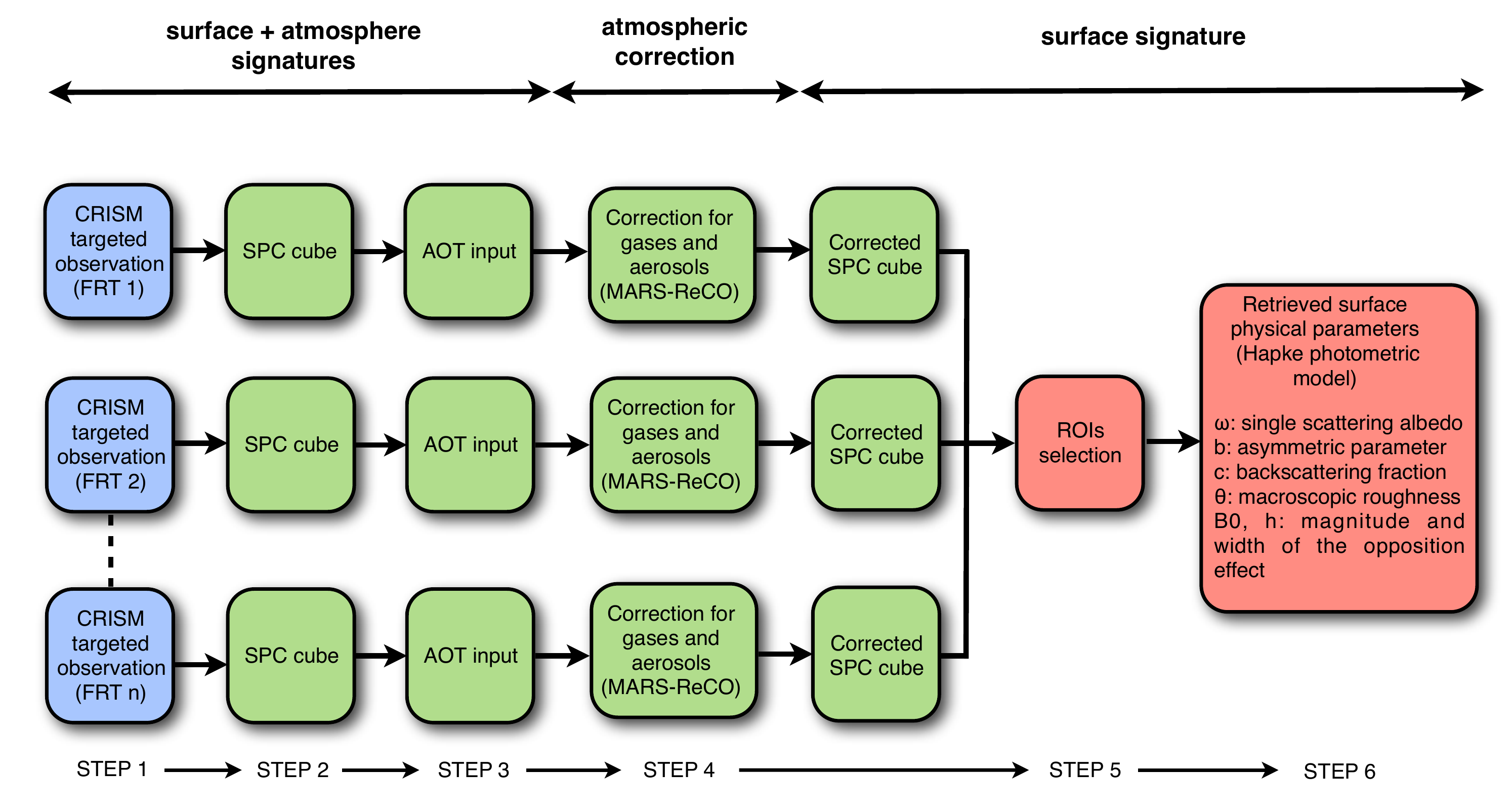}
\par\end{raggedright}

\caption{Detailed scheme of the estimation of surface photometric properties.
The blue blocks represent the initial FRT CRISM observations (from
$i$=1 to \textit{n}). The green blocks represent the aerosol optical
thickness retrieval and the correction for atmospheric contribution
for the determination of the surface bidirectional reflectance, which
is carried out by the methodology described in the companion paper
\citep{Ceamanos2012}. The red blocks correspond to the work presented
in this article, that is the estimation of surface photometric properties.
\label{scheme method} }
\end{figure}

This article and its companion take advantage of the multi-angular
capabilities of the CRISM instrument to correct for atmospheric contribution
in order to estimate the surface bidirectional reflectance \citep{Ceamanos2012}
and to determine the surface photometric parameters (this work). The
approach presented in this article includes the following steps: (i)
the selection of appropriate CRISM observations at both MER landing sites for the photometric
study, (ii) the determination of the surface bidirectional reflectance
by correcting for aerosol contributions, (iii) the combination
of several CRISM observations for a better sampling of the surface
bidirectional reflectance, and (iv) the estimation of the associated
surface photometric parameters. The detailed scheme of the procedure
is illustrated in Figure \ref{scheme method}. One should note that,
in order to test the performance of the method presented throughout
this article and its companion paper, the study is only conducted
at one wavelength and for some spatial pixels. We choose to work at 750 nm where (i) the contribution
of gases is minimal and thus the retrieval of photometric properties
is likely to be more accurate and (ii) in situ photometric measurements
from Pancam instrument are available for the comparison to the estimated
photometric parameters.

\subsection{CRISM data sets}

\subsubsection{The CRISM instrument and targeted observations \label{sub:The-CRISM-instrument}}

The CRISM instrument on-board MRO is a visible and infrared hyperspectral
imager (i.e. 362 to 3920 nm at 6.55 nm/channel) that operates from
a sun-synchronous, near-circular (255 x 320 km altitude), near-polar
orbit since November 2006. The appropriate mode to estimate surface
spectrophotometric properties is the so-called targeted mode providing
Full Resolution Targeted (FRT) observations consisting of a sequence
of eleven hyperspectral images from a single region acquired at different
emission angles. The solar incidence angle is almost constant during
the MRO flyby of a targeted observation. A typical targeted sequence
is composed of a nadir image ($\sim$10x10 km) at high spatial resolution
(15-19 m/pixel) and ten off nadir images with a 10x-spatial binning
(resulting in a resolution of 150-200 m/pixel) taken before and after
the nadir image. The latter sequence constitutes the so-called Emission
Phase Function (EPF) sequence. The pointing of CRISM can
rotate (gimbal) \textpm{} 60\textdegree{} \citep{murchie2007}.

\subsubsection{Selection of targeted observations \label{sub:Selection-of-FRT}}

As explained in the companion paper \citep{Ceamanos2012}, the accuracy
of the surface reflectance provided by MARS-ReCO when dealing with
a single targeted observation highly depends on the combination of
a moderate atmospheric opacity (i.e. aerosol optical thickness less than or equal to 2), 
reasonable illumination conditions (i,e. incidence angle less than or equal 60\textdegree{}),
an appropriate phase domain (i.e. significant difference between the available maximum and minimum phase angles up to 40\textdegree{}) 
and on the number and diversity of angular measurements. The combination of several targeted observations, since
it enables a better sampling of the bidirectional reflectance, could
therefore significantly improve the reflectance estimation as it provides 
more regular angular sampling of the surface target. \citet{pinet2005} and \citet{Jehl2008} proved the benefits
of using different spaceborne observations under varied illumination
conditions (OMEGA and HRSC). The principal requirement to combine targeted observations
is the absence of seasonal changes among the selected observations.

Several CRISM observations have been acquired over the MER landing
sites since the beginning of the mission. In particular, up to sixteen
and ten CRISM full targeted observations (FRT) are available in the
MER-Spirit and MER-Opportunity landing sites, respectively. In this
article, we select CRISM observations according to several criteria:
(i) the quality of overlap among the observations (above 70\%), (ii) the variation
of the solar incidence angle implying a widening of the phase angle
domain (note that only the variation of the seasonal solar longitude
($Ls$) can provide different incidence angles due to the sun-synchronous
orbit of MRO, and (iii) the absence of surface changes (e.g., seasonal
phenomena) as they can jeopardize the determination 
of the surface photometric properties. Taking into account
these criteria, three CRISM observations acquired over Gusev Crater
(i.e., FRT3192, FRT8CE1 and FRTCDA5) and over Meridiani Planum (i.e.,
FRT95B8, FRT334D and FRTB6B5) are selected, respectively. We note
that the selected observations have quite different phase angle ranges
as shown in Table \ref{tab:Selected-CRISM-observations}.

\begin{table}
\caption{Selected CRISM observations focused on the Spirit and Opportunity
landing sites, respectively at Gusev Crater and Meridiani Planum.
$Ls$, stands for the solar longitude, $\theta_{0}$ is the incidence
angle, $g$ is the phase angle range. AOT$\mathsf{_{mineral}}$ stands
for the mineral Aerosol Optical Thickness at 1 $\mu m$
from Wolff\textquoteright{}s estimates \citep{wolff2009} (personal
communication of Michael Wolff) and AOT$\mathsf{_{water}}$ is the
water Aerosol Optical Thickness at 320 nm by MARs Color Imager instrument
(personal communication of Michael Wolff). \label{tab:Selected-CRISM-observations}}

\raggedright{}\centering{\scriptsize }%
\begin{tabular}{cccc|ccc}
\hline 
 & \multicolumn{3}{c}{{\scriptsize Gusev Crater (MER-Spirit)}} & \multicolumn{3}{c}{{\scriptsize Meridiani Planum (MER-Opportunity)}}\tabularnewline
\hline 
 & {\scriptsize FRT3192} & {\scriptsize FRT8CE1} & {\scriptsize FRTCDA5} & {\scriptsize FRT95B8} & {\scriptsize FRT334D} & {\scriptsize FRTB6B5}\tabularnewline
\multicolumn{1}{c}{{\footnotesize Acquisition date}} & {\footnotesize 2006-11-22} & {\footnotesize 2007-12-17} & {\footnotesize 2008-10-07} & {\footnotesize 2008-01-11} & {\footnotesize 2006-11-30} & {\footnotesize 2008-07-08}\tabularnewline
{\scriptsize $Ls$ (degree)} & {\scriptsize 139.138} & {\scriptsize 4.040} & {\scriptsize 138.333} & {\scriptsize 16.223} & {\scriptsize 142.975} & {\scriptsize 96}\tabularnewline
{\scriptsize $\theta_{0}$ (degree)} & {\scriptsize 60.4} & {\scriptsize 40.02} & {\scriptsize 62.8} & {\scriptsize 39.3} & {\scriptsize 55.4} & {\scriptsize 56.4}\tabularnewline
{\scriptsize $g$ (degree)} & {\scriptsize $\sim$56-112} & {\scriptsize $\sim$41-90} & {\scriptsize $\sim$46-106} & {\scriptsize $\sim$41-86} & {\scriptsize $\sim$41-106} & {\scriptsize $\sim$40-106}\tabularnewline
{\scriptsize AOT$\mathsf{_{\mathrm{mineral}}}$ (1$\mu m$)} & {\scriptsize 0.33$\pm$0.04} & {\scriptsize 0.98$\pm$0.15} & {\scriptsize 0.32$\pm$0.04} & {\scriptsize 0.56$\pm$0.09} & {\scriptsize 0.35$\pm$0.04} & {\scriptsize 0.35$\pm$0.04}\tabularnewline
{\scriptsize AOT$_{\mathsf{\mathrm{water}}}$ (320 nm)} & {\scriptsize 0.08$\pm$0.03} & {\scriptsize 0.07$\pm$0.03} & {\scriptsize 0.03$\pm$0.03} & {\scriptsize 0.12$\pm$0.05} & {\scriptsize 0.12$\pm$0.03} & {\scriptsize 0.14$\pm$0.03}\tabularnewline
\hline 
\end{tabular}
\end{table}

Targeted observations are archived in the Planetary Data System (PDS)
and are composed of: (i) Targeted Reduced Data Records (TRDR), which
store the calibrated data in units of I/F (radiance factor - RADF, 
see Table \ref{tab:Different-photometric-units}), the ratio of measured intensity to solar flux,  
and (ii) Derived Data Records (DDR), which store the ancillary data
such as the spatial coordinates (latitude and longitude) and the geometric
configurations of each pixel by means of the incidence, emission and
phase angles. In the present study, CRISM products are being released 
with the TRDR2 version of CRISM calibration (TRR2 for brevity).

\begin{table}
\caption{Photometric units derived from the bidirectional reflectance $r$.
($I$: intensity, $F$: solar flux, $ $: $\theta_{0}$: incidence
angle, $\theta$: emergence angle, $g$: phase angle). The CRISM observations
are released in RADF unit and the SPC cubes in BRF units. \label{tab:Different-photometric-units}}

\raggedright{}\centering{\scriptsize }%
\begin{tabular}{ccccc}
\hline 
 & {\scriptsize Unit} & {\scriptsize Symbol} & {\scriptsize Name} & {\scriptsize Expression}\tabularnewline
\hline 
\multirow{4}{*}{{\scriptsize reflectance}} & \multirow{4}{*}{{\scriptsize $sr^{-1}$}} & {\scriptsize $r$} & {\scriptsize bidirectional reflectance} & {\scriptsize $r\left(\theta_{0},\theta,g\right)=\frac{I}{\pi\times F}$}\tabularnewline
 &  & {\scriptsize $BRDF$} & {\scriptsize bidirectional reflectance distribution function} & {\scriptsize $BRDF=r(\theta_{0},\theta,g)/\cos\left(\theta_{0}\right)$}\tabularnewline
 &  & {\scriptsize $RADF$} & {\scriptsize Radiance factor} & {\scriptsize $RADF=\pi\times r(\theta_{0},\theta,g)=\frac{I}{F}$}\tabularnewline
 &  & {\scriptsize $BRF$} & {\scriptsize bidirectional reflectance factor} & {\scriptsize $\rho=BRF=\pi\times r(\theta_{0},\theta,g)/\cos\left(\theta_{0}\right)$}\tabularnewline
\hline 
\end{tabular}
\end{table}

\subsubsection{SPC cubes: Integrated multi-angle product \label{sub:SPC cube}}

To facilitate the access to the multi-angular information pertaining
to each terrain unit, the eleven hyperspectral images corresponding
to a single targeted observation were spatially rearranged into data
set named SPC (spectro-photometric curve) cube (see \citet{Ceamanos2012} for more detail).
The SPC cube is composed of: (i) $x$ dimension (horizontal) corresponding
to angular configurations (up to eleven) grouping all reflectance values at different geometric views, (ii) $y$ dimension (vertical)
corresponding to the spatial coordinate defining a super-pixel. The super-pixels are rearranged as function of the number of available angular configurations and, 
(iii) $z$ dimension corresponding to the spectral sampling.
The photometric curves, stored in a SPC cube are in units of RADF (see Table \ref{tab:Different-photometric-units}), correspond
to the signals TOA. In the present work, all selected FRT observations are binned at 460 meters per pixel (the spatial resolution of each super-pixel).
This is done (i) to cope with the geometric deformation inaccuracies in the case of an oblique view with pointing errors, (ii) to minimize on poorly known topography (a CRISM pixel is smaller than MOLA resolution), (iii) to reduce local seasonal variations and (iv) to minimize local slopes effects. We found that binning at 460 meters is a good compromise.

\subsection{Estimation of surface bidirectional reflectance: Correction for atmospheric
contribution}

The radiative transfer in the Martian atmosphere 
is dominated by CO$_{2}$ and H$_{2}$O gases, mineral and ice
aerosols which are an obstacle for the studies of the surface properties.
Indeed the extinction of radiative fluxes, in particular the solar
irradiance in the wavelength range of CRISM observations, is mostly
due to absorption by gases and scattering by aerosols. Also, aerosols
produce an additive signal by scattering of the solar light. As a
result, a spectrum collected by the CRISM instrument at the Top Of-Atmosphere 
(TOA) is a complex signal determined by both the surface
and the atmospheric components (gases and aerosols). The atmospheric
correction chain proposed for CRISM observations 
is composed of: (i) the retrieval of Aerosol Optical Thickness (AOT),
(ii) the correction for gases, and (iii) the correction for aerosols
resulting in the estimation of the surface reflectance. The last step
is carried out by the technique referred to as the MARS-ReCO  \citep{Ceamanos2012}. In the
following, a short summary of the methodology is provided.

\subsubsection{Retrieval of AOT}

The AOT is defined as the aerosol optical depth along the vertical
of the atmosphere layer and is related to the aerosol content.

The retrieval of AOT from orbit is difficult because of the coupling
of signals between the aerosols and the surface. Over the last decades,
Emission Phase Function (EPF) sequences of Mars were obtained and
were used to separate the atmospheric and surface contributions. Significant
progresses have resulted from the pioneer works of \citet{Clancy1991}
based on Viking orbiter InfraRed Thermal Mapper (IRTM) EPF observations
and \citet{Clancy2003} from Mars Global Surveyor (MGS) Thermal Emission
Spectrometer (TES) EPF observations ten years later. 

For our study, we decided to use Michael Wolff's AOT estimates for
atmospheric correction purposes (personal communication). This parameter
is available for each CRISM observation and is derived from \citet{wolff2009}'s
work. This method is based on the analysis of CRISM EPF sequences
combined with information provided by ``ground truth'' results at
both MER landing sites which allow to isolate the single scattering
albedo. This method carries out a minimization of the mean square
error between measured and predicted TOA radiance based on the previously
estimated aerosol single scattering albedo and scattering phase function. 

Some assumptions regarding the aerosol properties (i.e., phase function,
mixing ratio, column optical depth, etc) and the surface properties
(i.e., phase function) must have been accounted for to separate the
atmospheric and surface contributions. Concerning the surface properties,
this method assumes a non-Lambertian surface to estimate the AOT by
using a set of surface photometric parameters that appears to describe
the surface phase function adequately for both MER landing sites \citep{Johnson2006a,Johnson2006b}.
This assumption is qualitative but reasonable for several reasons
enumerated by \citet{wolff2009}. For both MER landing sites it seems
from \citet{wolff2009}'s work that the AOT retrievals are overall
consistent with optical depths returned by the Pancam instrument (available
via PDS). Consequently, the Wolff's AOT estimates are suitable for
our study at both MER landing sites. Concerning the aerosol properties,
some uncertainties may exist especially for the aerosol scattering
phase function which is related to the aerosol particle size and shape.
First, this method assumes a mean particle aerosol but variations
are observed as a function of solar longitude, spatial location (latitude,
longitude and altitude) \citep{Wolff2006}. Second the mean aerosol
particle size is derived from CRISM observations acquired during planet-encircling
dust event in 2007. During a large dust event, aerosol particle size
are larger than those found under clear atmospheric conditions. Third,
the hypothesis of the particle shape used in \citet{wolff2009}'s study contain
a wrong backscattering part in the scattering phase function \citep{Wolff2010}.
Thus all points previously enumerated, may bias the AOT estimation
and must be taken into account during the analysis of the robustness
of results (the surface bidirectional reflectance and the surface
photometric parameters). However, the performance of MARS-ReCO is sensitive 
to the accuracy of the AOT estimate \citep{Ceamanos2012} (see \ref{sub: atmospheric correction}).

Note that the AOT is calculated at 1 $\mu m$ where
the absorption of gases is almost null.

\subsubsection{Correction for gases and aerosols \label{sub: atmospheric correction}}

In order to test the performances of the methodology presented in
this article and its companion paper, the present study is only conducted
at a single wavelength. We choose to work exclusively at 750 nm where
the contribution of gases is low and thus the retrieval of photometric
properties is likely to be sufficiently accurate. Furthermore, photometric
properties retrieved from in situ measurements taken by Pancam are
available at this wavelength and our photometric
properties can be validated. Note, however, that the presented methodology
can be applied to any CRISM wavelength provided the contribution of gases is corrected previously.

MARS-ReCO is devised to compensate for mineral aerosol effects considering
the anisotropic scattering properties of the surface and the aerosols.
This method is suitable for any CRISM multi-angle observation within
some atmospheric and geometrical constraints (AOT$\leq$2, incidence angle $\theta_{0}$<60\textdegree{}, 
phase angle range $g{}_{max}-g_{min}$>40\textdegree{}). MARS-ReCO is based on a coupled surface-atmosphere radiative transfer formulation using
a kernel-driven scattering model for the surface and a Green\textquoteright{}s
function to model the diffuse response of the atmosphere (please refer to \citet{Ceamanos2012} for more detail). The AOT of each observation is an input of MARS-ReCO. Table \ref{tab:Selected-CRISM-observations} 
presents the AOT$_{water}$ and AOT$_{mineral}$  for each selected CRISM observation. We can note that 
 AOT$_{water}$  is negligible in front of the AOT$_{mineral}$ and consequently, the photometric effects 
from aerosol water ice can be considered negligible in this study. 

The uncertainties pertaining to the AOT estimates (personal
communication of Michael Wolff, \citep{wolff2009}) and to the TOA
measurements by CRISM are integrated and propagated in the estimation
of the surface bidirectional reflectance in BRF units (cf. Table \ref{tab:Different-photometric-units}).

Besides the retrieval of the surface bidirectional reflectance, MARS-ReCO
also provides an indicator of the quality of the estimated solution
in a standard deviation sense, noted by parameter $\sigma_{\rho}$ (in
BRF units) and computed as $\sigma_{\rho}=\sqrt{\frac{1}{N_{g}}\sum_{j=1}^{Ng}\mathrm{tr}\left(\mathbf{C}_{\rho p}\right)}$,
where $\mathrm{tr}\left(\mathbf{C}_{\rho p}\right)$ is the a posterior
covariance matrix and $Ng$ the available geometries \citep{Ceamanos2012}.
In systematic test presented in the companion article, the parameter $\sigma_{\rho}$ has proved
to be highly correlated with the bidirectional reflectance error from MARS-ReCO and thus
provides us with reliable information on the accuracy of the estimated
surface bidirectional reflectance.

\subsection{Estimation of surface photometric properties: Bayesian inversion
based on a Hapke\textquoteright{}s photometric model}

\subsubsection{Defining regions of interest and selection \label{sub:Defining-regions-of}}

\begin{figure}
\begin{centering}
\includegraphics[scale=0.33]{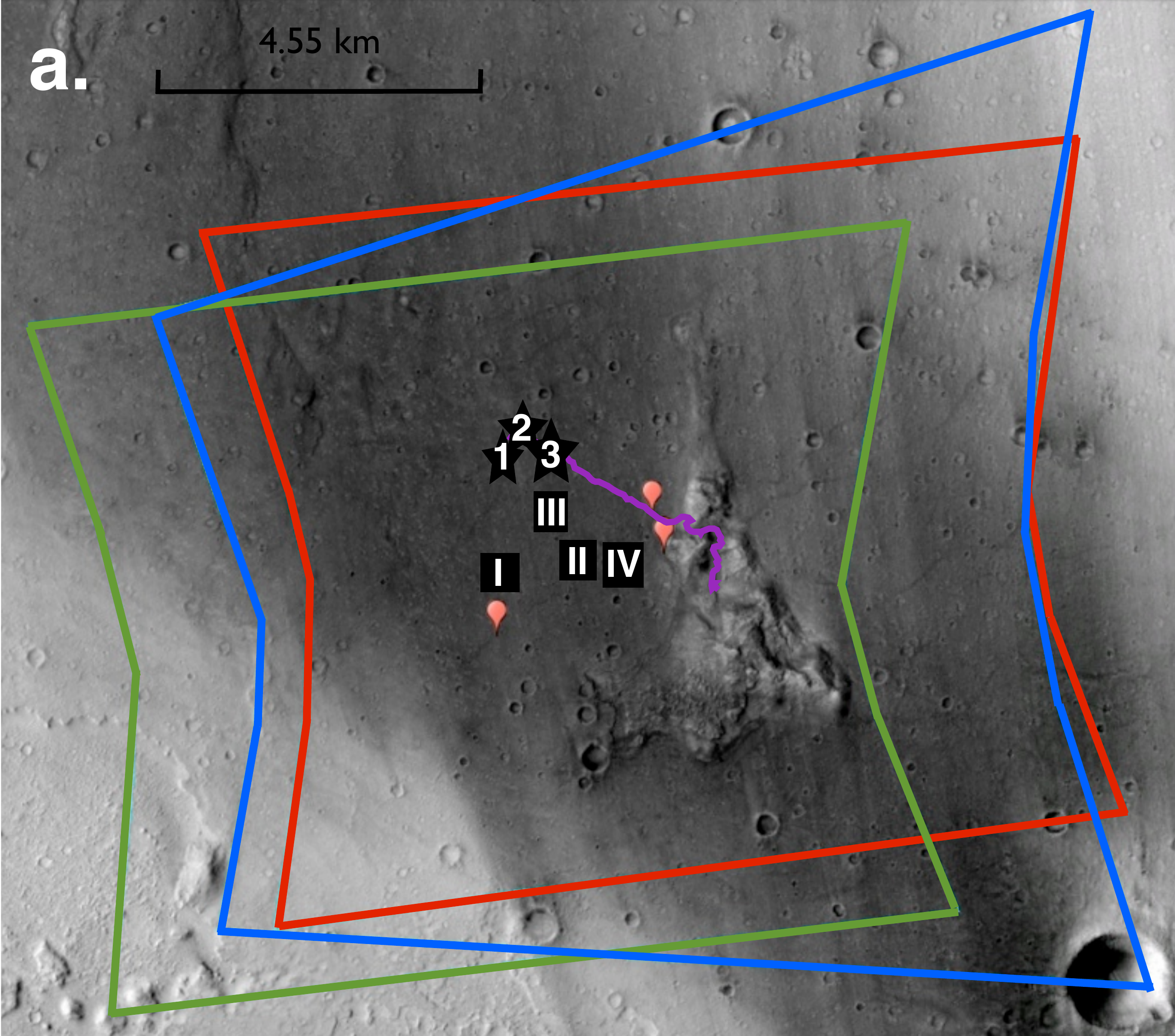} \includegraphics[scale=0.31]{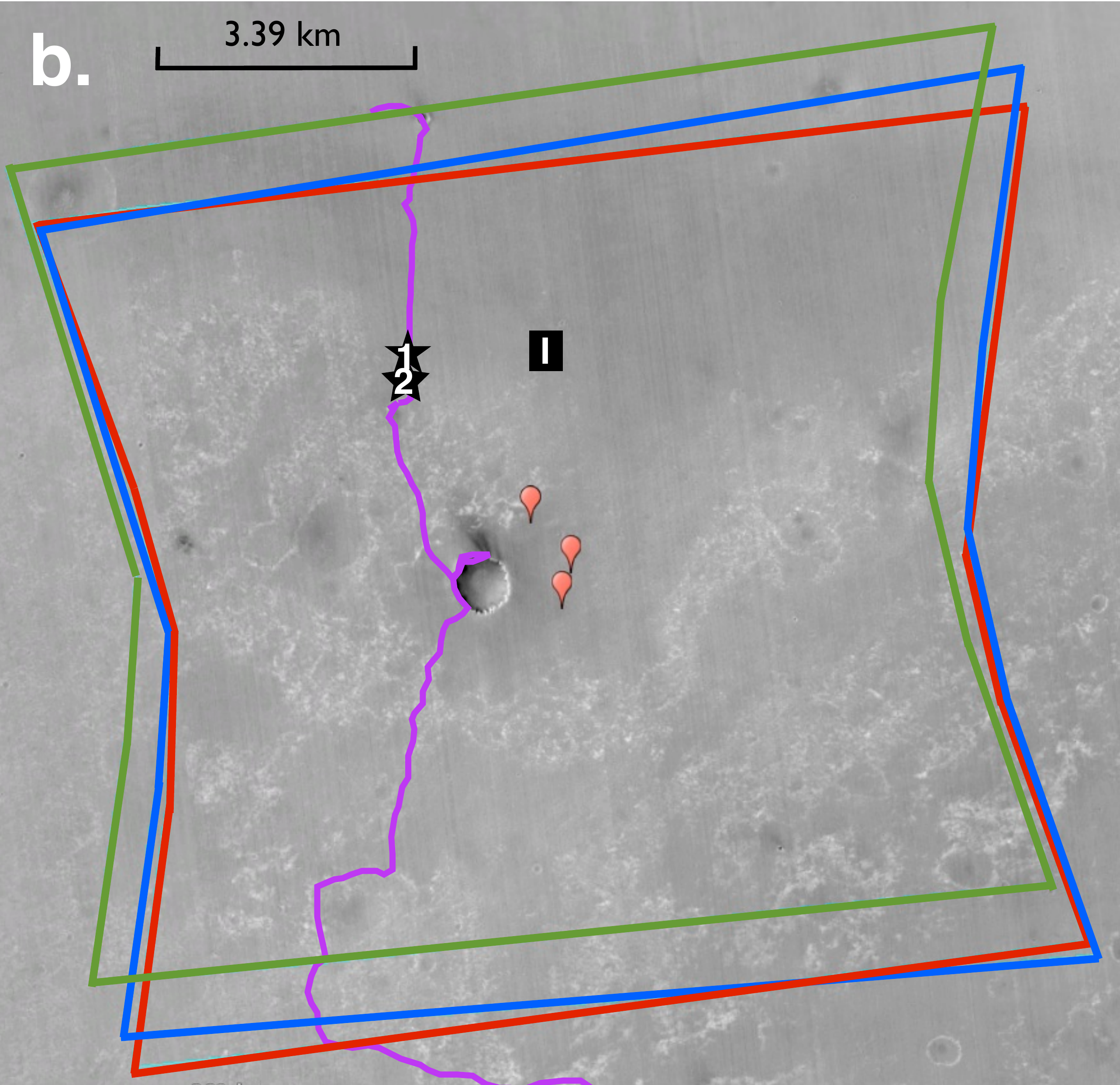}
\par\end{centering}

\caption{a. Image from Context camera (CTX) of the Spirit
landing site at Gusev Crater with the rover path across
the plain up to the Columbia hill in purple. The footprint of each selected 
CRISM observations (only the nadir image) is represented here (blue: 
FRT3192, green: FRT8CE1 and red: FRTCDA5). Full black stars point to 
the locations of the photometric measurements in the Gusev plains
(1. Landing Site, 2. Bonneville Rim, 3. NW of Missoula) taken by instrument onboard Spirit 
\citep{Johnson2006a}. Full black squares represent the four ROIs
that have been selected for our photometric study (ROI from I to IV). b. Image from CTX camera of the Opportunity landing site at Meridiani Planum with
the rover path in purple. The footprint of each selected 
CRISM observations (only the nadir image) is represented here (blue: 
FRT95B8, green: FRT334D and red: FRTB6B5). Full black stars represent
the locations of photometric measurements (1. South of Voyager, 2.
Purgatory region) taken by Pancam instrument onboard Opportunity
\citep{Johnson2006b}. Full black squares represent the selected ROI
for our photometric study (ROI I). \label{fig:HiRISE}}
\end{figure}

We now discuss the improvement that results when combining different CRISM targeted
observations based on their spatial coherence and thus a better sampling of the surface bidirectional reflectance and thus a maximization of the phase angle range. We defined in Subsection \ref{sub:SPC cube} a super-pixel 
as the combination of all reflectance value corresponding
to a same location unit coming from an individual CRISM sequence or SPC cube. The combination of each super-pixel from each selected CRISM targeted observation is performed when their central coordinates (latitude
and longitude) differ less than a half super-pixel size (460/2=230 meters). This is done
to ensure maximum overlap. We choose same combined super-pixels called regions of interest (ROI) in following for the photometric 
study using the several criteria. First, the different ROIs
must be located close to the MER Spirit and Opportunity rover's path,
specifically to the location of spectrophotometry measurements by
Pancam and in the same geological unit (i.e. presenting same materials). 
Second, the local topography makes the photometric study more
challenging when it is poorly known because it controls to a large
extent the incidence, emergence and azimuth local angles. Besides
in the case of an oblique illumination (i.e., up to 70\textdegree{}),
shadows decrease the signal/noise ratio. In this study, ROIs were
therefore selected only in flat areas. Third, ROIs are chosen to have
the richest angular configurations and the best angular sampling in
terms of phase angle range in order to constrain as much as possible
the photometric properties. Figure \ref{fig:HiRISE} presents the selected ROIs for this
photometric study at Gusev Crater and Meridiani Planum. Four ROIs
are selected for Gusev Crater study while only one is chosen for Meridiani
Planum study. The limited number of selected ROIs is explained by
the fact that few ROIs in both cases respect the combination of criteria
previously mentioned. In order to improve the number of ROIs, an improved 
pointing of each CRISM targeted observation could be
envisaged.

\subsubsection{Direct surface model: Hapke\textquoteright{}s photometric model \label{sub:Direct-surface-model:}}

Models describing the photometry of discrete granular media have been
proposed to express the surface bidirectional reflectance using semi-empirical analytical approaches \citep[e.g., ][]{hapke1981,hapke11981,hapke1984,hapke1986,Hapke1993,hapke2002}
and numerical approaches \citep[e.g., ][]{cheng2000,doute1998,mishchenko1999}.

Hapke's modeling \citep{Hapke1993} is widely used in the planetary
community due to the simplicity of its expression and due to the fact
that it introduces photometric parameters claimed to have physical
relevance. Previous Martian photometric studies have been
conducted based on orbital HRSC measurements \citep{Jehl2008} and
based on the analysis of in situ MER measurements \citep{Johnson2006a,Johnson2006b} using the Hapke's 1993 version model \citep{Hapke1993}. For this study, since we use the in situ investigation as ``ground truth'' in order to validate the MARS-ReCO approach, we have to use Hapke's 1993 version model. 

More recent Hapke models \citep{hapke2002,hapke2008} are available improving the original formulation \citep{hapke1981,Hapke1993} and can be used for future photometric studies. First, the model \citep{hapke2002} includes: (1) a more accurate analytic approximation for isotropic scatterers, (2) a better estimation of the bidirectional reflectance when the scatterers are anisotropic, and (3) the incorporation of 
coherent backscattering. Second, the model \citep{hapke2008} overcomes the limitations of the original model in order to predict porosity dependence of the bidirectional reflectance, 
in case of a particulate medium such as a planetary regolith. For that purpose the treatment of light propagating through the particle spacing.

Following \citet{Johnson2006a,Johnson2006b}'s works and for the sake of the comparison coherence we use the expression of \citet{Hapke1993} as follows: 

\begin{equation}
r\left(\theta_{0},\theta,g\right)=\frac{\omega}{4\pi}\,\frac{\mu_{0e}}{\left(\mu_{0e}+\mu_{e}\right)}\,\left\{ \left[1+B(g)\right]P(g)+H(\mu_{0e})H(\mu_{e})-1\right\} S(\theta_{0},\theta,g),\label{eq:Hapke}
\end{equation}

\begin{itemize}

\item Geometry $\theta_{0}$, $\theta$, and $g$: The incidence, emergence
and phase angles, respectively. . 

\item Single scattering albedo $\omega$: Factor $\omega$
($0\le\omega\le1$) depends on wavelength and
represents the fraction of scattered light to incident light by a
single particle \citep{chandrasekhar1960}.

\item Particle scattering phase function $P(g)$: Function $P(g)$ characterizes
the angular distribution of an average particle. The empirical 2-term
Henyey-Greenstein function (hereafter referred to as HG2) is used
commonly for studying planetary surfaces \citep[e.g., ][]{cord2003,hartman1998,Jehl2008,Johnson2006a,Johnson2006b,souchon2011}.
It reads: 

\begin{equation}
P(g)=\left(1-c\right)\,\frac{1-b^{2}}{\left(1+2b\cos\left(g\right)+b^{2}\right)^{3/2}}+c\,\frac{1-b^{2}}{\left(1-2b\cos\left(g\right)+b^{2}\right)^{3/2}}\label{eq:HG2}
\end{equation}

where the asymmetric parameter $b$ ($0\le b\le1$) characterizes
the anisotropy of the scattering lobe (from $b=0$, which corresponds
to the isotropic case, to $b=1$, which corresponds to a particle
which diffuses light in a single direction). The backscattering fraction
$c$ ($0\le c\le1$) characterizes the main direction of the diffusion 
($c<0.5$ corresponding to forward scattering and $c>0.5$ corresponding
to backward scattering).

\item Multiple scattering function $H(x)$: The exact values of the $H$
function for isotropic scatterers was given by \citet{chandrasekhar1960}.
For consistency we use the approximation of isotropic scattering
named H93 in the following \citep{Hapke1993} as was done for the
estimation of the surface photometric parameters by \citet{Johnson2006a,Johnson2006b}. 
The H93 differs by a relative error on $H$ lower than 1\% and a relative
error on the bidirectional reflectance of a regolith lower than 2\%
\citep{cheng2000}. Defining $y=(1-\omega^{1/2}$,
the multiple scattering function becomes:

\begin{equation}
H(x)=\left\{ 1-\left[1-y\right]x\left[\left(\frac{1-y}{1+y}\right)+\left(1-\frac{1}{2}\left(\frac{1-y}{1+y}\right)-x\left(\frac{1-y}{1+y}\right)\right)\ln\left(\frac{1+x}{x}\right)\right]\right\} ^{-1}\label{eq:H1993}
\end{equation}

\item Shadow Hiding Opposition Effect (SHOE) and Coherent Backscatter Opposition
Effect (CBOE): $B(g)$ is a function designed to model the sharp increase
of brightness around the zero phase angle often observed in the case
of particulate media, the so-called opposition effect. The $B(g)$
function is given by \citet{Hapke1993} as follows:

\begin{equation}
B(g)=\frac{B_{0}}{1+\frac{1}{h}\tan\left(\frac{g}{2}\right)}.
\end{equation}

Parameters $h$ and $B_{0}$ are respectively the angular width and
the amplitude of the opposition effect. Factor $h$ (ranging from
0 to 1) is physically related to compaction and particle size distribution
and $B_{0}$ (ranging from $0$ to $1$) is an empirical parameter
which is related to the particle transparency \citep{Hapke1993}.
It is important to mention that the Ross-Thick Li-Sparse (RTLS) model
for the surface reflectivity employed by MARS-ReCO \citep{Ceamanos2012}
is able to describe a backscattering lobe by means of its geometrical
kernel. Nevertheless its angular width is more characteristic of photometric
effects linked with shadows cast by macroscopic roughness than those
occurring at the grain scale such as the SHOE and the CBOE \citep{Lucht2000}.
Furthermore, the radiative transfer algorithm that is used to calculate
the atmospheric quantities at the core of the MARS-ReCO procedure
cannot propagate properly through the atmosphere the narrow backscattering
lobes of the SHOE and CBOE. Finally, CRISM orbital measurements never
reach the small phase angle domain ($\lesssim5\text{\textdegree}$)
where the previous phenomena are expressed. Consequently, MARS-ReCO
is never in the position of retrieving accurate values for $B_{0}$
and $h$.

\item Macroscopic roughness factor $S$: We note that planetary regoliths
present roughness driven by grain clusters to the pixel scale. In
the Hapke's surface model this phenomena is described by a Gaussian
distribution of slopes at a single spatial scale under the pixel size
which is not explicitly given. The mean slope angle $\bar{\theta}$
is the only required parameter \citep{Hapke1993}. Surface roughness
involves several radiative phenomena: (i) multiple reflection of light
between facets, (ii) shadows depending on the geometry, (iii) bias
on the incidence and emergence angles, and (iv) increase of the multiple
scattering component. In order to quantify their influence on the
bidirectional reflectance, Hapke's model introduces a simple multiplicative
factor $S$ depending on $\mu_{0e}$ and $\mu_{e}$ whose expressions are given in \citet{Hapke1993}.

\end{itemize}

\subsubsection{Bayesian inversion of the surface model\label{sub:Bayesian-inversion-of}}

\begin{figure}
\centering\includegraphics[scale=0.18]{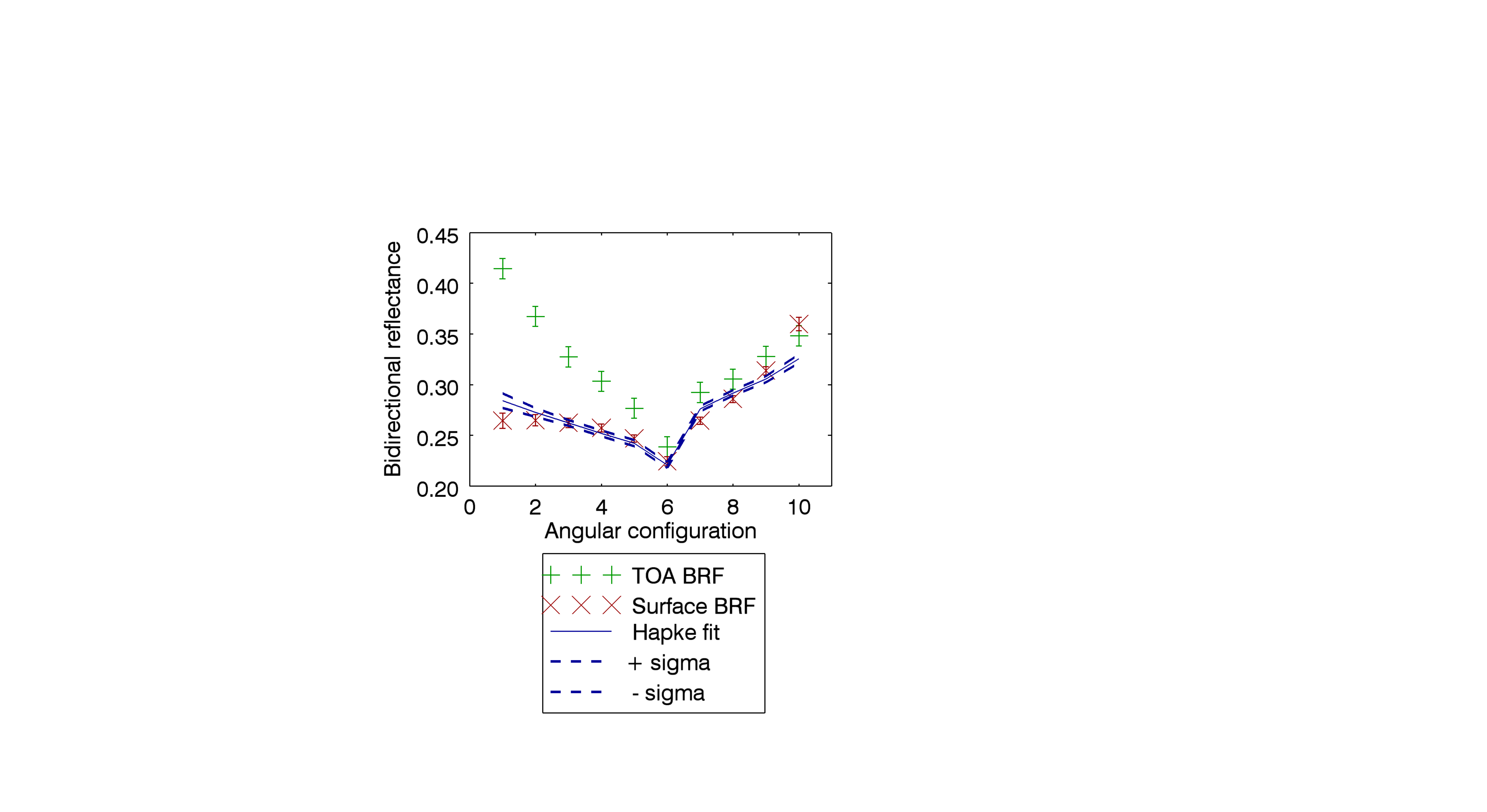}

\caption{Reflectance values (in BRF units) corresponding to the photometric
curve of the ROI I from FRT3192 observation composed of ten angular configurations
at the TOA (green plus), and at the surface after the correction for
atmospheric contribution carried out by MARS-ReCO (red crosses). In
both cases the $1\sigma$ uncertainty is expressed as a function of
angular configuration (the eleven geometric views of one CRISM EPF scans  
from 1st to 5th corresponds to the inbound direction, 6th is 
the central scan and from 7th to 11th are related to the outbound direction). 
The latter data are used in the Bayesian inversion. The blue solid and dashed lines represent
the Hapke's best fit and their $1\sigma$ uncertainties, respectively.
\label{fig:BRF scheme}}
\end{figure}

The ``inverse problem'' consists in estimating the model
parameters that best explain the observations. Unfortunately, inverse
problems do not have a unique solution if the direct model is non-linear, as does the Hapke's model. \citet{tarantola1982} proposed
to solve inverse problems in a general non-linear case based on the
concept of the state of information which is characterized by a Probability
Density Function (PDF). The PDF is defined over both the parameter space
and the observed space. The formalism of a PDF is used to define the
initial state of information (i.e., a priori knowledge on the parameters,
the uncertainties on the observation and on the model). To infer the
solution, the Bayes' theorem is applied. Key points concerning 
the bayesian inversion concept and framework assumptions are presented in the following:

\begin{itemize}

\item Data, model parameters and theoretical relationship: The direct model
consists of computing the simulated data $d$, from model parameters
$m$: 

\begin{equation}
d=F(m)
\end{equation}

\item Prior information on the model: The prior
information on model parameters $\rho_{m}(m)$
in the parameter space $(M)$ is independent with the data and corresponds
to the state of null information. For the Hapke model parameters
$\omega$, $b$, $c$, $\bar{\theta}$, $B_{0}$
and $h$, we consider a uniform PDF, different from zero on an interval
that insures their physical relevance (between $0$ and $1$ for $\omega$,
$b$, $c$, $B_{0}$ and $h$ and between $0{^\circ}$ and $90{^\circ}$
for $\bar{\theta}$). Outside the intervals, the PDF is null, avoiding
unphysical solution to appear. As discussed in Subsection \ref{sub:Direct-surface-model:}
in real planetary situations, the SHOE and the CBOE phenomena are
only expressed for phase angles $g$ $\lesssim5\text{\textdegree}$ out
of the range encompassed by typical CRISM observations ($g$ $\gtrsim30\text{\textdegree}$).
Thus a priori, neglecting both phenomena should not influence the
retrieval of the other parameters . However following \citet{souchon2011},
the model can still be profitably inverted on the data by keeping
parameters $B_{0}$ and $h$ since they will compensate for discrepancies
between the model and the measurements in some situations. Consequently, in the present work, 
we first chose to invert the parameters $B_{0}$ and $h$ 
to systematically control if they are constrained or not by the CRISM data set. However in a second phase we also tested the inversion by setting $B_{0}$ and $h$ to zero and no change was observed on the determination of the other parameters 
($\omega$, $\bar{\theta}$, $b$, $c$).

\item Prior information on the data: The prior information on
data $\rho_{d}(d)$ in the observation space
$(D)$ is assumed to be a Gaussian PDF according to the MARS-ReCO
formalism and retrieval strategy. Note that the error $\sigma$ on a CRISM
measurement at one geometry is assumed to be independent on the state
of the surface and on the other geometries (i.e., the PDF has a diagonal
covariance matrix with elements $\sigma_{1}^{2},\ldots,\sigma_{N_{g}}^{2}$, 
where $N_{g}$ is the number of available geometries, up to 11). At 750 nm, the signal to noise ratio was estimated before launch to be equal to 450 
\citep{murchie2007} but due to additional artifacts such as spikes, and calibration issues 
\citep{Seelos2011}  we evaluated the uncertainty $\sigma$ of the reflectance measurement ($\rho$) 
at each geometry $j$ to be of the order of $\sigma_{j}=(1/50)\times\rho_{j}$ 
 where $j$=1,...,$N_{g}$ and $\rho_{j}$ is the CRISM dataset at the $j^{th}$ angular configuration \citep{Ceamanos2012}. 
Moreover, MARS-ReCO takes into account the uncertainty of the AOT
input. Figure \ref{fig:BRF scheme} presents a typical TOA photometric
curve collected by the CRISM instrument (green plus) and the bidirectional
reflectance curve produced by the MARS-ReCO algorithm (red crosses).
For each geometry the bidirectional reflectance value is accompanied
by its $1\sigma$ uncertainty. Those means and root mean square errors
are used to build $\rho_{d}(d)$ that serves
as an input PDF of the Bayesian inversion.

\item Posterior probability density function and resolution of inverse problems:
Inversion problems correspond to the particular case where information
from the data space $(D)$ is translated into the model space $(M)$.
The posterior PDF in the model space $\sigma_{M}(m)$
as defined by \citet{tarantola1982} reads

\begin{equation}
\sigma_{M}(m)=k\,\rho_{M}(m)\, L(m),
\end{equation}

where $k$ is a constant and $L(m)$ is the likelihood function

\begin{equation}
L(m)=\int_{D}\partial d\,\frac{\rho_{D}(d)\,\theta(d\mid m)}{\mu_{D}(d)},
\end{equation}

where $\theta(d\mid m)$ is the theoretical relationship of the PDF
for $d$ given $m$, and $\mu_{D}(d)$ is null
information PDF for the data $d$. The solution of the general inverse
problem is given by the PDF $\sigma_{M}(m)$
from which any information on the model parameter can be obtained such as
mean values, uncertainty bars, etc . Please refer to \citet{tarantola1982} for
more detail. 

\item Sampling of solutions to inverse problems: In our case, the relationship
between model parameters and observed data through Hapke's modeling
is non-linear. While it is not possible to describe the posterior
PDF analytically, it can be sampled using a Monte Carlo approach using
a Markov chain \citep{Mosegaard1995}. After a sufficient number of
steps, the state of the chain corresponds to the desired distribution.
According our tests, the best trade-off between computation time and
accuracy is a burn-in phase (phase in which the Markov chain approaches 
a stationary state after a certain number of runs) of 500 steps. The next 500 iterations
are used to estimate the posterior PDF allowing the determination
of the mean and standard deviation of each parameter. Note that a
posteriori PDF of a retrieved parameter is not necessarily a gaussian
distribution but can be a square function or multi-modal as seen in
Figures \ref{fig:Probability-Density-Function gusev} and \ref{fig:Probability-Density-Function meridiani}. To describe the results of each parameter, we choose to compute the mean of the posterior PDF. To describe the uncertainties, we choose to compute the standard deviation. We warn the reader because sometimes these estimators can be inappropriate to describe the PDF. In the following graphs, $2\sigma$ error bars are plotted to describe more accurately the PDFs.  

\item Root Mean Square residual $RMS$: In order to estimate the
difference between the fit and the observed bidirectional 
reflectance, the Root Mean Square residual (noted  $RMS$) is given for each Bayesian inversion as follows:

\begin{equation}
RMS=\sqrt{\frac{\sum_{N_{g}}\left(\rho_{obs}-\rho_{mod}\right)^{2}}{N_{g}}}
\end{equation}

where $N$ is the available geometric configurations, $\rho_{obs}$
the CRISM bidirectional reflectance corrected for atmosphere and $\rho_{mod}$
the modeled bidirectional reflectance taken as the mean of the 500
iterations used to estimate the posterior PDF.

\item Non-uniformity criterion $k$: Photometric parameters \emph{$m$}
are constrained if their marginal posterior PDF differs from the prior
state of information (i.e., a null information taken as a uniform distribution,
in our case). In order to distinguish if a given parameter has a solution
we perform a statistical test leading to a non-uniformity criterion
$k$ (see Appendix A). For $k>0.5$, the marginal posterior PDF is
considered to be non-uniform and thus we consider that the mean and
standard deviation of the PDF satisfactorily describe the solution(s). 

\end{itemize}

\section{Analysis of retrieved photometric parameters \label{sec:Results-of-retrieved}}

As mentioned in Subsection \ref{sub:Selection-of-FRT}, we selected
three CRISM targeted observations from Gusev Crater and from Meridiani
Planum (cf. Table \ref{tab:Selected-CRISM-observations}). All observations
were individually corrected for aerosols using MARS-ReCO,
knowing the respective AOT values (cf. Table \ref{tab:Selected-CRISM-observations}).
Based on the corrected bidirectional reflectance, we use the proposed
methodology described in Subsection \ref{sub:Bayesian-inversion-of}
to estimate the photometric parameters of the materials encompassed
by four ROIs in the case of Gusev Crater (ROI I to ROI IV) and
by one ROI in the case of Meridiani Planum (ROI I). 

For each ROI, we determine the surface photometric parameters at 750
nm from (i) a single CRISM FRT observation or (ii) a combination of
all selected FRT observations (i.e., FRT3192, FRT8CE1 and FRTCDA5
in the case of Gusev Crater, and FRT95B8, FRT334D and FRTB6B5 in the
case of Meridiani Planum). Remember that the chosen targeted observations
are complementary in terms of phase angle range (cf. Table \ref{tab:Selected-CRISM-observations}).

For each parameter of the photometric Hapke's model (i.e., $\omega$,
$b$, $c$, $\bar{\theta}$, $B_{0}$ and $h$) we determine its mean
value and standard deviation by running the proposed Bayesian inversion
procedure. The non-uniform criterion noted $k$ is conjointly computed
and detailed in Tables \ref{tab:Retrieved-photometric-parameter gusev1} and \ref{tab:Retrieved-photometric-parameter gusev2}
for Gusev Crater and Table \ref{tab:Retrieved-photometric-parameter meridiani}
for Meridiani Planum. In the following, we have two criteria that
help us estimate the existence and the quality of a solution. On the
one hand, the non-uniform criteria $k$ allows us to reject posterior
PDF's that do not carry a solution. On the other hand, the standard
deviation allows us to quantify to which extent a given parameter
is constrained by the current solution.

\subsection{Results on Gusev Crater \label{sub:On-Gusev-Crater}}

\begin{table}
\begin{raggedright}
\caption{Retrieved Hapke's parameters ($\omega$, $b$, $c$, $\bar{\theta}$ in degree, $B_0$, $h$) and their standard deviation in parentheses 
(the underconstrained parameters are indicated by "(-)") corresponding to the ROI I and II
from CRISM measurement at 750 nm using our Bayesian inversion as well as the non-uniform criterion $k$ (second line of each entry) after MARS-ReCO. This analysis is done for each CRISM FRT observation (i.e.,
FRT3192, FRT8CE1 and FRTCDA5) or for the combination of the three
observations after correction for the atmospheric effects by MARS-ReCO.
The standard deviation {$\sigma_{\rho}$} (in BRF units)
is a parameter provided by MARS-ReCO estimating the quality of the
retrieved surface bidirectional reflectance. The absolute quadratic residual of the fit $RMS$ (in 
BRF units) evaluates the deviation of the fit from observed bidirectional reflectance. ($nb$: number of angular configurations  $g$: phase angle range in degree) \label{tab:Retrieved-photometric-parameter gusev1} }
\centering{\scriptsize }%
\begin{tabular}{ccccccccccccc}
\hline 
{\scriptsize FRT} & {\scriptsize Lat. } & {\scriptsize Lon.} & \multicolumn{1}{c}{{\scriptsize AOT}} & {\scriptsize $\sigma_{\rho}$} & {\scriptsize $\omega$} & {\scriptsize $b$} & {\scriptsize $c$} & {\scriptsize $\bar{\theta}$ } & {\scriptsize $B_{0}$} & {\scriptsize $h$} & {\scriptsize $nb$} & {\scriptsize $g$}\tabularnewline
 & {\scriptsize (\textdegree{}N)} & {\scriptsize{} (\textdegree{}E)} & {\scriptsize} & {\scriptsize $RMS$}  & {\scriptsize $k$} & {\scriptsize $k$} & {\scriptsize $k$} & {\scriptsize $k$} & {\scriptsize $k$} & {\scriptsize $k$} &  & {\scriptsize}\tabularnewline
\hline 
\multirow{2}{*}{{\scriptsize 3192}} & \multirow{2}{*}{{\scriptsize -14.603}} & \multirow{2}{*}{{\scriptsize 175.478}} & \multirow{2}{*}{{\scriptsize 0.33$\pm$0.04}} & {\scriptsize 0.01} & {\scriptsize 0.68 (0.06)} & {\scriptsize 0.17 (0.20)} & {\scriptsize 0.62 (0.20)} & {\scriptsize 11.62 (3.98)} & {\scriptsize 0.52 (-)} & {\scriptsize 0.52 (-)} & \multirow{2}{*}{{\scriptsize 10}} & \multirow{2}{*}{{\scriptsize $\sim$56-112}}\tabularnewline
 &  &  &  & {\scriptsize 0.01} & {\scriptsize 1.00} & {\scriptsize 1.92} & {\scriptsize 0.96} & {\scriptsize 1.00} & {\scriptsize 0.24} & {\scriptsize 0.12} &  & \tabularnewline
\cline{5-13} 
\multirow{2}{*}{{\scriptsize 8CE1}} & \multirow{2}{*}{{\scriptsize -14.606}} & \multirow{2}{*}{{\scriptsize 175.478}} & \multirow{2}{*}{{\scriptsize 0.98$\pm$0.15}} & {\scriptsize 0.03} & {\scriptsize 0.77 (0.07)} & {\scriptsize 0.30 (0.28)} & {\scriptsize 0.42 (0.27)} & {\scriptsize 17.23 (9.64)} & {\scriptsize 0.47 (-)} & {\scriptsize 0.50 (-)} & \multirow{2}{*}{{\scriptsize 9}} & \multirow{2}{*}{{\scriptsize $\sim$37-79}}\tabularnewline
 &  &  &  & {\scriptsize 0.01} & {\scriptsize 1.00} & {\scriptsize 1.47} & {\scriptsize 0.60} & {\scriptsize 0.98} & {\scriptsize 0.20} & {\scriptsize 0.12} &  & \tabularnewline
\cline{5-13} 
\multirow{2}{*}{{\scriptsize CDA5}} & \multirow{2}{*}{{\scriptsize -14.604}} & \multirow{2}{*}{{\scriptsize 175.481}} & \multirow{2}{*}{{\scriptsize 0.32$\pm$0.04}} & {\scriptsize 0.01} & {\scriptsize 0.78 (0.08)} & {\scriptsize 0.44 (0.32)} & {\scriptsize 0.54 (0.23)} & {\scriptsize 16.71 (4.64)} & {\scriptsize 0.51 (-)} & {\scriptsize 0.52 (-)} & \multirow{2}{*}{{\scriptsize 8}} & \multirow{2}{*}{{\scriptsize $\sim$46-107}}\tabularnewline
 &  &  &  & {\scriptsize 0.00} & {\scriptsize 0.99} & {\scriptsize 0.91} & {\scriptsize 0.70} & {\scriptsize 1.00} & {\scriptsize 0.05} & {\scriptsize 0.30} &  & \tabularnewline
\cline{5-13} 
\multirow{2}{*}{{\scriptsize ROI I - 3 FRTs}} & \multirow{2}{*}{{\scriptsize -}} & \multirow{2}{*}{{\scriptsize -}} & \multirow{2}{*}{{\scriptsize -}} & {\scriptsize -} & {\scriptsize 0.71 (0.05)} & {\scriptsize 0.22 (0.18)} & {\scriptsize 0.54 (0.18)} & {\scriptsize 15.78 (2.65)} & {\scriptsize 0.49 (-)} & {\scriptsize 0.54 (-)} & \multirow{2}{*}{{\scriptsize 27}} & \multirow{2}{*}{{\scriptsize $\sim$37-112}}\tabularnewline
 &  &  &  & {\scriptsize 0.01} & {\scriptsize 1.00} & {\scriptsize 1.45} & {\scriptsize 0.95} & {\scriptsize 1.00} & {\scriptsize 0.28} & {\scriptsize 0.23} &  & \tabularnewline
\hline 
\end{tabular}
\par\end{raggedright}{\scriptsize \par}

\begin{raggedright}
\centering{\scriptsize }%
\begin{tabular}{ccccccccccccc}
\hline 
{\scriptsize FRT} & {\scriptsize Lat. } & {\scriptsize Lon.} & \multicolumn{1}{c}{{\scriptsize AOT}} & {\scriptsize $\sigma_{\rho}$} & {\scriptsize $\omega$} & {\scriptsize $b$} & {\scriptsize $c$} & {\scriptsize $\bar{\theta}$ } & {\scriptsize $B_{0}$} & {\scriptsize $h$} & {\scriptsize $nb$} & {\scriptsize $g$}\tabularnewline
 & {\scriptsize (\textdegree{}N)} & {\scriptsize{} (\textdegree{}E)} & {\scriptsize} & {\scriptsize $RMS$} & {\scriptsize $k$} & {\scriptsize $k$} & {\scriptsize $k$} & {\scriptsize $k$} & {\scriptsize $k$} & {\scriptsize $k$} &  & {\scriptsize}\tabularnewline
\hline 
\multirow{2}{*}{{\scriptsize 3192}} & \multirow{2}{*}{{\scriptsize -14.599}} & \multirow{2}{*}{{\scriptsize 175.499}} & \multirow{2}{*}{{\scriptsize 0.33$\pm$0.04}} & {\scriptsize 0.01} & {\scriptsize 0.72 (0.08)} & {\scriptsize 0.29 (0.31)} & {\scriptsize 0.59 (0.20)} & {\scriptsize 12.32 (5.26)} & {\scriptsize 0.52 (-)} & {\scriptsize 0.52 (-)} & \multirow{2}{*}{{\scriptsize 9}} & \multirow{2}{*}{{\scriptsize $\sim$55-112}}\tabularnewline
 &  &  &  & {\scriptsize 0.01}  & {\scriptsize 1.00} & {\scriptsize 1.69} & {\scriptsize 0.92} & {\scriptsize 1.00} & {\scriptsize 0.23} & {\scriptsize 0.16} &  & \tabularnewline
\cline{5-13} 
\multirow{2}{*}{{\scriptsize 8CE1}} & \multirow{2}{*}{{\scriptsize -14.603}} & \multirow{2}{*}{{\scriptsize 175.500}} & \multirow{2}{*}{{\scriptsize 0.98$\pm$0.15}} & {\scriptsize 0.03} & {\scriptsize 0.76 (0.08)} & {\scriptsize 0.34 (0.29)} & {\scriptsize 0.44 (-)} & {\scriptsize 17.58 (9.91)} & {\scriptsize 0.48 (-)} & {\scriptsize 0.48 (-)} & \multirow{2}{*}{{\scriptsize 7}} & \multirow{2}{*}{{\scriptsize $\sim$36-90}}\tabularnewline
 &  &  &  &  {\scriptsize 0.02} & {\scriptsize 0.99} & {\scriptsize 1.42} & {\scriptsize 0.41} & {\scriptsize 0.98} & {\scriptsize 0.08} & {\scriptsize 0.09} &  & \tabularnewline
\cline{5-13} 
\multirow{2}{*}{{\scriptsize CDA5}} & \multirow{2}{*}{{\scriptsize -14.602}} & \multirow{2}{*}{{\scriptsize 175.498}} & \multirow{2}{*}{{\scriptsize 0.32$\pm$0.04}} & {\scriptsize 0.01} & {\scriptsize 0.74 (0.07)} & {\scriptsize 0.37 (0.28)} & {\scriptsize 0.65 (0.20)} & {\scriptsize 14.62 (4.97)} & {\scriptsize 0.53 (-)} & {\scriptsize 0.51 (-)} & \multirow{2}{*}{{\scriptsize 8}} & \multirow{2}{*}{{\scriptsize $\sim$46-107}}\tabularnewline
 &  &  &  & {\scriptsize 0.01} & {\scriptsize 1.00} & {\scriptsize 1.06} & {\scriptsize 0.81} & {\scriptsize 1.00} & {\scriptsize 0.14} & {\scriptsize 0.10} &  & \tabularnewline
\cline{5-13} 
\multirow{2}{*}{{\scriptsize ROI II - 3 FRTs}} & \multirow{2}{*}{{\scriptsize -}} & \multirow{2}{*}{{\scriptsize -}} & \multirow{2}{*}{{\scriptsize -}} & {\scriptsize -} & {\scriptsize 0.72 (0.05)} & {\scriptsize 0.27 (0.15)} & {\scriptsize 0.56 (0.16)} & {\scriptsize 15.62 (2.43)} & {\scriptsize 0.49 (-)} & {\scriptsize 0.52 (-)} & \multirow{2}{*}{{\scriptsize 24}} & \multirow{2}{*}{{\scriptsize $\sim$36-112}}\tabularnewline
 &  &  &  & {\scriptsize 0.01} & {\scriptsize 1.00} & {\scriptsize 1.18} & {\scriptsize 0.99} & {\scriptsize 1.00} & {\scriptsize 0.05} & {\scriptsize 0.23} &  & \tabularnewline
\hline 
\end{tabular}
\par\end{raggedright}{\scriptsize \par}

\raggedright{}\centering
\end{table}

\begin{table}
\caption{Same as Table \ref{tab:Retrieved-photometric-parameter gusev1} but for the ROI III and ROI IV \label{tab:Retrieved-photometric-parameter gusev2}}

\begin{raggedright}
\centering{\scriptsize }%
\begin{tabular}{ccccccccccccc}
\hline 
{\scriptsize FRT} & {\scriptsize Lat. } & {\scriptsize Lon.} & \multicolumn{1}{c}{{\scriptsize AOT}} & {\scriptsize $\sigma_{\rho}$} & {\scriptsize $\omega$} & {\scriptsize $b$} & {\scriptsize $c$} & {\scriptsize $\bar{\theta}$ } & {\scriptsize $B_{0}$} & {\scriptsize $h$} & {\scriptsize $nb$} & {\scriptsize $g$}\tabularnewline
 & {\scriptsize (\textdegree{}N)} & {\scriptsize{} (\textdegree{}E)} & {\scriptsize} & {\scriptsize $RMS$} & {\scriptsize $k$} & {\scriptsize $k$} & {\scriptsize $k$} & {\scriptsize $k$} & {\scriptsize $k$} & {\scriptsize $k$} &  & {\scriptsize}\tabularnewline
\hline 
\multirow{2}{*}{{\scriptsize 3192}} & \multirow{2}{*}{{\scriptsize -14.593}} & \multirow{2}{*}{{\scriptsize 175.496}} & \multirow{2}{*}{{\scriptsize 0.33$\pm$0.04}} & {\scriptsize 0.01} & {\scriptsize 0.72 (0.08)} & {\scriptsize 0.32 (0.29)} & {\scriptsize 0.58 (0.20)} & {\scriptsize 10.53 (5.81)} & {\scriptsize 0.50 (-)} & {\scriptsize 0.50 (-)} & \multirow{2}{*}{{\scriptsize 7}} & \multirow{2}{*}{{\scriptsize $\sim$56-108}}\tabularnewline
 &  &  &  & {\scriptsize 0.01}  & {\scriptsize 1.00} & {\scriptsize 1.12} & {\scriptsize 0.94} & {\scriptsize 1.00} & {\scriptsize 0.15} & {\scriptsize 0.10} &  & \tabularnewline
\cline{5-13} 
\multirow{2}{*}{{\scriptsize 8CE1}} & \multirow{2}{*}{{\scriptsize -14.596}} & \multirow{2}{*}{{\scriptsize 175.494}} & \multirow{2}{*}{{\scriptsize 0.98$\pm$0.15}} & {\scriptsize 0.03} & {\scriptsize 0.77 (0.09)} & {\scriptsize 0.35 (0.29)} & {\scriptsize 0.44 (-)} & {\scriptsize 19.13 (10.52)} & {\scriptsize 0.49 (-)} & {\scriptsize 0.51 (-)} & \multirow{2}{*}{{\scriptsize 7}} & \multirow{2}{*}{{\scriptsize $\sim$37-84}}\tabularnewline
 &  &  &  & {\scriptsize 0.01} & {\scriptsize 0.99} & {\scriptsize 1.36} & {\scriptsize 0.39} & {\scriptsize 0.98} & {\scriptsize 0.02} & {\scriptsize 0.06} &  & \tabularnewline
\cline{5-13} 
\multirow{2}{*}{{\scriptsize CDA5}} & \multirow{2}{*}{{\scriptsize -14.594}} & \multirow{2}{*}{{\scriptsize 175.497}} & \multirow{2}{*}{{\scriptsize 0.32$\pm$0.04}} & {\scriptsize 0.01} & {\scriptsize 0.72 (0.07)} & {\scriptsize 0.26 (0.25)} & {\scriptsize 0.65 (0.20)} & {\scriptsize 14.21 (5.59)} & {\scriptsize 0.51 (-)} & {\scriptsize 0.54 (-)} & \multirow{2}{*}{{\scriptsize 7}} & \multirow{2}{*}{{\scriptsize $\sim$46-106}}\tabularnewline
 &  &  &  & {\scriptsize 0.01} & {\scriptsize 1.00} & {\scriptsize 1.67} & {\scriptsize 0.85} & {\scriptsize 1.00} & {\scriptsize 0.15} & {\scriptsize 0.24} &  & \tabularnewline
\cline{5-13} 
\multirow{2}{*}{{\scriptsize ROI III - 3 FRTs}} & \multirow{2}{*}{{\scriptsize -}} & \multirow{2}{*}{{\scriptsize -}} & \multirow{2}{*}{{\scriptsize -}} & {\scriptsize -} & {\scriptsize 0.69 (0.04)} & {\scriptsize 0.19 (0.14)} & {\scriptsize 0.66 (0.18)} & {\scriptsize 13.96 (4.40)} & {\scriptsize 0.50 (-)} & {\scriptsize 0.53 (-)} & \multirow{2}{*}{{\scriptsize 21}} & \multirow{2}{*}{{\scriptsize $\sim$37-108}}\tabularnewline
 &  &  &  & {\scriptsize 0.01} & {\scriptsize 1.00} & {\scriptsize 1.43} & {\scriptsize 0.87} & {\scriptsize 1.00} & {\scriptsize 0.08} & {\scriptsize 0.11} &  & \tabularnewline
\hline 
\end{tabular}
\par\end{raggedright}{\scriptsize \par}

\raggedright{}\centering{\scriptsize }%
\begin{tabular}{ccccccccccccc}
\hline 
\multirow{1}{*}{{\scriptsize FRT}} & {\scriptsize Lat. } & {\scriptsize Lon.} & \multicolumn{1}{c}{{\scriptsize AOT}} & {\scriptsize $\sigma_{\rho}$} & {\scriptsize $\omega$} & {\scriptsize $b$} & {\scriptsize $c$} & {\scriptsize $\bar{\theta}$ } & {\scriptsize $B_{0}$} & {\scriptsize $h$} & \multicolumn{1}{c}{{\scriptsize $nb$}} & {\scriptsize $g$}\tabularnewline
 & {\scriptsize (\textdegree{}N)} & {\scriptsize{} (\textdegree{}E)} & {\scriptsize{}} & {\scriptsize $RMS$} & {\scriptsize $k$} & {\scriptsize $k$} & {\scriptsize $k$} & {\scriptsize $k$} & {\scriptsize $k$} & {\scriptsize $k$} &  & {\scriptsize}\tabularnewline
\hline 
\multirow{2}{*}{{\scriptsize 3192}} & \multirow{2}{*}{{\scriptsize -14.600}} & \multirow{2}{*}{{\scriptsize 175.507}} & \multirow{2}{*}{{\scriptsize 0.33$\pm$0.04}} & {\scriptsize 0.01} & {\scriptsize 0.68 (0.06)} & {\scriptsize 0.17 (0.21)} & {\scriptsize 0.61 (0.22)} & {\scriptsize 11.74 (4.43)} & {\scriptsize 0.53 (-)} & {\scriptsize 0.56 (-)} & \multirow{2}{*}{{\scriptsize 9}} & \multirow{2}{*}{{\scriptsize $\sim$56-112}}\tabularnewline
 &  &  &  & {\scriptsize 0.01} & {\scriptsize 1.00} & {\scriptsize 1.80} & {\scriptsize 0.97} & {\scriptsize 1.00} & {\scriptsize 0.28} & {\scriptsize 0.39} &  & \tabularnewline
\cline{5-13} 
\multirow{2}{*}{{\scriptsize 8CE1}} & \multirow{2}{*}{{\scriptsize -14.602}} & \multirow{2}{*}{{\scriptsize 175.508}} & \multirow{2}{*}{{\scriptsize 0.98$\pm$0.15}} & {\scriptsize 0.04} & {\scriptsize 0.78 (0.09)} & {\scriptsize 0.38 (0.29)} & {\scriptsize 0.47 (-)} & {\scriptsize 20.40 (11.41)} & {\scriptsize 0.51 (-)} & {\scriptsize 0.50 (-)} & \multirow{2}{*}{{\scriptsize 6}} & \multirow{2}{*}{{\scriptsize $\sim$41-84}}\tabularnewline
 &  &  &  & {\scriptsize 0.01} & {\scriptsize 0.99} & {\scriptsize 0.80} & {\scriptsize 0.24} & {\scriptsize 0.97} & {\scriptsize 0.22} & {\scriptsize 0.12} &  & \tabularnewline
\cline{5-13} 
\multirow{2}{*}{{\scriptsize CDA5}} & \multirow{2}{*}{{\scriptsize -14.600}} & \multirow{2}{*}{{\scriptsize 175.509}} & \multirow{2}{*}{{\scriptsize 0.32$\pm$0.04}} & {\scriptsize 0.01} & {\scriptsize 0.74 (0.07)} & {\scriptsize 0.35 (0.27)} & {\scriptsize 0.65 (0.20)} & {\scriptsize 14.55 (5.48)} & {\scriptsize 0.51 (-)} & {\scriptsize 0.50 (-)} & \multirow{2}{*}{{\scriptsize 8}} & \multirow{2}{*}{{\scriptsize $\sim$46-107}}\tabularnewline
 &  &  &  & {\scriptsize 0.01} & {\scriptsize 1.00} & {\scriptsize 1.10} & {\scriptsize 0.79} & {\scriptsize 1.00} & {\scriptsize 0.14} & {\scriptsize 0.09} &  & \tabularnewline
\cline{5-13} 
\multirow{2}{*}{{\scriptsize ROI IV - 3 FRTs}} & \multirow{2}{*}{{\scriptsize -}} & \multirow{2}{*}{{\scriptsize -}} & \multirow{2}{*}{{\scriptsize -}} & {\scriptsize -} & {\scriptsize 0.79 (0.07)} & {\scriptsize 0.59 (0.27)} & {\scriptsize 0.56 (0.16)} & {\scriptsize 10.88 (5.11)} & {\scriptsize 0.50 (-)} & {\scriptsize 0.48 (-)} & \multirow{2}{*}{{\scriptsize 23}} & \multirow{2}{*}{{\scriptsize $\sim$41-112}}\tabularnewline
 &  &  &  & {\scriptsize 0.01} & {\scriptsize 1.00} & {\scriptsize 0.53} & {\scriptsize 0.97} & {\scriptsize 1.00} & {\scriptsize 0.09} & {\scriptsize 0.05} &  & \tabularnewline
\hline 
\end{tabular}
\end{table}

The quality of the surface bidirectional reflectance estimated by
MARS-ReCO is given by the standard deviation $\sigma_{\rho}$.
This quality parameter is available for each ROI of each CRISM targeted
observation of the present study (cf. Tables \ref{tab:Retrieved-photometric-parameter gusev1}
and \ref{tab:Retrieved-photometric-parameter gusev2}). The
highest {\footnotesize $\sigma_{\rho}$} value is observed for FRT8CE1
which can be explained by a wrong AOT estimation (highest AOT value) in this case (i.e.,
FRT8CE1: $\sigma_{\rho}$ = 0.03-0.04 for AOT = 0.98). The positive correlation of this uncertainty and the AOT is also observed in the sensitivity study led by \citet{Ceamanos2012}.
Indeed the error computed for 
synthetic reference data mimicking the photometric properties of the planet Mars increases 
with AOT (i.e., for AOT = 1, $\sigma_{\rho}\sim$ 0.05 and error is $\sim$10\% while for 
AOT = 1.5, $\sigma_{\rho}\sim$ 0.10 and 
error is $\sim$20\%, see \citet{Ceamanos2012} for more detail).

Tables \ref{tab:Retrieved-photometric-parameter gusev1} and 
\ref{tab:Retrieved-photometric-parameter gusev2} present results
regarding the Hapke's model parameters for the different ROIs of Gusev Crater.

The goodness of fit is estimated through the absolute quadratic residual $RMS$ value 
(cf. Tables \ref{tab:Retrieved-photometric-parameter gusev1} and \ref{tab:Retrieved-photometric-parameter gusev2}). For all Bayesian inversions, the estimates are less than 0.02 which mean that the inversions are deemed satisfactory.

\begin{figure}
\centering\includegraphics[scale=0.48]{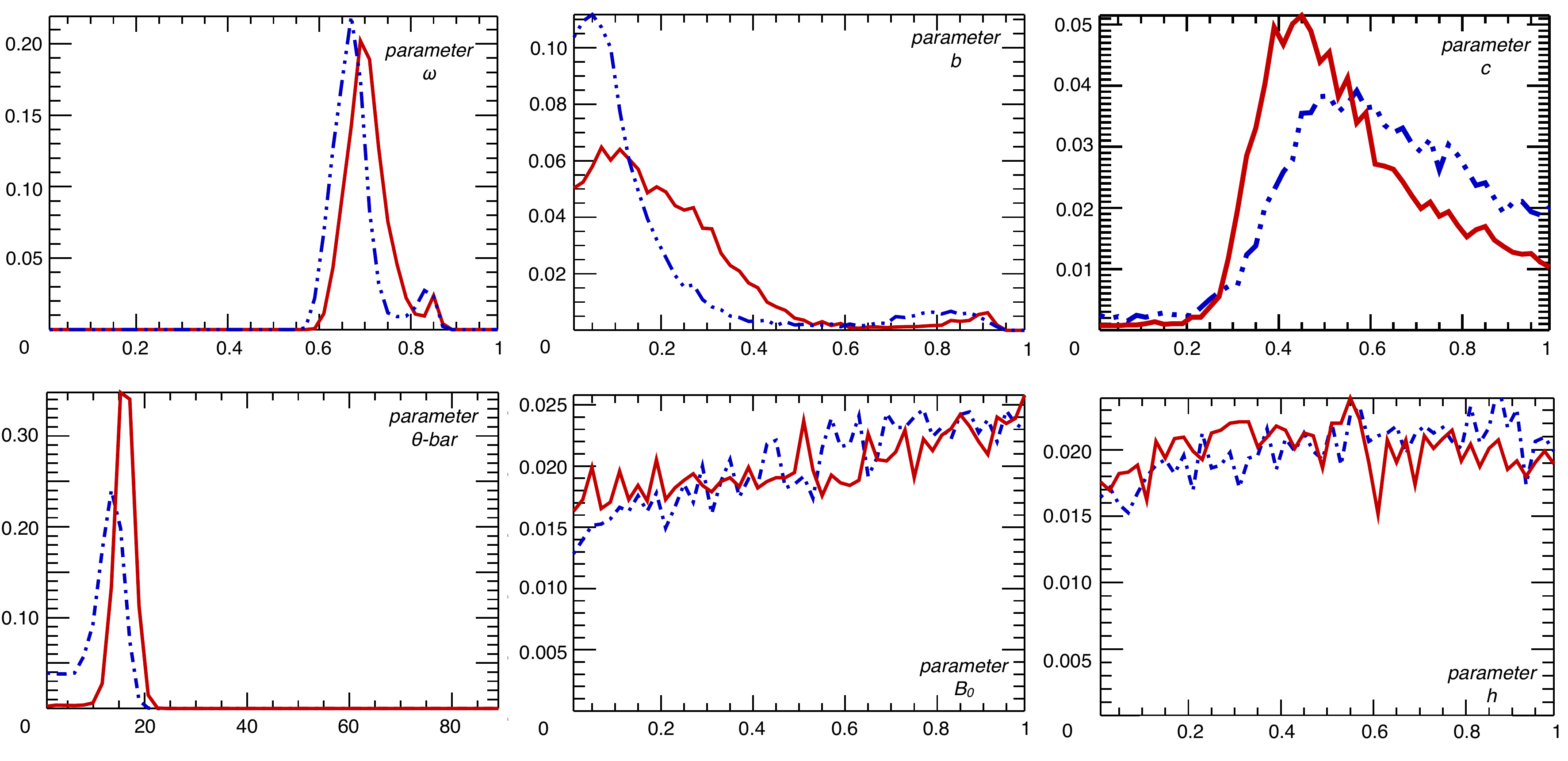}

\caption{The PDF of each of the six physical parameters retrieved for ROI I.
Two cases are considered: the inversion of a single CRISM FRT observation
(i.e., FRT3192 in blue line) and the inversion of three combined CRISM
FRT observations (in red line). \label{fig:Probability-Density-Function gusev}}
\end{figure}

Figure \ref{fig:Probability-Density-Function gusev} represents the
PDF of each parameter considering the inversion of a single FRT observation
as well as the inversion of the three combined observations. According
to results, a solution exists for parameters $\omega$, $b$, $c$
and $\bar{\theta}$ (because the PDF is non-uniform) whereas no solution exists for parameters $B_{0}$ and $h$ (uniform PDF).
Several conclusions can be drawn when using a single targeted observation:

\begin{itemize}

\item Solutions exist for parameter $\omega$ in all
cases ($k\sim1$). The standard deviation shows that the single scattering
albedo is the most constrained parameter (i.e., 0.06 < $\sigma$ < 0.09). Examples of posteriori PDFs are plotted in Figures \ref{fig:Probability-Density-Function gusev} and \ref{fig:gusev-pancam w}.

\item Although solutions exist for parameter $b$ in all cases ($k$ >> 0.5), the 
standard deviation shows that it is poorly constrained (i.e., 0.25 < $\sigma$ < 0.32) 
except for the ROI I and ROI IV using the CRISM observation FRT3192. 
This discrepancy can be explained by the higher number of available 
geometric configurations, respectively 10 and 9. Examples of posteriori PDFs are plotted in Figures \ref{fig:Probability-Density-Function gusev} and \ref{fig:gusev-pancam b/c}.

\item We find meaningful values for parameter $c$ only when we use FRT3192
or FRTCDA5 ($k$ > 0.5). The standard deviation is relatively low (i.e.,
0.20 < $\sigma$ < 0.23) in this case. In the case
of FRT8CE1, no solution is found for parameter $c$ ($k$ < 0.5)
for ROI II, III and IV. For the ROI I, however, a solution exists
but it is poorly constrained (i.e., $\sigma$ = 0.27). Examples of posteriori PDFs are plotted in Figures \ref{fig:Probability-Density-Function gusev} and \ref{fig:gusev-pancam b/c}.

\item While parameter $\bar{\theta}$ has a solution in all cases ($k\sim1$),
we distinguish two types of results: (i) using FRT3192 or FRTCDA5,
we note that the standard deviation is relatively low (i.e., 3.98 < $\sigma$ < 5.81),
and (ii) using FRT8CE1, the standard deviation is relatively high
(i.e., 9.64 < $\sigma$ < 20.40) .Examples of posteriori PDFs are plotted in Figures \ref{fig:Probability-Density-Function gusev} and \ref{fig:gusev-pancam theta}.

\item No solutions are found for parameters $B_{0}$ and $h$ in any case
($k$ << 0.5). As mentioned in Subsection \ref{sub:Direct-surface-model:},
nor the phase domain covered by the CRISM observations nor the capabilities
of MARS-ReCO allow to constrain the opposition effect \citep{Ceamanos2012}.
Consequently, accurate values for $B_{0}$ and $h$ cannot be retrieved
and no physical interpretation can be done in this case. Examples of posteriori PDFs are plotted in Figure \ref{fig:Probability-Density-Function gusev}.

\end{itemize}

Regarding the processing of a single CRISM observation, we note that
parameters $c$ and $\bar{\theta}$ are, respectively, non-constrained
and poorly constrained only when treating data from FRT8CE1. The reason
to explain such a difference may be that the available maximum phase
angle range is less than 90\textdegree{} for FRT8CE1. Indeed, \citet{Helfenstein1988} 
underlined the necessity to have observations which extend from small phase 
angles out to phase angles above 90\textdegree{} for an accurate determination of the photometric roughness. 
By contrast, the phase angle range expands by more than 100\textdegree{} for the
other observations (cf. Table \ref{tab:Selected-CRISM-observations}).
We note that parameter $b$ is poorly constrained when treating data
from all available observations. This result may be explained again
by the available phase angle range (cf. Table \ref{tab:Selected-CRISM-observations}).
In conclusion, the presented study clearly demonstrates that the existence and
quality of a solution for parameters $b$, $c$ and $\bar{\theta}$
is dependent on the available phase angle range. The bidirectional
reflectance curve from a single CRISM observation does not contain enough phase angle information. 

\begin{figure}
\begin{centering}
\includegraphics[scale=0.40]{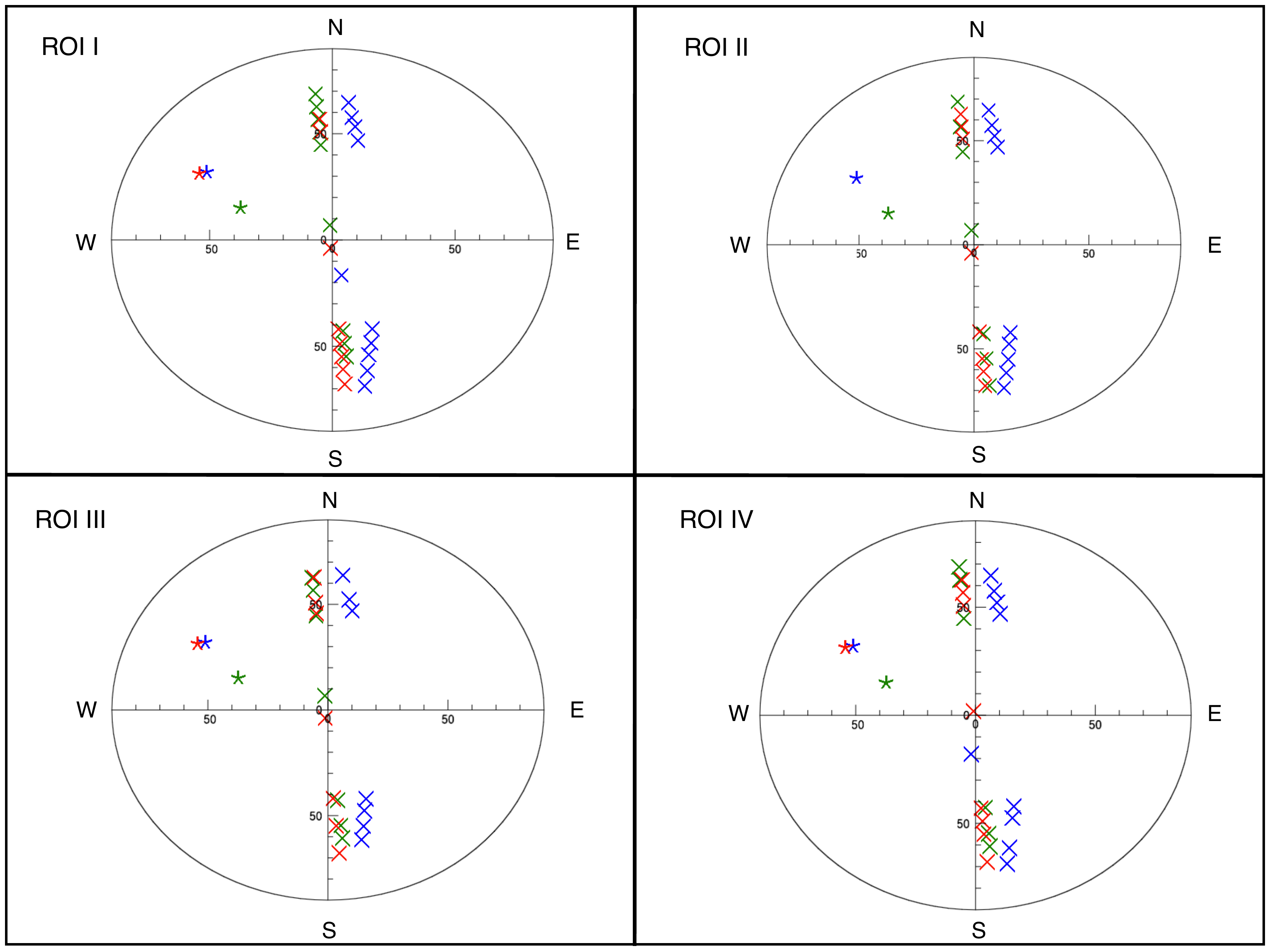}
\par\end{centering}

\caption{North projection of geometric conditions (stars: incidence rays and crosses: emergence
directions) of the four selected ROIs for the Gusev Crater study (blue: FRT3192, green: FRT8CE1 and red: FRTCDA5). \label{fig:North projection (gusev)}}
\end{figure}

By contrast, the combination of three targeted observations provides improved constrains on
all Hapke's parameters except for $B_{0}$ and $h$.
Indeed, the standard deviation of each estimated parameter is lower
than those obtained when using only a single targeted observation.
We note that for ROI IV, the Hapke's parameters are less constrained
than for the other ROIs, especially for parameter $b$ for which no
solutions were found. This result can be explained by the limitation in the bidirectional sampling. 
In fact, we can note in Figure \ref{fig:North projection (gusev)} which represents the north 
projection of geometric conditions for the four selected ROIs that the ROI IV misses a near 
nadir geometry from the observation FRT8CE1. The latter shows a more different incidence angles 
(nearly 40\textdegree{}) than the two other CRISM FRT observations (nearly 60\textdegree{}) 
(see also Table \ref{tab:Selected-CRISM-observations}) thus explaining that the parameters 
are less constrained for ROI IV than the other ROI. 

To conclude, the quality of these
results show the benefits of combining several targeted observations in constraining photometric parameters.

\subsection{Results on Meridiani Planum \label{sub:On-Meridiani-Planum}}

\begin{table}
\caption{Same as Table \ref{tab:Retrieved-photometric-parameter gusev1} but
for Meridiani Planum with CRISM FRT observations FRT95B8, FRT334D
and FRTB6B5.\label{tab:Retrieved-photometric-parameter meridiani}}

\raggedright{}\centering{\scriptsize }%
\begin{tabular}{ccccccccccccc}
\hline 
{\scriptsize FRT} & {\scriptsize Lat. } & {\scriptsize Lon.} & \multicolumn{1}{c}{{\scriptsize AOT}} & {\scriptsize $\sigma_{\rho}$} & {\scriptsize $\omega$} & {\scriptsize $b$} & {\scriptsize $c$} & {\scriptsize $\bar{\theta}$ } & {\scriptsize $B_{0}$} & {\scriptsize $h$} & {\scriptsize $nb$} & {\scriptsize $g$}\tabularnewline
 & {\scriptsize (\textdegree{}N)} & {\scriptsize{} (\textdegree{}E)} & {\scriptsize} & {\scriptsize $RMS$}  & {\scriptsize $k$} & {\scriptsize $k$} & {\scriptsize $k$} & {\scriptsize $k$} & {\scriptsize $k$} & {\scriptsize $k$} &  & {\scriptsize}\tabularnewline
\hline 
\multirow{2}{*}{{\scriptsize 95B8}} & \multirow{2}{*}{{\scriptsize -2.000}} & \multirow{2}{*}{{\scriptsize -5.484}} & \multirow{2}{*}{{\scriptsize 0.56$\pm$0.09}} & {\scriptsize 0.02} & {\scriptsize 0.72 (0.09)} & {\scriptsize 0.34 (0.30)} & {\scriptsize 0.42 (-)} & {\scriptsize 17.32( 10.33)} & {\scriptsize 0.48 (-)} & {\scriptsize 0.50 (-)} & \multirow{2}{*}{{\scriptsize 5}} & \multirow{2}{*}{{\scriptsize $\sim$41-86}}\tabularnewline
 &  &  &  & {\scriptsize 0.01} & {\scriptsize 0.99} & {\scriptsize 1.40} & {\scriptsize 0.44} & {\scriptsize 0.98} & {\scriptsize 0.12} & {\scriptsize 0.16} &  & \tabularnewline
\cline{5-13} 
\multirow{2}{*}{{\scriptsize 334D}} & {\scriptsize -2.001} & \multirow{2}{*}{{\scriptsize -5.482}} & \multirow{2}{*}{{\scriptsize 0.35$\pm$0.04}} & {\scriptsize 0.01} & {\scriptsize 0.67 (0.09)} & {\scriptsize 0.38 (0.32)} & {\scriptsize 0.49 (0.25)} & {\scriptsize 16.65 (7.82)} & {\scriptsize 0.49 (-)} & {\scriptsize 0.48 (-)} & \multirow{2}{*}{{\scriptsize 5}} & \multirow{2}{*}{{\scriptsize $\sim$50-98}}\tabularnewline
 &  &  &  & {\scriptsize 0.01} & {\scriptsize 0.99} & {\scriptsize 1.16} & {\scriptsize 0.61} & {\scriptsize 1.00} & {\scriptsize 0.18} & {\scriptsize 0.16} &  & \tabularnewline
\cline{5-13} 
\multirow{2}{*}{{\scriptsize B6B5}} & \multirow{2}{*}{{\scriptsize -2.004}} & \multirow{2}{*}{{\scriptsize -5.482}} & \multirow{2}{*}{{\scriptsize 0.35$\pm$0.04}} & {\scriptsize 0.01} & {\scriptsize 0.65 (0.07)} & {\scriptsize 0.19 (0.20)} & {\scriptsize 0.52 (-)} & {\scriptsize 18.09 (5.82)} & {\scriptsize 0.50 (-)} & {\scriptsize 0.50 (-)} & \multirow{2}{*}{{\scriptsize 6}} & \multirow{2}{*}{{\scriptsize $\sim$41-106}}\tabularnewline
 &  &  &  & {\scriptsize 0.01} & {\scriptsize 1.00} & {\scriptsize 2.06} & {\scriptsize 0.41} & {\scriptsize 1.00} & {\scriptsize 0.12} & {\scriptsize 0.14} &  & \tabularnewline
\cline{5-13} 
\multirow{2}{*}{{\scriptsize ROI I - 3 FRTs}} & \multirow{2}{*}{-} & \multirow{2}{*}{-} & \multirow{2}{*}{{\scriptsize -}} & {\scriptsize -} & {\scriptsize 0.68 (0.08)} & {\scriptsize 0.36 (0.26)} & {\scriptsize 0.41 (0.21)} & {\scriptsize 18.37 (5.20)} & {\scriptsize 0.51 (-)} & {\scriptsize 0.52 (-)} & \multirow{2}{*}{{\scriptsize 16}} & \multirow{2}{*}{{\scriptsize $\sim$41-106}}\tabularnewline
 &  &  &  & {\scriptsize 0.02} & {\scriptsize 0.99} & {\scriptsize 0.82} & {\scriptsize 0.95} & {\scriptsize 1.00} & {\scriptsize 0.15} & {\scriptsize 0.24} &  & \tabularnewline
\hline 
\end{tabular}
\end{table}

Similar to the photometric study on the Gusev Crater, the standard
deviation {\footnotesize $\sigma_{\rho}$} determined by MARS-ReCO
is given for each CRISM observation and the ROI I used for the present
study (cf. Table \ref{tab:Retrieved-photometric-parameter meridiani}).
Note that $\sigma_{\rho}$ values are acceptable (0.01 < $\sigma_{\rho}$ < 0.02) meaning that estimated surface bidirectional reflectance are accurate. 

\begin{figure}
\centering\includegraphics[scale=0.48]{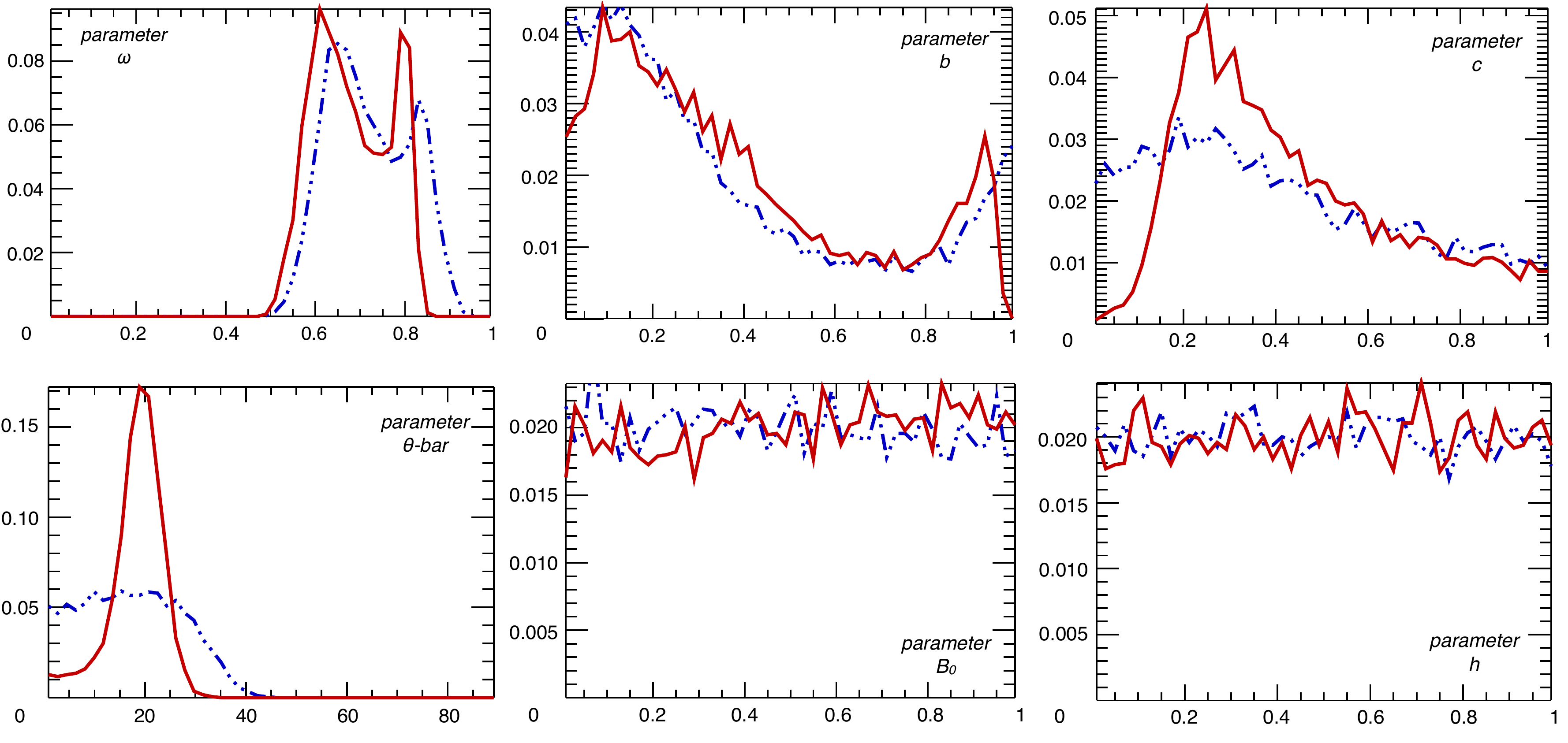}

\caption{The PDF of each of the six physical parameters retrieved for ROI I of Meridiani Planum.
Two cases are considered: the inversion of a single CRISM FRT observation
(i.e., FRT95B8 in blue line) and the inversion of three combined CRISM
FRT observations (in red line).\label{fig:Probability-Density-Function meridiani}}
\end{figure}

Table \ref{tab:Retrieved-photometric-parameter meridiani} presents
the results obtained on Meridiani Planum. Similar to Gusev Crater
study, Figure \ref{fig:Probability-Density-Function meridiani} shows
that a solution exists for the parameters $\omega$, $b$, $c$ and
$\bar{\theta}$ (i.e., non-uniform PDF) when a single CRISM FRT and
in the case of three combined observations, except for parameters
$B_{0}$ and $h$. 

The goodness of fit is estimated through the absolute quadratic residual $RMS_{abs}$ value 
(cf. Table \ref{tab:Retrieved-photometric-parameter meridiani}). 
For all bayesian inversions, the estimates are less than 0.02 which mean that the inversions are deemed satisfactory.

We note that for the parameter $\omega$, two maximum
are visible at nearly 0.6 and 0.8. In order to understand the origin of the bimodal 
distribution for the parameter $\omega$ in this case, the reflectance of typical Martian 
material was generated by using realistic photometric properties determined 
by the Pancam instrument aboard the MER Opportunity site in Meridiani 
Planum (i) at the same geometric configurations as the FRT95B8 observations, and 
(ii) at the same combined geometric configurations when merging the three 
selected observations. In both cases, the posterior PDF for the parameter $\omega$ 
shows a bimodal distribution. If the geometric sampling is broader (i.e., with 
varied incidence, emergence and phase angles), the posterior PDF of the parameter 
$\omega$ becomes a gaussian with a single peak. The presence of two possible solutions is the consequence of the limitation of a sufficient geometric diversity in our selection of CRISM observations for the Meridiani Planum study to constrain the parameter 
$\omega$, which is otherwise the best-constrained parameter in photometric modeling. 

Several conclusions can be drawn when using a single CRISM targeted observation:

\begin{itemize}

\item Solutions exist for parameter $\omega$ in all
cases ($k\sim1$). The standard deviation shows that the single scattering
albedo is the most constrained parameter (i.e., 0.07 < $\sigma$ < 0.15). Examples of posteriori PDFs are plotted in Figures \ref{fig:Probability-Density-Function meridiani}
and \ref{fig:meridiani-pancam w}.

\item We find meaningful values for parameter $b$ only when for FRT95B8
or FRT334D or FRTB6B5 ($k$ > 0.5). Albeit the standard deviation is
relatively high for FRT95B8 or FRT334D (i.e., 0.30 < $\sigma$ < 0.32),
the standard deviation becomes relatively low for FRTB6B5 (i.e.,
$\sigma$ = 0.20). Examples of posteriori PDFs are plotted in Figures \ref{fig:Probability-Density-Function meridiani} and \ref{fig:meridiani-pancam b/c}.

\item Solutions only exist for parameter $c$ when using FRT334D ($k$ > 0.5).
However, the standard deviation is relatively high ($\sigma$ = 0.25). Examples of posteriori PDFs are plotted in Figures \ref{fig:Probability-Density-Function meridiani}
and \ref{fig:meridiani-pancam b/c}.

\item While parameter $\bar{\theta}$ has a solution in all cases ($k\sim1$),
we distinguish two types of results: (i) using FRT334D or FRTB6B5,
we note that the standard deviation is relatively low (i.e., 5.82 < $\sigma$ <7.82) 
and (ii) the standard deviation is relatively high (i.e., $\sigma$ = 10.33)
when using FRT95B8. Examples of posteriori PDFs are plotted in Figures \ref{fig:Probability-Density-Function meridiani} and \ref{fig:meridiani-pancam theta}.

\item Similar to the Gusev study, no solution is found for parameters $B0$
and $h$ in any case ($k$ >> 0.5). Examples of posteriori PDFs are plotted in Figure \ref{fig:Probability-Density-Function meridiani}.

\end{itemize}

\begin{figure}
\begin{centering}
\includegraphics[scale=0.45]{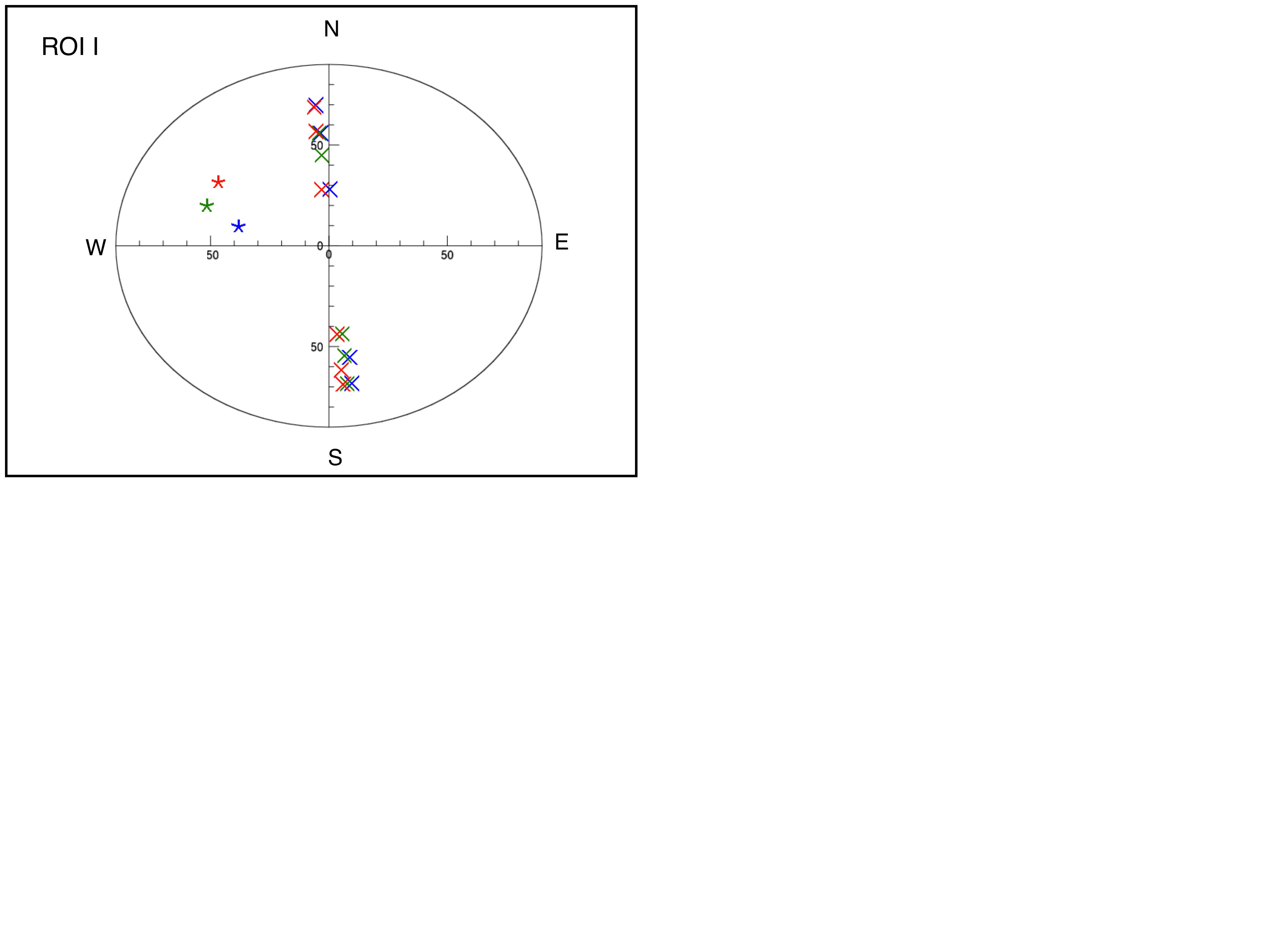}
\par\end{centering}

\caption{North projection of geometric conditions (stars: incidence directions and crosses: emergence
directions) of the selected ROI for the Meridiani Planum study (blue: FRT95B8, green: FRT334D and red: FRTB6B5). \label{fig:North projection (meridiani)}}
\end{figure}

Dealing with single observations, we note that parameter $\bar{\theta}$
is poorly constrained when treating data from FRT95B8 and FRT334D
whereas it is highly constrained when using FRTB6B5 which can be explained 
by a maximum phase angle below 90\textdegree{} in case of FRT95B8 and FRT334D (cf. Table 
\ref{tab:Selected-CRISM-observations}). \citet{Helfenstein1988} underlined 
the necessity to have observations which extend from small phase 
angles out to phase angles above 90\textdegree{} for an accurate determination 
of the photometric roughness. Note that parameter $b$
and $c$ are non-constrained or poorly constrained even through solutions
exist in all cases. This outcome can be explained by a worse quality of 
the Meridiani Planum bidirectional reflectance sampling. Indeed, we note in 
Figure \ref{fig:North projection (meridiani)} which represents the north 
projection of geometric conditions of each selected CRISM FRT observations that the three selected CRISM FRT (i) miss near 
nadir geometries (close to emergence equal 0) and (ii) present lower number of available 
angular configurations compared to Gusev Crater work (lower than 6).

We then improve the bidirectional reflectance sampling by combining the three selected targeted observations.
We observe that parameters $\omega$, $b$, $\bar{\theta}$
become significantly more constrained. Indeed, the standard deviation of each
estimated parameter is lower than those obtained when using only a
single observation. However, the parameters are less constrained than those obtained 
for the Gusev Crater study, which can be explained by the lack of 
near nadir geometry for this case (cf. Figure \ref{fig:North projection (meridiani)}). 

Again, results underline the benefits of combining several observations.

\section{Validation \label{sec:Validation}}

This section focuses on the validation of the estimated photometric
parameters by comparing them to previous photometric studies based
on experimental, in situ and orbital photometric studies. As the PDF 
of each parameter is not necessary a gaussian, the estimated mean 
and the standard deviation are not always representative of the entire distribution. 
Consequently for each parameter, we use the PDF built from the last 500 iterations of the Markov chain process (see \ref{sub:Defining-regions-of}) instead of the previous statistical estimators for the comparison to previous studies.

\subsection{Validation of results from Gusev crater}

\subsubsection{Comparison to experimental measurements on artificial and natural
samples}

\citet{mcGuire1995} studied the scattering properties of different
isolated artificial particles which have different structure types
(sphere/rough particles, clear/irregular shape particle, with/without
internal scatterers, ...). Their study showed that the Henyey-Greenstein
function with two parameters, HG2 (backscattering fraction $c$ and
asymmetric parameter $b$) provided the best description of their
laboratory bidirectional reflectance measurements. In short, their
study shows that smooth clear spheres exhibit greater forward scattering
(low values of $c$) and narrower scattering lobes (high values of
$b$) whereas particles characterized by their roughness or internal
scatterers exhibit greater backward scattering (high values of $c$)
and broader scattering lobe (low values of $b$). In a graph mapping
the $b$ and $c$ parameter space, the results exhibit a ``L-shape''
from particles with high density of internal scatterers to smooth,
clear, spherical particles. For their study, \citet{mcGuire1995}
used centimeter sized artificial particles which are larger than the
light wavelength and far than typical constituents of the planetary
regoliths. In order to examine the impact of particle size on the
scattering phase function, \citet{hartman1998} observed that there
are no significant variations of the latter in \citet{mcGuire1995}
measurements when the particle size is similar to typical planetary
regoliths particles. \citet{hartman1998} concluded that \citet{mcGuire1995}
results could be considered to be representative of their respective
particle structure types independent of particle size. However, recent
experimental works have questioned the initial interpretation of the
Hapke\textquoteright{}s parameters. Indeed, the Hapke's parameters seem to be
more sensitive to the organization of the surface material (effects
of close packing, micro-roughness) than to the optical properties
depending on individual particles \citep{cord2003,shepard2007}. Similarly,
\citet{souchon2011} measured for a comprehensive set of geometries
the reflectance factor of natural granular surfaces composed of volcanic
materials differing by their grain size (from the micron-scale to
the millimeter-scale), shapes, surface aspect, and mineralogy (including
glass and minerals). Thus the main novelty of \citet{souchon2011}'s
experimental study compared to \citet{mcGuire1995}'s works is the
determination of the parameters $b$ and $c$ for planetary analogs
of basaltic granular surfaces and not isolated artificial particles.
\citet{souchon2011} compared the scattering parameters retrieved by 
inversion of Hapke's model with results on artificial materials \citep{mcGuire1995}
and a similar trend was found, though with some variations and new
insights. Granular surfaces even with a moderate proportion of isolated
translucent monocrystals and/or fresh glass exhibit strongly forward
scattering properties and a new part of the L-shape domain in the
$b$ and $c$ parameter space is explored.

\begin{figure}
\centering\includegraphics[scale=0.40]{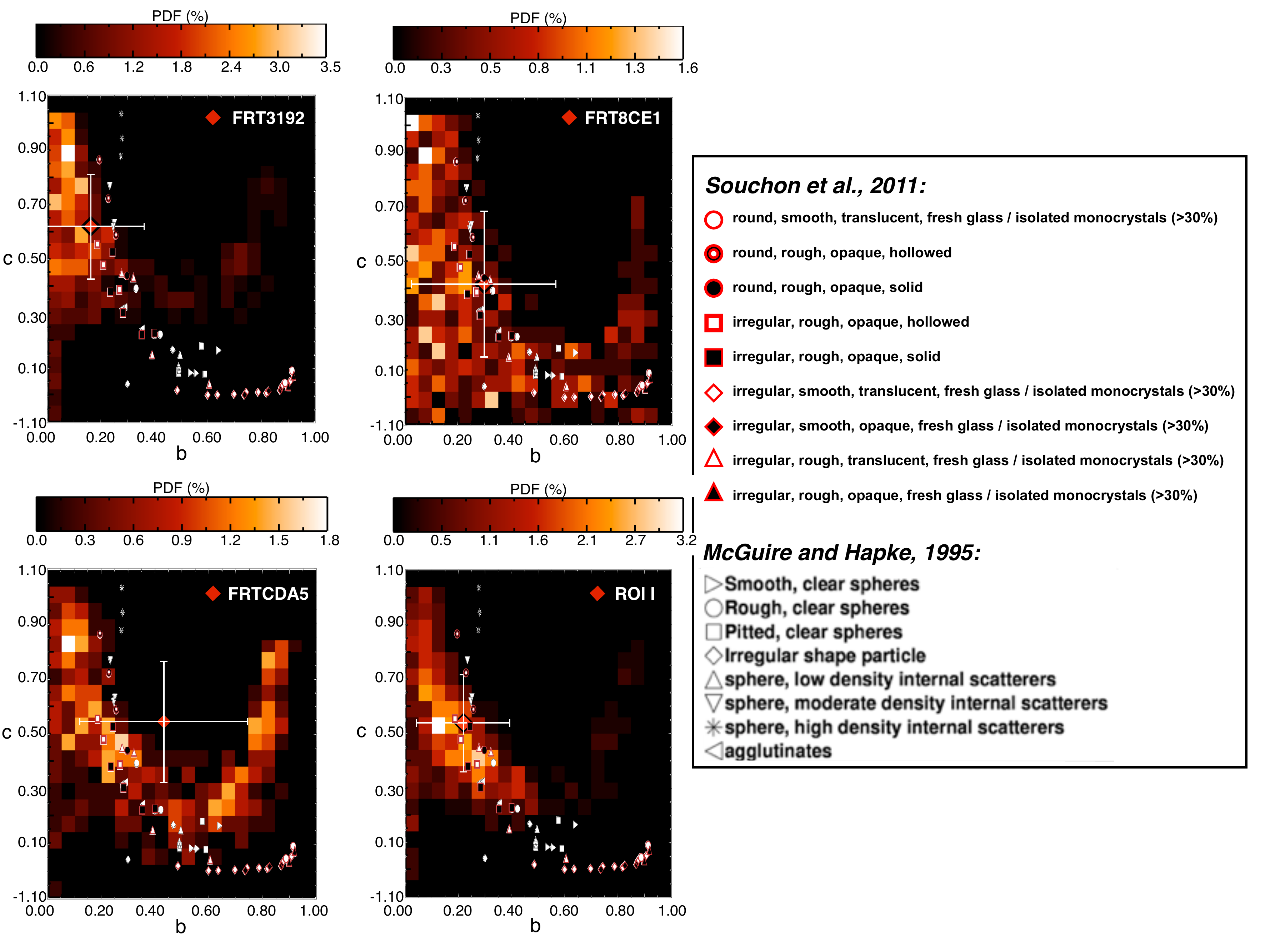}

\caption{Probability density map of the asymmetric parameter $b$ (horizontal axis) versus the backscattering fraction
$c$ (vertical axis) solutions retrieved at 750 nm estimated from the last 500 iterations of the 
Bayesian inversion for FRT3192, FRT8CE1 and
FRTCDA5 observations and their combination (ROI I). The grid is divided into 24 (vertical) x 20 (horizontal) square bins. The coloring gives the probability corresponding to each bin. Means and 2$\sigma$ uncertainties
(red rhombus) are plotted too. The inversion solutions are plotted
against the experimental $b$ and $c$ values pertaining to artificial
particules measured by \citet{mcGuire1995} and to natural particles
measured by \citet{souchon2011}. \label{fig:gusev-b vs c experimental study}}
\end{figure}

In Figure \ref{fig:gusev-b vs c experimental study} that is given as an example of similar figures, the scattering parameters (i.e., backscattering fraction
$c$ and asymmetric parameter $b$) of the ROI I from CRISM data (cf. Table \ref{tab:Retrieved-photometric-parameter gusev1}) 
are plotted along with the scattering parameters obtained from laboratory 
measurements of artificial \citep{mcGuire1995}
and natural samples \citep{souchon2011}.
Results shows that parameters $b$ and $c$ retrieved from the inversion
of combined CRISM FRT observations are consistent with the laboratory studies. 
The combination of three FRTs is necessary to constrain satisfactorily the $b$
and $c$ values with acceptable error bars.  As it can be seen,
surface materials of ROI I have high values of $c$ and low values
of $b$, which indicate broad backscattering properties related to
artificial materials composed of spheres with a moderate density of
internal scatterers and close to irregular or round rough, opaque
and solid natural particles (cf. Figure \ref{fig:gusev-b vs c experimental study}). 

\subsubsection{Comparison to in situ measurements taken by Pancam/MER-Spirit \label{sub:pancam gusev}}

The Pancam instrument on-board MER-Spirit acquired several spectrophotometric
observations along the rover's traverse paths to determine the surface physical and chemical properties
of rocks and soils encountered at the Gusev Crater. \citet{Johnson2006a} evaluated the parameters of Hapke's
photometric model \citep{Hapke1993} using Equations \ref{eq:Hapke},
\ref{eq:HG2} and \ref{eq:H1993} from measured radiance for several
identified units (i.e., ``Gray'' rocks, which are free of airfall-deposit
dust or other coatings, ''Red'' rocks, which have coating, and ''Soil''
unit, corresponding to unconsolidated materials). The measured radiance at the ground was first corrected for
diffuse sky illumination \citep{Johnson2006a} and local surface facet
orientations \citep{soderblom2004}.

The difference of spatial resolution between CRISM and Pancam instruments
must be considered prior to comparison. Indeed, a direct comparison
between both sets of photometric parameters must be handled with care
as Pancam provides local measurements (at a centimetric scale
such as soils and rocks), whereas orbital instruments such as CRISM
measure extended areas integrating different geologic units (at a
pluri-decametric scale such as a landscape). In the latter case, measurements
may be dominated by unconsolidated materials such as soils. As mentioned
in Subsection \ref{sub:Defining-regions-of}, the four selected ROIs
are chosen close to the MER-Spirit's path. The retrieved photometric
parameters for each ROI from the combination of three CRISM observations
were compared to those extracted from three sequences taken by Pancam
in the Gusev plains. These measurements are located near to the four
ROIs in Landing site (Sol 013), Bonneville Rim (Sol 087-088) and Missoula (Sol 102-103) to 
the northwest side of Columbia Hill 
(cf. Figure \ref{fig:HiRISE}) \citep{Johnson2006a}.
In this case the combination of three CRISM observations were treated
using the H93 version of the Hapke's model in order to match the same
model used in \citet{Johnson2006a}. Pancam results are summarized
in Table \ref{tab:Retrieved-Hapke-parameters}. 

\begin{table}[H]
\caption{Retrieved Hapke\textquoteright{}s parameters and their standard deviation
(underconstrained parameters are indicated by \textquotedblleft{}(+,-)\textquotedblright{}) from the 2-term model 
for Sol 13, Sol 87-88 and Sol 102-103 and each of the three considered
units (Gray Rock, Red Rock and Soil) at 753 nm (except for the parameter $c$ of Red Rock unit for Landing Site which corresponds to the 754 nm model results) from Pancam onboard
Spirit \citep{Johnson2006a} \label{tab:Retrieved-Hapke-parameters}}

\raggedright{}\centering{\scriptsize }%
\begin{tabular}{cccccccc}
\hline 
{\scriptsize Site} & {\scriptsize Unit} & {\scriptsize $\omega$} & {\scriptsize $\bar{\theta}$ (deg.)} & {\scriptsize $b$} & {\scriptsize $c$} & {\scriptsize Number} & {\scriptsize $g$ (deg.)}\tabularnewline
\hline 
{\scriptsize Landing Site } & {\scriptsize Bright Soil} & {\scriptsize 0.75 (+0.01, -0.00)} & {\scriptsize 2 (+, -)} & {\scriptsize 0.243 (+0.020, -0.017)} & {\scriptsize 0.625 (+0.012, -0.013)} & {\scriptsize 21} & {\scriptsize $\sim$30-120}\tabularnewline
{\scriptsize (Sol 013)} & {\scriptsize Gray Rock} & {\scriptsize 0.83 (+0.01, -0.01)} & {\scriptsize 7 (+3, -4)} & {\scriptsize 0.931 (+0.045, -0.044)} & {\scriptsize 0.065 (+0.058, -0.058)} & {\scriptsize 66} & {\scriptsize $\sim$30-120}\tabularnewline
 & {\scriptsize Red Rock} & {\scriptsize 0.79 (+0.03, -0.02)} & {\scriptsize 20 (+3, -3)} & {\scriptsize 0.187 (+0.026, -0.031)} & {\scriptsize 0.720 (+0.084, -0.087)} & {\scriptsize 68} & {\scriptsize $\sim$30-120}\tabularnewline
 & {\scriptsize Soil} & {\scriptsize 0.76 (+0.01, -0.01)} & {\scriptsize 15 (+2, -1)} & {\scriptsize 0.262 (+0.010, -0.010)} & {\scriptsize 0.715 (+0.029, - 0.032)} & {\scriptsize 51} & {\scriptsize $\sim$30-120}\tabularnewline
\hline 
{\scriptsize Bonneville Rim } & {\scriptsize Gray rock} & {\scriptsize 0.72 (+0.04, -0.04)} & {\scriptsize 23 (+3, -4)} & {\scriptsize 0.434 (+0.035, -0.037)} & {\scriptsize 0.359 (+0.048, -0.050)} & {\scriptsize 63} & {\scriptsize $\sim$25-120}\tabularnewline
{\scriptsize (Sol 087-088)} & {\scriptsize Red rock} & {\scriptsize 0.70 (+0.01, -0.01)} & {\scriptsize 15 (+3, -3)} & {\scriptsize 0.219 (+0.017, -0.020)} & {\scriptsize 1.000 (+, -)} & {\scriptsize 72} & {\scriptsize $\sim$25-120}\tabularnewline
 & {\scriptsize Soil} & {\scriptsize 0.66 (+0.00, -0.00)} & {\scriptsize 7 (+1, -1)} & {\scriptsize 0.170 (+0.008, -0.008)} & {\scriptsize 0.823 (+0.025, -0.024)} & {\scriptsize 54} & {\scriptsize $\sim$25-120}\tabularnewline
\hline 
{\scriptsize NW of Missoula } & {\scriptsize Gray rock} & {\scriptsize 0.70 (+0.02, -0.02)} & {\scriptsize 13 (+2, -3)} & {\scriptsize 0.406 (+0.018, -0.032)} & {\scriptsize 0.206 (+0.019, -0.025)} & {\scriptsize 225} & {\scriptsize $\sim$0-125}\tabularnewline
{\scriptsize (Sol 102-103)} & {\scriptsize Red rock} & {\scriptsize 0.83 (+0.02, -0.02)} & {\scriptsize 19 (+1, -2)} & {\scriptsize 0.450 (+0.023, -0.064)} & {\scriptsize 0.255 (+0.032, -0.072)} & {\scriptsize 141} & {\scriptsize $\sim$0-125}\tabularnewline
 & {\scriptsize Soil} & {\scriptsize 0.69 (+0.01, -0.00)} & {\scriptsize 11 (+1, -1)} & {\scriptsize 0.241 (+0.011, -0.009)} & {\scriptsize 0.478 (+0.036, -0.022)} & {\scriptsize 132} & {\scriptsize $\sim$0-125}\tabularnewline
\hline 
\end{tabular}
\end{table}

\begin{itemize}

\item Single scattering albedo: Figure \ref{fig:gusev-pancam w}
plots means and uncertainties as well as the PDF of single scattering albedo $\omega$ estimated
for the four selected ROIs against values extracted for each unit
defined by \citet{Johnson2006a} (i.e., Gray rocks, Red rocks and
Soil) at the Landing site, Bonneville rim and NW of Missoula. The bimodality is identical to Meridiani ROI I (see subsection \ref{sub:On-Meridiani-Planum}) and is the consequence of the lack of geometric diversity in the CRISM dataset. The estimated values of $\omega$ from CRISM are consistent
with the Pancam outputs for the Soil unit.

\begin{figure}
\centering\includegraphics[scale=0.33]{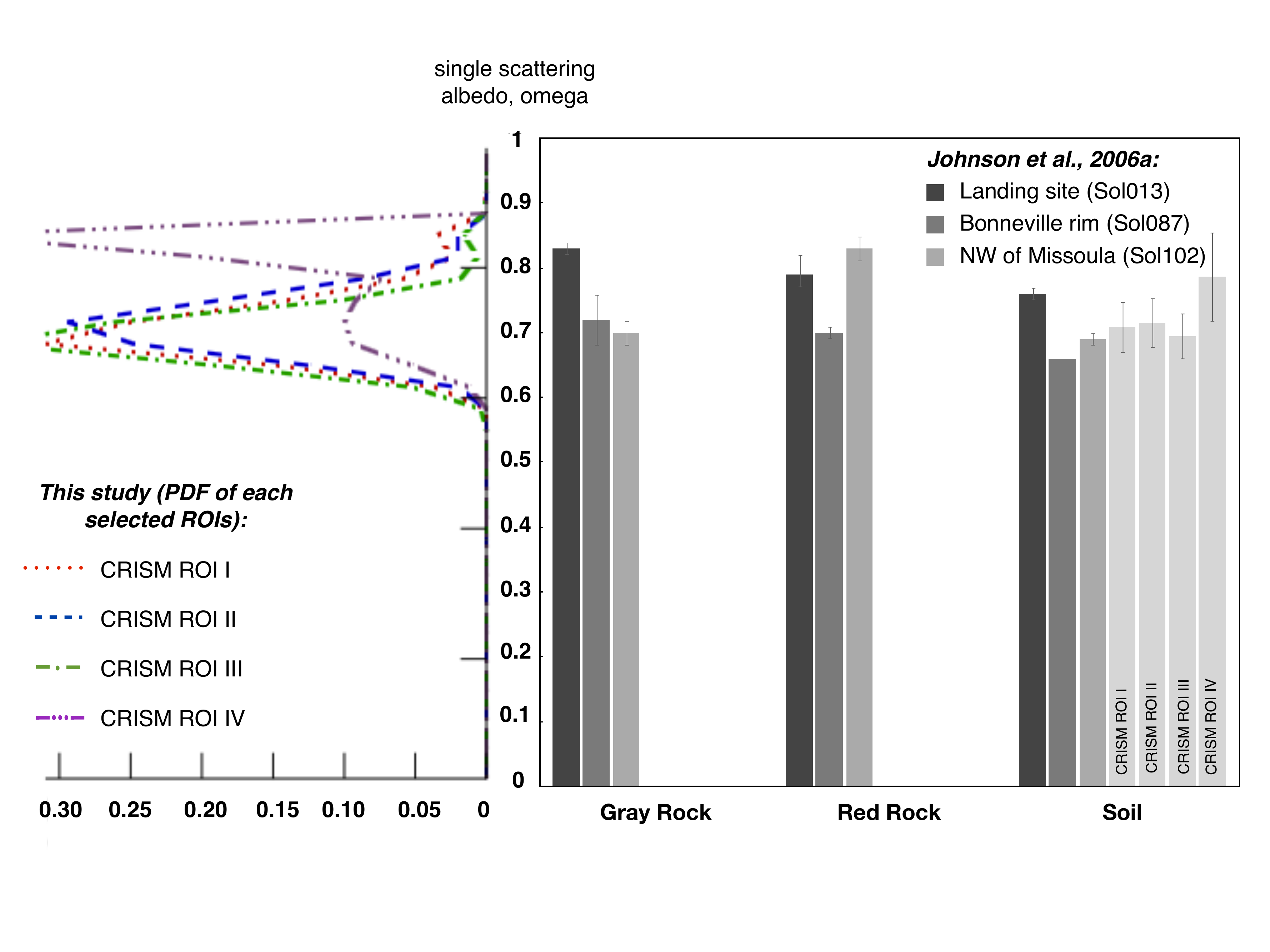}

\caption{Mean and uncertainties of the single scattering albedo $\omega$ estimated
from Pancam measurements at 753 nm for different geological units
at Landing Site (Sol 013), Bonneville Rim (Sol 087-088) and NW of
Missoula (Sol 102-103) \citep{Johnson2006a} compared to those estimated
from CRISM measurements at 750 nm (note that the error bar represents
2$\sigma$ uncertainties) derived from 2-term HG models. The PDFs of
the parameter $\omega$ estimated from the last 500 iterations of
the bayesian inversion are also represented on the left side. This is helpful
when mean and uncertainties are not entirely representative of the PDF (such
as the PDF of the parameter $\omega$ for the ROI IV which shows a
bimodal distribution). \label{fig:gusev-pancam w}}
\end{figure}

\item Phase function: Figure \ref{fig:gusev-pancam b/c} plots
the asymmetric parameter $b$ versus the backscattering fraction $c$ solutions provided by the Bayesian inversion
for each ROI in comparison to the photometric parameters obtained
by \citet{Johnson2006a}. In the case of Red Rock unit for Landing Site, the parameter $c$ is determined at 754 nm instead of usual 753 nm. Moreover, the parameter $c$ is not constrained at both wavelengths in the case of Red Rock at Bonneville rim. Consequently the parameters $b$ and $c$ 
are not plotted here. The $b$ and $c$ pair from CRISM observations are most consistent with the Soil
unit for all the studied Spirit sites and the Red Rock unit from the Landing Site which exhibit broad backscattering properties. 
Again, we note that for ROI IV, the Hapke's parameters are less constrained than for the other
ROIs, especially for parameter $b$ for which no solutions were found.

\begin{figure}
\centering\includegraphics[scale=0.40]{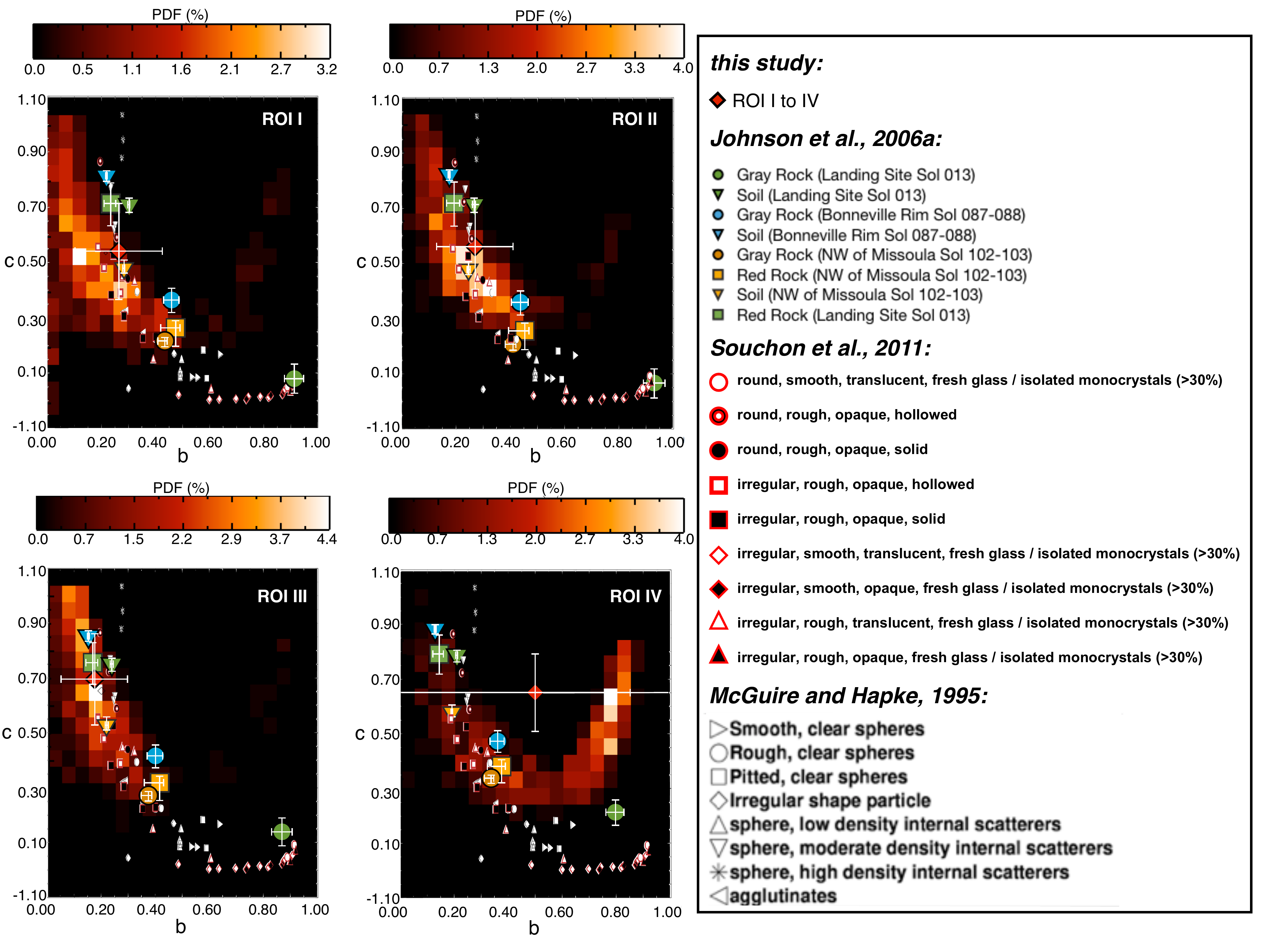}
\caption{Probability density map of the asymmetric parameter $b$ (horizontal axis) and the backscattering fraction
$c$ (vertical axis) solutions from the last 500 iterations of the Bayesian inversion
estimated at 750 nm obtained for the four selected ROIs. The grid is divided into 24 (vertical) x 20 (horizontal) square bins. The coloring gives the probability corresponding to each bin. Means and
2$\sigma$ uncertainties (red rhombus) are plotted too. The inversion
solutions are plotted against the experimental $b$ and $c$ values
pertaining to artificial particles measured by \citet{mcGuire1995}
and to natural particles measured by \citet{souchon2011}. The Gray
Rock (circle), Red Rock (square) and Soil (triangle) units from Pancam
sequences at Landing Site (green), Bonneville Rim (blue) and NW of
Missoula (orange) estimated at 753 nm (except for the Red Rock unit
at the Landing Site which is constrained at 754 nm) \citep{Johnson2006a}
are plotted here. \label{fig:gusev-pancam b/c}}
\end{figure}

\item Macroscopic roughness: Means and uncertainties as well as the PDF of 
the macroscopic roughness $\bar{\theta}$ obtained
for each selected ROI are presented in  Figure \ref{fig:gusev-pancam theta}.
The parameter $\bar{\theta}$ of all selected ROIs, estimated from CRISM 
measurements is consistent with the Soil unit found at the
Landing Site and at NW of Missoula (cf. Figure \ref{fig:gusev-pancam theta}).
By contrast, this parameter indicates a rougher surface than the soil
unit found at Bonneville Rim which is consistent
with \citet{Johnson2006a}. At intercrater plains
where the Landing site is situated, the greater proportion of small clasts in the soil 
(compared to the soils near Bonneville rim and Missoula area) may explain 
the rougher surface texture \citep{ward2005}.

\begin{figure}
\centering\includegraphics[scale=0.33]{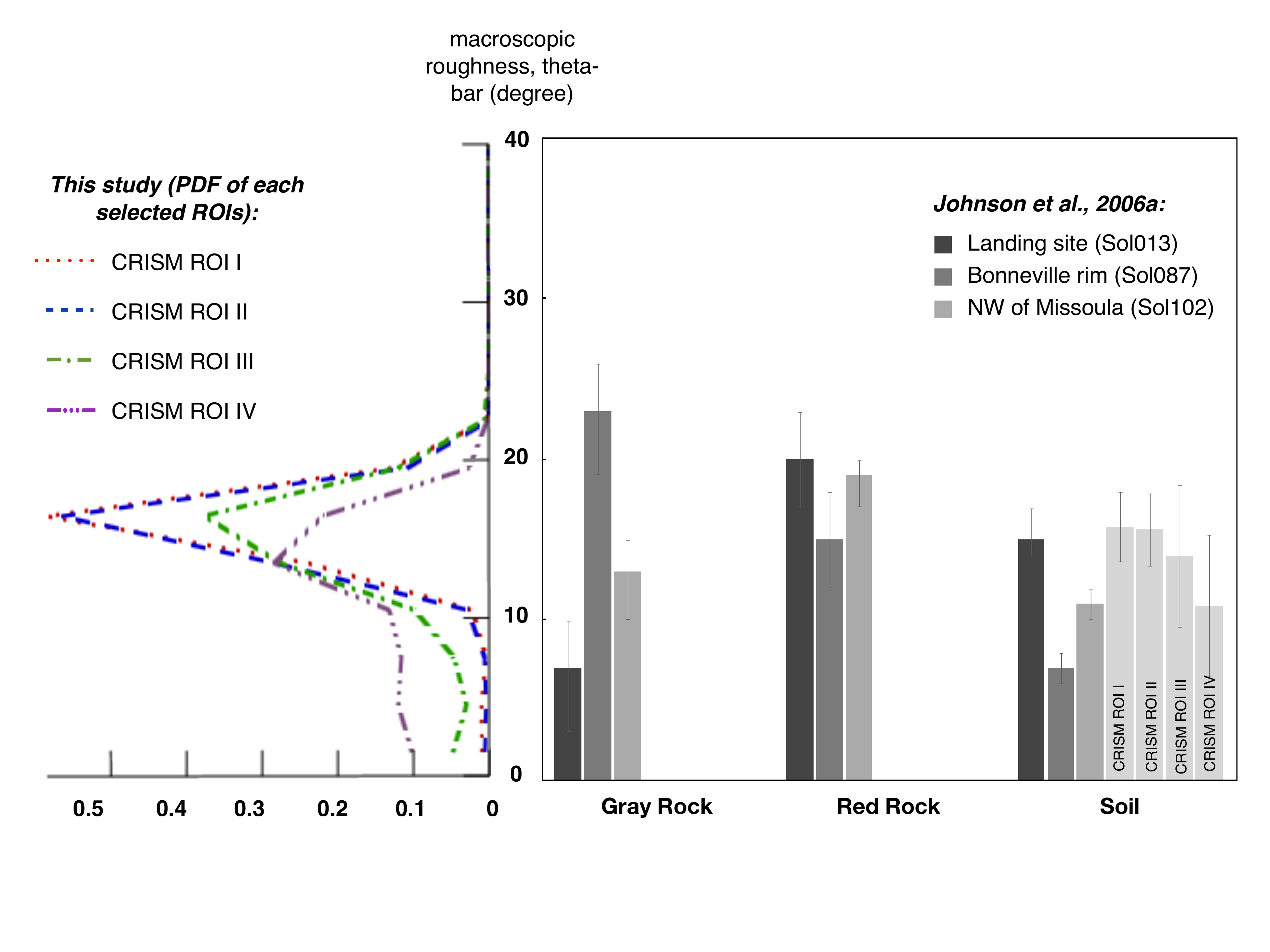}

\caption{Mean and uncertainties of the macroscopic roughness $\bar{\theta}$
estimated from Pancam measurements at 753 nm for different geological
units at Landing Site (Sol 013), Bonneville Rim (Sol 087-088) and
NW of Missoula (Sol 102-103) \citep{Johnson2006a}  compared to those
estimated from CRISM measurements at 750 nm (note that the error bar
represents 2$\sigma$ uncertainties) derived from 2-term HG models.
The PDFs of the parameter $\bar{\theta}$ estimated from the last 500
iterations of the Bayesian inversion are also represented on the left
side. This is helpful when mean and uncertainties are not representative of
the PDF. \label{fig:gusev-pancam theta}}
\end{figure}

\end{itemize}

Two principal points from the previous comparison can be outlined.
First, at the effective spatial scale achieved from space by CRISM
for the photometric characterization (460 m), the surface behaves
like the Soil units defined by \citet{Johnson2006a} and particularly
like the Soil unit observed at the Landing Site. Second, the values
of $\omega$ and $(b,c)$ resulting from our analysis are approximately
inside the range of variation seen in Figures \ref{fig:gusev-pancam w}
and \ref{fig:gusev-pancam b/c} for the Soil units
at the three locations. This is not the case for the macroscopic roughness which is 
close to (within the error bars) the Soil unit values only for the Landing Site and 
Missoula area (cf. Figure \ref{fig:gusev-pancam theta}).
Consequently all photometric parameters retrieved following the methodology
proposed in this article are consistent with the results arising from
the analysis of Pancam measurements at Gusev provided that ROIs are
associated to the proper soil unit. That means that the two independent 
investigations (the present work and the \citet{Johnson2006a}'s study) 
are cross-validated in the Gusev case.

\subsubsection{Comparison with photometric estimates derived from orbital measurements
by HRSC/MEx }

Using HRSC imagery, which provides multi-angle data sets (up to five angular
configurations by orbit) \citep{mcCord2007}, \citet{Jehl2008} determined
the regional variations of the photometric properties at the kilometer
spatial scale across Gusev Crater and the south flank of Apollinaris
Patera using several observations in order to cover a phase angle
range from 5\textdegree{} to 95\textdegree{}. The photometric study
was carried out without any atmospheric correction but ensuring that
the atmospheric contribution was limited by selecting HRSC observations
with AOT lower than 0.9. \citet{Jehl2008} applied an inversion procedure
developed by \citet{cord2003} based on the H93 version of Hapke's
model (equation \ref{eq:Hapke}). In \citet{Jehl2008}'s study, even photometric units 
were determined using a principal component analysis at 675 nm in which 
one of them corresponds to the Spirit landing site area. There is a robust 
overall first-order consistency between these photometric orbital estimates 
independently retrieved from HRSC and CRISM observations, particularly for 
ROIs I, II and III (see Table \ref{tab:HRSC results}).

\begin{table}
\caption{Retrieved Hapke's parameters and their standard deviation (the underconstrained parameters are indicated by "(-)") from CRISM
measurement at 750 nm (present results) compared to those from HRSC
measurements at 675 nm \citep{Jehl2008}. The yellow unit is associated
to the Spirit landing site. The case 1 is the Hapke's parameters determination
when the opposition effect parameters are set free and the case 2,
when the opposition effect is neglected, with phase angles larger
than 20\textdegree{} (see \citet{Jehl2008} for more details). \label{tab:HRSC results}}

\raggedright{}\centering{\scriptsize }%
\begin{tabular}{cccccccc}
\hline 
{\scriptsize Instrument} & {\scriptsize ROI or unit} & {\scriptsize $\omega$} & {\scriptsize $b$} & {\scriptsize $c$} & {\scriptsize $\bar{\theta}$ } & {\scriptsize $B_{0}$} & {\scriptsize $h$}\tabularnewline
 &  &  &  &  & {\scriptsize (deg.)} &  & \tabularnewline
\hline 
\multirow{4}{*}{{\scriptsize CRISM}} & \multirow{1}{*}{{\scriptsize ROI I}} & {\scriptsize 0.71 (0.05)} & {\scriptsize 0.22 (0.18)} & {\scriptsize 0.54 (0.18)} & {\scriptsize 15.78 (2.65)} & {\scriptsize 0.49 (-)} & {\scriptsize 0.54 (-)}\tabularnewline
 & \multirow{1}{*}{{\scriptsize ROI II}} & {\scriptsize 0.72 (0.05)} & {\scriptsize 0.27 (0.15)} & {\scriptsize 0.56 (0.16)} & {\scriptsize 15.62 (2.43)} & {\scriptsize 0.49 (-)} & {\scriptsize 0.52 (-)}\tabularnewline
 & \multirow{1}{*}{{\scriptsize ROI III}} & {\scriptsize 0.69 (0.04)} & {\scriptsize 0.19 (0.14)} & {\scriptsize 0.66 (0.18)} & {\scriptsize 13.96 (4.40)} & {\scriptsize 0.50 (-)} & {\scriptsize 0.53 (-)}\tabularnewline
 & \multirow{1}{*}{{\scriptsize ROI IV}} & {\scriptsize 0.79 (0.07)} & {\scriptsize 0.59 (0.27)} & {\scriptsize 0.56 (0.16)} & {\scriptsize 10.88 (5.11)} & {\scriptsize 0.50 (-)} & {\scriptsize 0.48 (-)}\tabularnewline
\hline 
\multirow{2}{*}{{\scriptsize HRSC}} & {\scriptsize case 1} & {\scriptsize 0.72$\pm$0.02} & {\scriptsize 0.06$\pm$0.02} & {\scriptsize 0.34$\pm$0.06} & {\scriptsize 18.5$\pm$1.5} & {\scriptsize 0.73$\pm$0.07} & {\scriptsize 0.75$\pm$0.13}\tabularnewline
 & {\scriptsize case 2} & {\scriptsize 0.80$\pm$0.02} & {\scriptsize 0.22$\pm$0.04} & {\scriptsize 0.41$\pm$0.06} & {\scriptsize 17.2$\pm$3.0} & {\scriptsize -} & {\scriptsize -}\tabularnewline
\hline 
\end{tabular}
\end{table}

The principal point from the comparison between both orbital photometric
results is that there is a robust overall first-order consistency between these photometric orbital
estimates reached independently from HRSC and CRISM observations,
particularly for ROIs I, II and III (see Table \ref{tab:HRSC results}). 
However, in detail, two differences can be noted for the parameters 
$\omega$ and $c$ (see Table \ref{tab:HRSC results}). In fact, 
the parameters $\omega$ and $c$ from CRISM measurements 
are respectively lower and higher than those determined from HRSC measurements. 
Moreover, the parameters $\omega$ and $c$ estimated from CRISM data set are more consistent 
with in situ measurements.These differences are explained by the use of an aerosol 
correction in the CRISM data whereas no correction was made in HRSC data. 
As a consequence, the contribution of bright and forward scattering aerosols 
in the HRSC measurements appeared to increase the apparent surface single scattering albedo 
and to decrease the surface backscattering fraction value.

\subsection{Validation of Meridiani Planum}

\subsubsection{Comparison to experimental measurements on artificial particles}

\begin{figure}
\centering\includegraphics[scale=0.40]{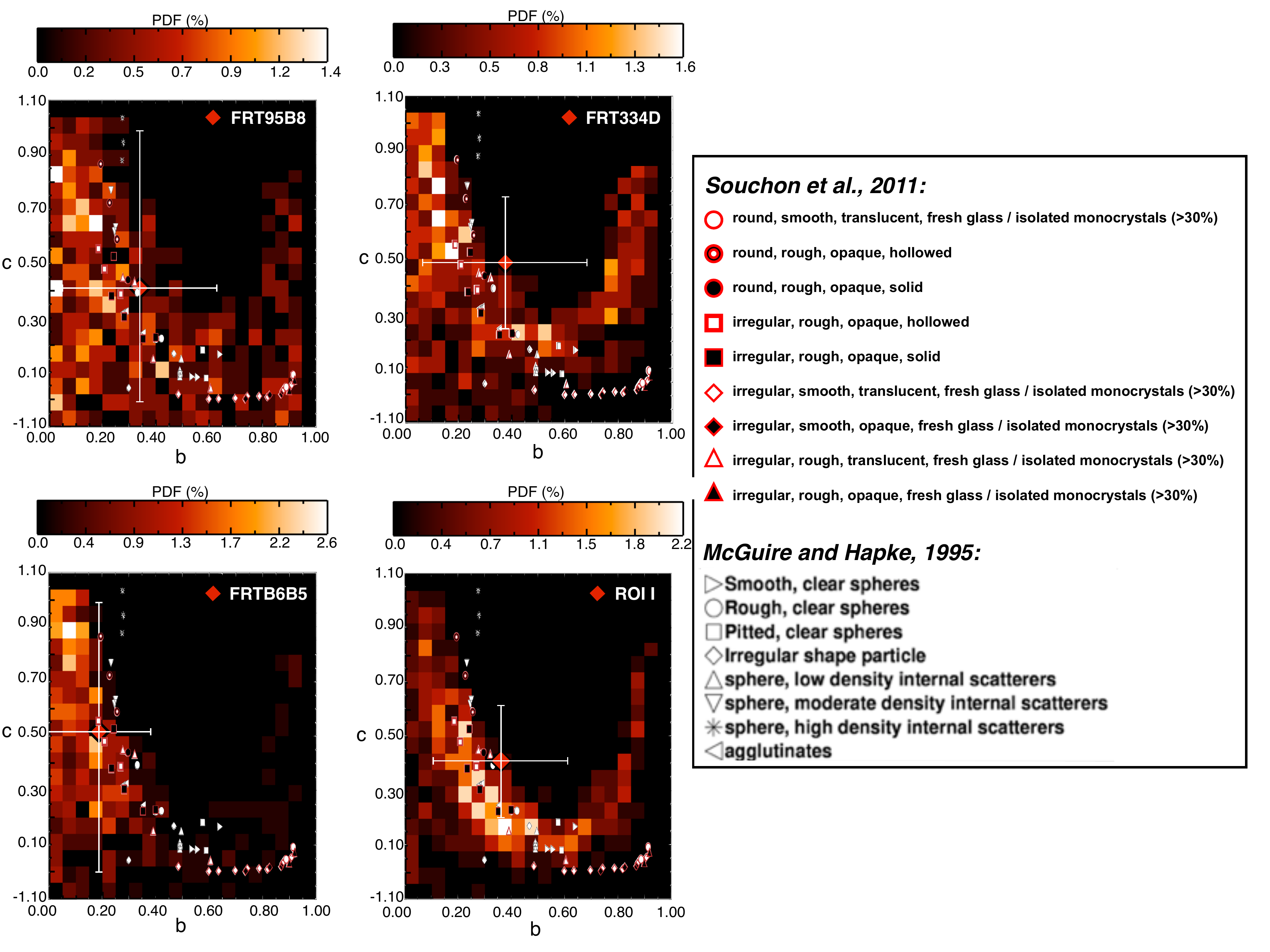}

\caption{Same as Figure \ref{fig:gusev-b vs c experimental study} for Meridiani
Planum using observations FRT95B8, FRT334D and FRTB6B5 and their combination
(ROI I). \label{fig:meridiani-experimental study}}
\end{figure}

Our investigation shows that parameters $b$ and $c$ estimated from
the inversion of both individual and combined (cf. Figure \ref{fig:meridiani-experimental study})
CRISM FRTs are consistent with the L-shape defined from laboratory studies.
However, the bidirectional reflectance sampling is not adequate
to provide accurate estimation of the scattering parameters, even by combining
up to three FRTs. In this case, according to parameters $b$ and $c$,
the surface represented by ROI I is slightly forward scattering with
a relatively broad lobe (low $b$ and $c$) which is consistent with artificial materials
composed of round and clear sphere or agglutinates and close to irregular
or round rough, opaque and solid natural particles (cf. Figure \ref{fig:meridiani-experimental study}). 

\subsubsection{Comparison to in situ measurements taken by Pancam/MER-Opportunity}

Similar to the work done at the Spirit's landing site, a photometric
study was done at Meridiani Planum by using the Pancam instrument
from Eagle crater to Purgatory ripple in \citep{Johnson2006b}. Units were defined, for example, as ``Spherule soil''
(typical soil with abundant spherules), ``Bounce marks'' (soil compressed
by the airbags at the landing site), and ``Outcrop rock'' (bedrock). Our results are
compared to two photometric sequences taken by Pancam that are near the 
selected ROI: South of Voyager (Sol 437-439) and Purgatory
region (Sol 449-473) (cf. Figure \ref{fig:HiRISE}).
In addition to the Spherule soil unit, \citet{Johnson2006b} defined
the soil class ``Striped soil'', which characterizes soils with
a striped appearance on the faces of some dune forms. Especially for
the Purgatory region, large ripples were present and a specific
class was defined. Bright soil deposits among the Striped soil unit
were defined as a Dusty soil unit which was limited in spatial extent. 
Results are summarized in Table \ref{tab:Retrieved-Hapke-parameters meridiani}. 
In this article the photometric results
are obtained from the combination of four FRT observations using the
H93 version of the Hapke's model. In this way, we use the same model
as \citet{Johnson2006b}. 

\begin{table}[H]
\raggedright{}\caption{Retrieved Hapke's parameters and their standard deviation (the unconstrained
parameters are indicated by ``(+,-)'') from the 2-term HG model for Sol 437-439 (Spherule
Soil and Striped soil) and Sol 449-473 units (Spherule Soil, Striped
Soil, Dusty Soil and Ripples) at 753 nm from Pancam onboard Opportunity
\citep{Johnson2006b}. \label{tab:Retrieved-Hapke-parameters meridiani}}
\centering{\scriptsize }%
\begin{tabular}{cccccccc}
\hline 
{\scriptsize Site} & {\scriptsize Unit} & {\scriptsize $\omega$} & {\scriptsize $\bar{\theta}$ (deg.)} & {\scriptsize $b$} & {\scriptsize $c$} & {\scriptsize Number} & {\scriptsize $g$ (deg.)}\tabularnewline
\hline 
{\scriptsize South of Voyager} & {\scriptsize Spherule soil} & {\scriptsize 0.53 (+0.02, -0.07)} & {\scriptsize 14 (+1, -2)} & {\scriptsize 0.249 (+0.022, -0.031)} & {\scriptsize 0.491 (+0.058, -0.057)} & {\scriptsize 144} & {\scriptsize $\sim$5-140}\tabularnewline
{\scriptsize (Sol 437-439)} & {\scriptsize Striped soil} & {\scriptsize 0.56 (+0.01, -0.01)} & {\scriptsize 15 (+1, -1)} & {\scriptsize 0.305 (+0.010, -0.022)} & {\scriptsize 0.353 (+0.033, -0.032)} & {\scriptsize 119} & {\scriptsize $\sim$5-140}\tabularnewline
\hline 
{\scriptsize Purgatory region} & {\scriptsize Spherule soil} & {\scriptsize 0.51 (+0.00, -0.01)} & {\scriptsize 10 (+0, -1)} & {\scriptsize 0.230 (+0.006, -0.007)} & {\scriptsize 0.761 (+0.023, -0.013)} & {\scriptsize 686} & {\scriptsize $\sim$0-135}\tabularnewline
{\scriptsize (Sol 449-473)} & {\scriptsize Striped soil} & {\scriptsize 0.52 (+0.24, -0.00)} & {\scriptsize 11 (+0, -1)} & {\scriptsize 0.117 (+0.011, -0.012)} & {\scriptsize 1.000 (+, -)} & {\scriptsize 104} & {\scriptsize $\sim$0-135}\tabularnewline
 & {\scriptsize Dusty soil} & {\scriptsize 0.66 (+0.02,-0.02)} & {\scriptsize 16 (+1,-1)} & {\scriptsize 0.449 (+0.011,-0.020)} & {\scriptsize 0.364 (+0.019,-0.053)} & {\scriptsize 64} & {\scriptsize $\sim$0-135}\tabularnewline
 & {\scriptsize Ripples} & {\scriptsize 0.52 (+0.01,-0.01)} & {\scriptsize 9 (+1,-1)} & {\scriptsize 0.255 (-0.008,-0.007)} & {\scriptsize 0.434 (+0.023,-0.014)} & {\scriptsize 234} & {\scriptsize $\sim$0-135}\tabularnewline
\hline 
\end{tabular}
\end{table}

\begin{itemize}

\item Single scattering albedo: Figure \ref{fig:meridiani-pancam w}
plots means and uncertainties as well as the PDF of the single scattering albedo $\omega$ of the selected ROI against
the values of the Spherule Soil and Striped Soil units defined by
\citet{Johnson2006b} at South of Voyager and the additional Dusty
soil and Ripples units in the Purgatory region. The retrieved values
of $\omega$ from CRISM are higher than those estimated from
Pancam measurements except for the spatially sparse Dusty soil unit (first maximum of the PDF around 0.65).

\begin{figure}
\centering\includegraphics[scale=0.35]{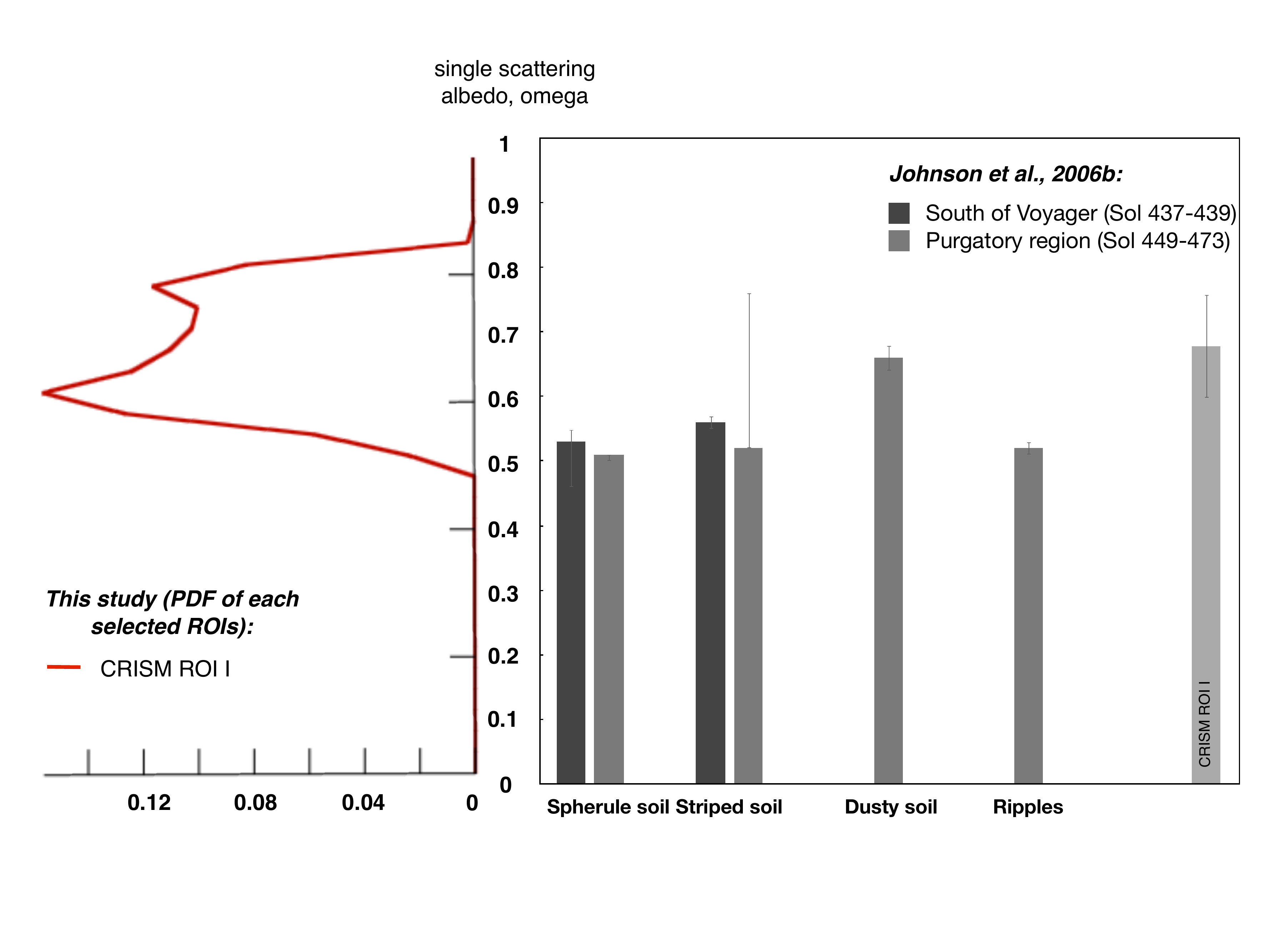}

\caption{Mean and uncertainties of the single scattering albedo $\omega$ estimated
from Pancam measurements at 753 nm for different geological units
at South of Voyager (Sol 437-439) and Purgatory region (Sol 449-473) \citep{Johnson2006b} compared to those estimated from CRISM measurements
at 750 nm (note that the error bar represents 2$\sigma$ uncertainties)
derived from 2-term HG models. The PDFs of the parameter $\omega$
estimated from the last 500 iterations of the bayesian inversion are
also represented on the left side. This is helpful when mean and uncertainties
are not representative of the PDF (such as the PDF of the parameter
$\omega$ for the ROI I which shows a bimodal distribution). \label{fig:meridiani-pancam w}}
\end{figure}

\item Phase function: Figure \ref{fig:meridiani-pancam b/c}
plots the asymmetric parameter $b$ and the backscattering fraction $c$ solutions of the Bayesian inversion
for the ROI I in comparison to results of two photometric
units (i.e., Spherule soil and Striped soil) from South of Voyager 
and three photometric units (Spherule soil, Dusty Soil 
and Ripples) for the Purgatory region. Note that
the parameter $c$ is underconstrained in the Striped Soil unit in the Purgatory region area because
of the relative lack of phase angle coverage ($\sim$45-110\textdegree{}). As mentioned in Subsection
\ref{sub:On-Meridiani-Planum}, parameters $b$ and $c$ are not well
constrained in the CRISM case but exhibit slightly more forward scattering
properties than at Gusev crater. This results is still potentially
consistent with all units in Purgatory region and South of Voyager,
except for the Spherule soil which exhibits more backscattering properties.

\begin{figure}
\centering\includegraphics[scale=0.45]{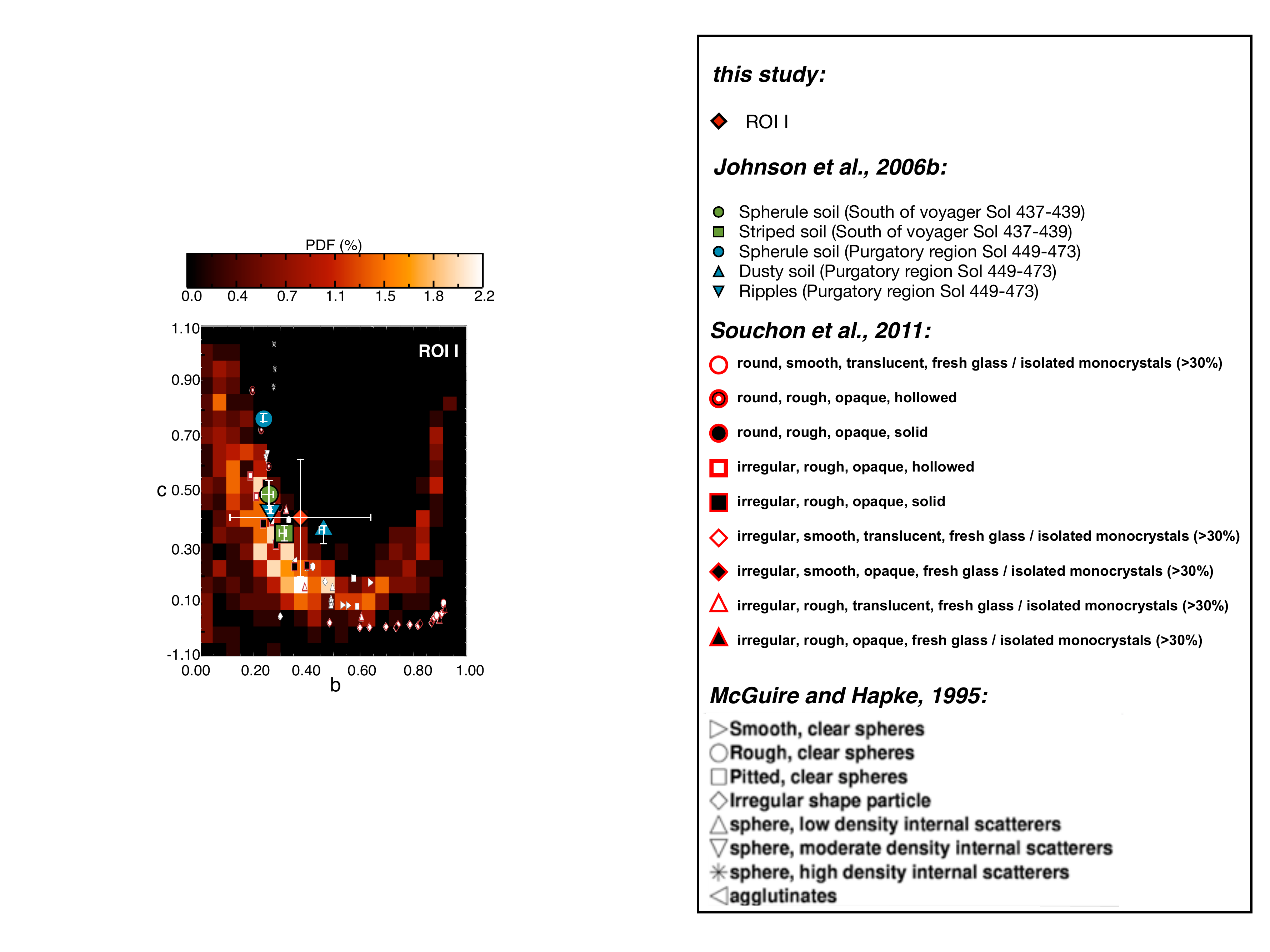}

\caption{Probability density map of the asymmetric parameter $b$ (horizontal axis) and the backscattering fraction
$c$ (vertical axis) solutions from the last 500 iterations of the Bayesian inversion
estimated at 750 nm obtained for the selected ROI. The grid is divided into 24 (vertical) x 20 (horizontal) square bins. The coloring gives the probability corresponding to each bin. Means and 2$\sigma$
uncertainties (red rhombus) are plotted too. The inversion solutions
are plotted against the experimental $b$ and $c$ values pertaining
to artificial particles measured by \citet{mcGuire1995} and to natural
particles measured by \citet{souchon2011}. The Spherule Soil (circle),
Striped Soil (square), Ripple (triangle) and Dusty Soil (triangle)
units from Pancam sequences at South of Voyager (green and Purgatory
region (blue) estimated at 753 nm \citep{Johnson2006b} are plotted
here. \label{fig:meridiani-pancam b/c}}
\end{figure}

\item Macroscopic roughness: Means and uncertainties as well as the PDF 
of the macroscopic roughness $\bar{\theta}$ modeled
from CRISM data is higher than that retrieved at the South of
Voyager and at Purgatory region (cf. Figure \ref{fig:meridiani-pancam theta}).

\begin{figure}
\centering\includegraphics[scale=0.35]{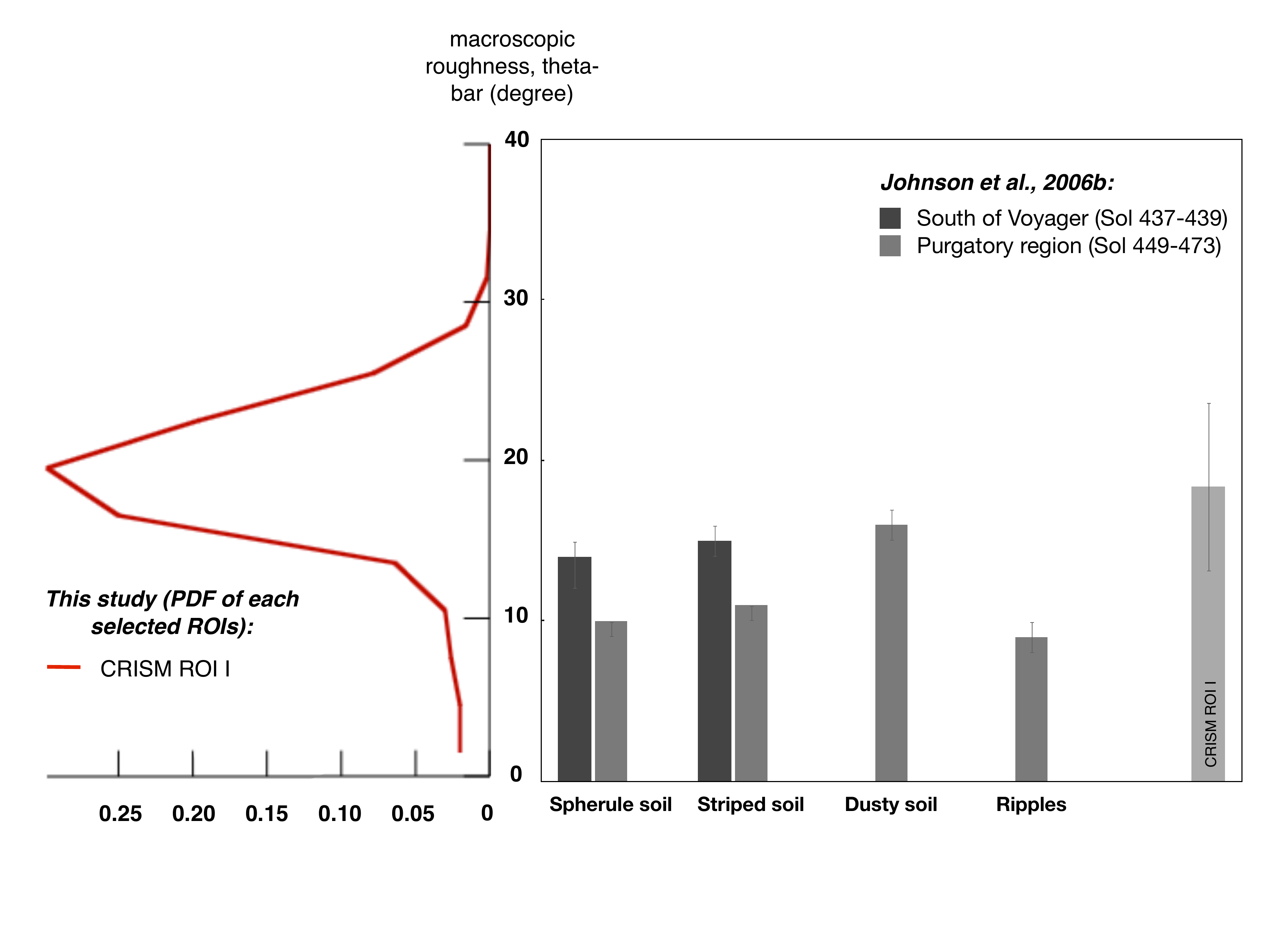}

\caption{Mean and uncertainties of the macroscopic roughness $\bar{\theta}$
estimated from Pancam measurements at 753 nm for different geological
units at South of Voyager (Sol 437-439) and Purgatory region (Sol
449-473) \citep{Johnson2006b} compared to those estimated from CRISM
measurements at 750 nm (note that the error bar represents 2$\sigma$
uncertainties) derived from 2-term HG models. The PDFs of the parameter
$\bar{\theta}$ estimated from the last 500 iterations of the Bayesian
inversion are also represented on the left side. This is helpful when mean
and uncertainties are not entirely representative of the PDF. \label{fig:meridiani-pancam theta}}
\end{figure}

\end{itemize}

Two principal points from the previous comparison can be outlined.
First, at the effective spatial scale achieved from space by CRISM
for the photometric characterization (460 m), the surface behaves
particularly like the Dusty Soil unit defined by \citet{Johnson2006b}
at the Purgatory region: higher single scattering albedo, more forward scattering materials,
and rougher surfaces. Second the photometric properties estimated 
from CRISM data are not strongly supported by those determined from Pancam. In fact, the selected ROI and the Pancam photometric sequences (South of Voyager and Purgatory region) are located several kilometers to the south of the Opportunity landing site where a geological transition region is observed. Indeed bright plains which exposed a smaller areal abundance of hematite, brighter fine-grained dust rich in nanophase iron oxides (such as in Gusev Crater plains) and a larger areal of bright outcrops are observed from orbital measurements (cf. Auxiliary Materials - Figure 1) compared to plains near the landing site \citep{arvidson2006b}. Inside the ROI I, even if the bright material unit does not seem to be the main geological unit at centimeter spatial scale, it seems to be more abundant from the orbit.

\section{Discussion \label{sec:Analysis-of-the}}

The previous results show that the surface photometric parameters
retrieved in this study are consistent with those derived from in
situ measurements. This cross-validation demonstrates that MARS-ReCO
is able to estimate accurately the surface bidirectional reflectance
of Mars. Nevertheless one could question the significance of these
estimates. Therefore, this section aims at testing the non-Lambertian
surface hypothesis used by MARS-ReCO and the influences of the surface
bidirectional reflectance sampling on the determination of the photometric
parameters.

\subsection{Non-Lambertian surface hypothesis \label{sub:Non-lambertian-hypothesis}}

The compensation for aerosol contribution represents a great challenge
for photometric studies. In planetary remote sensing, the Lambertian
surface assumption is often adopted by the radiative atmospheric calculations
that allow retrieval of surface reflectance \citep{Vincendon2007,McGuire2008,Brown2009,Wiseman2012}.
The surface reflectance is then assumed to be independent on geometry
(i.e., the variation of TOA radiance with geometry is supposed to
exclusively relate to aerosol properties). This hypothesis is generally
used as it simplifies the radiative transfer modeling. However, it
has been proved that most surface materials (e.g., minerals and ices)
have an anisotropic non-Lambertian scattering behaviors \citep{Grenier1995,Pinet2001,Johnson2006a,Johnson2006b,Johnson2008,Lyapustin2010}.
Consequently, considering a Lambertian hypothesis can potentially
create biases in the determination of the surface reflectance. 

For the AOT estimate some assumptions regarding the surface properties
were taken in \citet{wolff2009}'s work. Indeed this method assumes
a non-Lambertian surface to estimate the AOT by using a set of surface
photometric parameters \citep{Johnson2006a, Johnson2006b} that appears
to describe the surface phase function adequately for both MER landing
sites (roughly associated to the brighter or dusty soils). This assumption
is qualitative but reasonable for several reasons enumerated by \citet{wolff2009}.
However bias in the estimation of surface bidirectional reflectance
can appear. For both MER landing sites it seems from \citet{wolff2009}'s
work that the AOT retrievals are overall consistent with optical depths
returned by the Pancam instrument (available via PDS). 

In order to test whether or not the non-adoption of a Lambertian hypothesis
improves the quality of the retrieved surface reflectance, we compare
the surface reflectance and the surface photometric parameters retrieved
by a radiative transfer-based atmospheric correction technique that
adopts a Lambertian assumption for the surface \citep{Doute2009}
and those given by MARS-ReCO \citep{Ceamanos2012}.

\begin{figure}
\centering\includegraphics[scale=0.18]{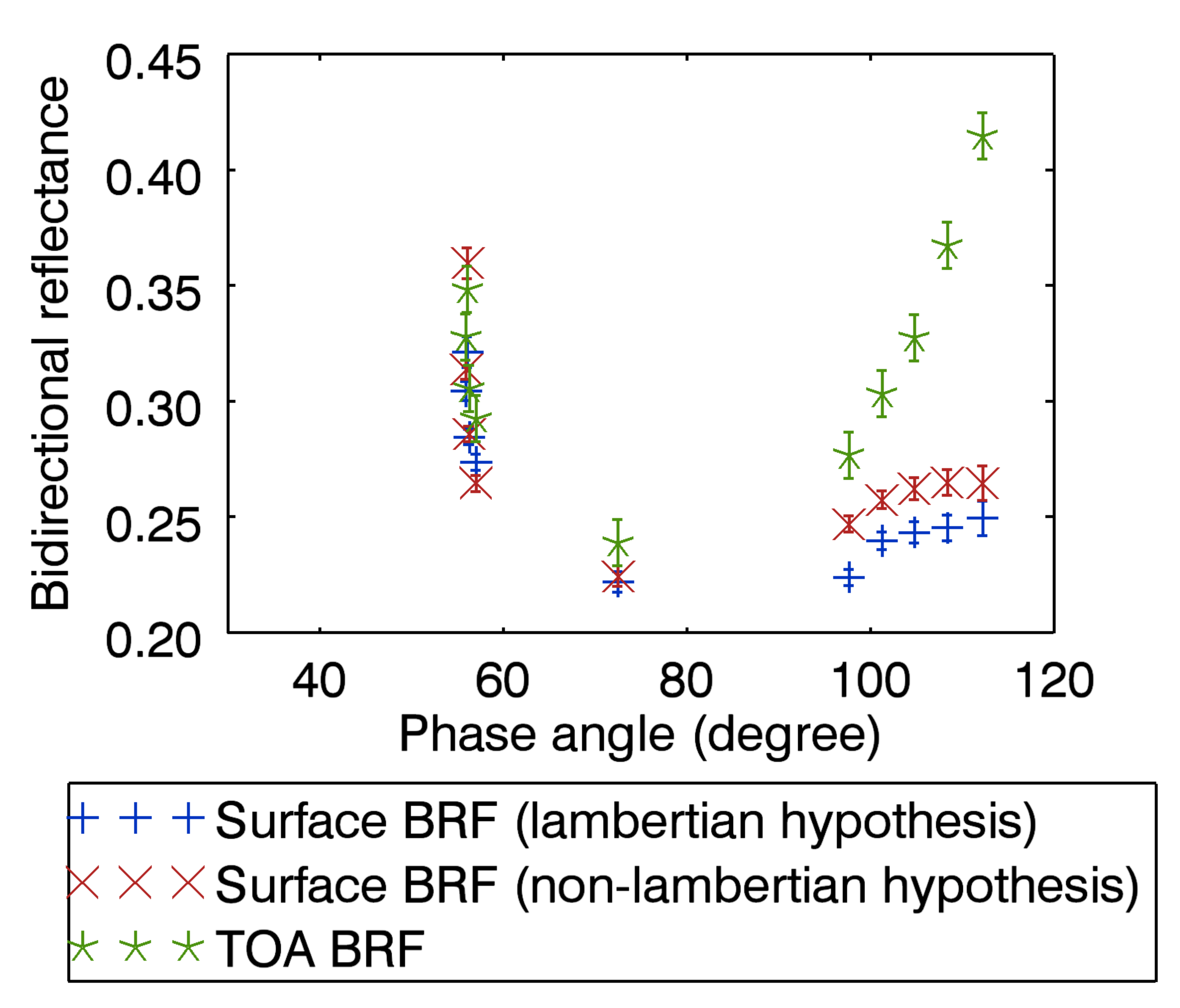}

\caption{Photometric curves corresponding to ROI I of the FRT3192
observation and composed of ten angular configurations. The reflectance values (in BRF units) are extracted at the TOA (green asterisk), and at the surface after correction
for atmospheric contribution. The latter takes into account a non-Lambertian
surface hypothesis using MARS-ReCO (red crosses) and a Lambertian
surface assumption (blue plus sign) (see details in Subsection \ref{sub:Non-lambertian-hypothesis}).
In the three cases we plot error bars corresponding to the $1\sigma$
uncertainty. \label{fig:lamb vs. non-lamb brdf}}
\end{figure}

Figure \ref{fig:lamb vs. non-lamb brdf} presents the initial TOA
reflectance and the surface reflectance derived from both atmospheric
correction techniques as a function of the phase angle. As can be
seen, the main difference is observed at high phase angle (greater
than 90\textdegree{}) where lower BRF are estimated in the case of
the Lambertian method. This difference can be explained by the non-Lambertian
assumption of the surface by MARS-ReCO. Table \ref{tab:lamb vs. non-lamb hypo} 
presents the photometric parameters retrieved from the surface bidirectional reflectance estimated from
both techniques. First, solutions are found for the parameters $\omega$,
$b$, $c$ and $\bar{\theta}$ with higher standard deviations from the Lambertian method.
Second, higher $b$ is observed from the Lambertian method which is not
compatible with the in situ photometric results. This shows 
that the consideration of a non-Lambertian surface in the correction
for aerosol contribution by MARS-ReCO is needed for the determination
of accurate surface bidirectional reflectance and thus for the estimation
of accurate surface photometric parameters.

\begin{table}
\caption{Retrieved Hapke's parameters ($\omega$, $b$, $c$, $\bar{\theta}$ in degree, $B_0$, $h$) and their standard deviation (the underconstrained parameters are indicated by "(-)") from CRISM
measurement at 750 nm using our Bayesian inversion assuming (i) a
non-Lambertian assumption (MARS-ReCO) \citep{Ceamanos2012} and (ii) a Lambertian
hypothesis \citep{Doute2009} for the surface. The Hapke's photometric
parameters corresponding to the ROI I are retrieved by Bayesian inversion
as well as the non-uniform criterion $k$. ($nb$: number of angular
configurations, $g$: phase angle range in degree) \label{tab:lamb vs. non-lamb hypo}}

\raggedright{}\centering{\scriptsize }%
\begin{tabular}{ccccccccccc}
\hline 
{\scriptsize ROI} & {\scriptsize FRT} & \multicolumn{1}{c}{{\scriptsize AOT}} & {\scriptsize $\omega$} & {\scriptsize $b$} & {\scriptsize $c$} & {\scriptsize $\bar{\theta}$ } & {\scriptsize $B_{0}$} & {\scriptsize $h$} & {\scriptsize $nb$} & {\scriptsize $g$}\tabularnewline
 &  &  &  {\scriptsize $k$} &  {\scriptsize $k$} &  {\scriptsize $k$} & {\scriptsize $k$} &  {\scriptsize $k$} &  {\scriptsize $k$}  &  & {\scriptsize }\tabularnewline
\hline 
\multirow{4}{*}{{\scriptsize ROI I}} & \multirow{2}{*}{{\scriptsize 3192 (non-lamb. hypo.)}} & \multirow{2}{*}{{\scriptsize 0.33$\pm$0.04}} & {\scriptsize 0.68 (0.06)} & {\scriptsize 0.17 (0.20)} & {\scriptsize 0.62 (0.20)} & {\scriptsize 11.62 (3.98)} & {\scriptsize 0.52 (-)} & {\scriptsize 0.52 (-)} & \multirow{2}{*}{{\scriptsize 10}} & \multirow{2}{*}{{\scriptsize $\sim$56-112}}\tabularnewline
 &  &  & {\scriptsize 1.00} & {\scriptsize 1.92} & {\scriptsize 0.96} & {\scriptsize 1.00} & {\scriptsize 0.24} & {\scriptsize 0.12} &  & \tabularnewline
\cline{4-11} 
 & \multirow{2}{*}{{\scriptsize 3192 (lamb. hypo.)}} & \multirow{2}{*}{{\scriptsize 0.33$\pm$0.04}} & {\scriptsize 0.74 (0.08)} & {\scriptsize 0.46 (0.32)} & {\scriptsize 0.64 (0.17)} & {\scriptsize 12.43 (4.86)} & {\scriptsize 0.52 (-)} & {\scriptsize 0.50 (-)} & \multirow{2}{*}{{\scriptsize 10}} & \multirow{2}{*}{{\scriptsize $\sim$56-112}}\tabularnewline
 &  &  & {\scriptsize 0.99} & {\scriptsize 1.12} & {\scriptsize 0.74} & {\scriptsize 0.98} & {\scriptsize 0.25} & {\scriptsize 0.03} &  & \tabularnewline
\hline 
\end{tabular}
\end{table}

\subsection{Influences of the surface bidirectional reflectance sampling on the
determination of the photometric parameters}

The accuracy of the determination of Hapke's parameters $\omega$, $b$, $c$, $\bar{\theta}$, $B_{0}$ and $h$ highly depends on the quality and the representativeness of the surface bidirectional reflectance used for their estimation. Indeed, for a given phase angle range, the estimation of all parameters is essentially controlled by the degree to which the angular space (incidence, emergence, azimuth and consequently phase angles) is covered rather than by the number of available angular configurations \citep{souchon2011}. Also, a large phase angle domain is required. In fact, certain photometric parameter are sensitive to the availability of low and high phase angle values. First, the opposition effect is visible only with phase angle values less than 20\textdegree{}. Second, for an accurate determination of the macroscopic roughness, observations that span from small phase angle out to phase angle above 90\textdegree{} is needed \citep{Helfenstein1988}. Finally, materials with  strong backscattering or forward scattering properties require the consideration of large phase angles, ideally greater than 140\textdegree{} \citep{Kamei2002,Shkuratov2007,Shepard2011,Helfenstein1991}.

As presented previously (Subsection \ref{sub:The-CRISM-instrument}), CRISM targeted observations are composed of eleven images which are characterized by a constant incidence angle and eleven different emergence angles. Two modes of relative azimuth are used corresponding to the inbound and outbound portions of the spacecraft trajectory. Based on laboratory experiments performed under a given incidence angle, \citet{Souchon2012} showed that the photometric parameters are determined with a good reliability under the condition of varied emergence and azimuth angles, resulting in varied phase angle values. However, each CRISM observation has intrinsically limited azimuthal and emergence coverage, which affects the determination of all the photometric parameters such as the opposition effect parameters, $B_{0}$ and $h$, the phase function parameters, $b$ and $c$, and the macroscopic roughness, $\bar{\theta}$. The solution used in the present study is to combine several targeted observations with different illumination conditions taken at different times in a year (because of the MRO sun-synchronous orbit) in order to enrich as much as possible the phase angle range by adding low and high phase angles. Nevertheless, low (less than 20\textdegree{}) and high (greater than 120\textdegree{}) phase angle values are not available here.

Thanks to Bayesian inversion, the shape of the a posteriori PDF is known and will inform us whether the bidirectional reflectance sampling is sufficient to estimate accurate photometric parameters. In fact, a uniform or bimodal PDF are the consequence of a lack of the bidirectional reflectance sampling as confirmed by synthetic tests.  

Note that the surface bidirectional sampling also influences the AOT
estimate notably through the determination of the single scattering
albedo. Large AOT error bars are related to CRISM targeted observations
presenting limited geometries \citep{wolff2009}. However, the surface
bidirectional reflectance estimated by MARS-ReCO is sensitive to the
accuracy of the AOT input \citep{Ceamanos2012}.

\subsection{Comparison of surface photometric results from Pancam versus CRISM
\label{sub:Comparison-of-surface}}

The differences regarding the acquisition geometries between Pancam and CRISM can influence the determination of the photometric properties. Ideally, in order to properly compare both results, the bidirectional reflectance sampling should be the same. However, this cannot be achieved because of technical constraints on the acquisition of the bidirectional reflectance
measurements in the case of a spaceborne instrument and an in situ
vehicle, respectively. In addition, the atmospheric contribution does
not affect the observation in the same manner. In the orbital 
case the light travels twice across the atmosphere while in the 
in situ case, it only travels once before reaching the sensor.

Comparison of orbital photometric results to in situ photometric results demonstrates two main differences arising from different acquisition modes. The first comes from the sampling of the bidirectional reflectance. Pancam acquires a photometric sequence by varying the illumination angles (observations acquired at several times of day) and by varying the local azimuth and emission angles among similar units to order to enrich the phase angle range. 

The second difference comes from the spatial scale that is
accessible. CRISM photometric curves are obtained at the hectometric
spatial scale on an extended area whereas Pancam observes at
the centimeter spatial scale. Some characteristics
of the surface can be observed at a hectometric scale but not at centimeter
scale and conversely. From the ground, Pancam is able to distinguish
rocks and soils with relatively high variability, whereas CRISM observes extended areas for which the variability of rocks and soils may appear relatively reduced. Despite these differences,
the choice is made here to consider the Pancam in situ photometric
results as the ground truth, taken as a reference.

\subsection{Interpretation of the retrieved photometric parameters from orbit}

Consistent trends were found with regard to the in situ photometric
results. At the hectometer-scale, the surface behaves like the Soil
unit defined by \citet{Johnson2006a}. The signal acquired by CRISM
mixes several contributions coming from distinct surface materials
occurring at the sub-pixel level. This mixing can be (i) linear or
(ii) non-linear. In the case of linear mixing (geographic mixing),
the measured signal is a linear sum of all the radiated energy curves
of materials making up the pixel. Consequently, the consistency
of our results with the Soil unit behavior can be due to their dominant
relative abundance within the CRISM pixel.
In the case of non-linear mixing, the signal measured is the result
of a non-linearly weighted combination of signals coming from different units. 
In both cases, the apparent consistency with the Soil unit could be coincidental. 

Previous orbital observations acquired at the MER landing sites gave
several clues concerning its physical and chemical properties.
This information can be used to validate the present photometric
results to provide a better understanding of the geological meaning of the
estimated photometric properties. First, we compare our results to
the albedo estimates by Thermal Emission Spectrometer (TES) instrument.
Spirit landed in a relatively dark streak feature which is composed of hundreds
of dark-toned, small, sub-parallel streaks created by dust devils, roughly
oriented NW-SE. This feature is characterized by low albedo values
(between 0.16 and 0.22) \citep{Martinez-Alonso2005}. Pancam in situ
analysis \citep{bell2004,Farrand2006} and coupled OMEGA measurements
and Spirit data \citep{Lichtenberg2007} show that the Spirit landing
site is mainly composed of soil deposits sprinkled
with small rocks interspersed with dust deposits composed of fine-grained
nanophase iron-oxide-rich. These results are consistent with
the intermediate single scattering albedo estimates modeled from CRISM observations ($\omega\sim0.68$). Meridiani Planum is characterized by a low albedo estimated at 0.09-0.19 \citep{Mellon2000}. Dust-free dark sand surface is mainly observed on the Meridiani Planum landing site. However the intermediate single scattering albedo value modeled from CRISM dataset is more consistent with bright plains which exposed a smaller areal abundance of hematite, brighter fine-grained dust rich in nanophase iron oxides (such as in Gusev Crater plains) and a larger areal of bright outcrops are observed several kilometers to the south of the Opportunity landing site from orbital measurements \citep{arvidson2006b}.  

Second, we compare our results to the thermal
inertia (TI) estimated by THermal EMission Imaging System (THEMIS)
images. This is a measure of the resistivity of surface materials
to a change in temperature which is function of the particle size,
bulk density and cohesion. The intermediate THEMIS-derived TI at Gusev
landing site (186-347 SI, 3x6 km resolution, \citep{Martinez-Alonso2005}) suggests that
the surface is dominated by unconsolidated materials (cohesionless
surface) which is consistent with the photometric estimates. This
shows that the surface dominantly behaves like the Soil unit, particularly those
observed at the Landing site which exhibit more small clasts in
soil. At Meridiani landing site, lower THEMIS-derived TI are found 
(i.e., 175 SI, 3x6 km resolution, \citep{Jakosky2006})
suggests that the surface is dominated by more fine-grained materials.
Those pieces of information are consistent with the CRISM-derived
photometric estimates showing that the area has similar properties
as the Soil unit. 

Even if consistent trends were found, one difference is observed with
regard to the in situ photometric results. The difference concerns
the macroscopic roughness $\bar{\theta}$. At Gusev Crater, the intermediate
values of $\bar{\theta}$ conjugated to the presence of more small
clasts on soil \citep{ward2005,Johnson2006a} suggests that the parameter $\bar{\theta}$ is mainly sensitive
to micro-structure. In contrast, slightly 
rougher soil was modeled from CRISM data at Meridiani
Planum. This outcome can be explained by (i) the presence of rocks
but very few rocks are observed \citep{Golombek2005} and/or
(ii) the presence of less than 1 cm-hematite-rich concretions \citep{Calvin2008}
and/or (iii) the presence of morphological structures such as cm- high ripples \citep{Golombek2010}.
The second and third explanations could be responsible for the roughness
measured at hectometer scale, suggesting that the parameter $\bar{\theta}$
is sensitive to both the specific local micro- and macro-structure.
At the Gusev plain, it seems that the surface roughness is dominated by small 
clasts in soils, whereas, at Meridiani Planum it is dominated by the soil particles 
(concretions) and ripples.

\section{Conclusions }

CRISM observations acquired over the MER-Spirit and MER-Opportunity
landing sites allow us to validate the accuracy of (i) the Martian
surface bidirectional reflectance estimated by the procedure named
MARS-ReCO developed by \citet{Ceamanos2012} and (ii) the determined
Martian photometric parameters, by comparing our results to the in
situ photometric results modeled from data taken by the Pancam instrument
on-board both rovers. Indeed, Hapke's parameters
($\omega$, $\bar{\theta}$, $b$ and $c$) estimated from CRISM
measurements (using a combination of CRISM multi-angle images after
individually by applying to each FRT observation the MARS-ReCO atmospheric
correction procedure), are mainly consistent with parameters modeled from
in situ measurements taken at the Spirit and Opportunity landing sites.
The innovative assumption of a non-Lambertian surface is used in our
methodology to accurately estimate intrinsic surface photometric properties
from space. Our results appear to improve (i) those achieved in HRSC-based
photometric studies, in which no aerosol correction was used, and (ii) those estimated from surface bidirectional reflectance derived from atmospheric correction assuming a Lambertian surface, as they better compare with the in situ Pancam results. This outcome of our
study shows that MARS-ReCO gives access to consistent surface bidirectional
reflectance. As a consequence, surface photometric parameters can
be reliably estimated from CRISM observations, provided that the atmospheric
conditions are not turbid. As presented previously, the AOT values derived from \citet{wolff2009}'s work are estimated with assumptions concerning the surface
properties, imposed to be similar as MER observations at both landing
sites. Further developments could lead to the joint estimation of AOT 
and surface bidirectional reflectance by considering the full spectral dimension.

This presented approach may suffer from intrinsic limitations due
to the scarcity of CRISM measurements with broad bidirectional reflectance
sampling. However the combination of several CRISM observations alleviates
this problem by improving the phase angle range thus better constraining
the determination of photometric parameters. Furthermore, Since September, 2010, the inbound segment in targeted observations is absent due to problems of the instrument gimbal \citep{Murchie2012}. Only 6 angular configurations are still available after this date. Consequently, more targeted observations must be combined in order to compensate this limitation.

The presented methodology opens the possibility to map the surface bidirectional 
reflectance thus the spatial variations of the photometric parameters. 
The determination of the physical state of the surface materials (i.e., mean grain 
size, grain sphericity and grain heterogeneity, mean surface porosity and surface 
brightness) through the study of their photometric properties (phase function of the particle, 
single scattering albedo, surface roughness, and opposition effects) provides 
complementary information to composition. It has implications for the characterization 
of different types of terrain over Mars (i.e., extrusive and intrusive volcanic terrains, 
sedimentary terrains, impact craters) and for identification of surface temporal 
modifications (i.e., aeolian process, space weathering process). Moreover, 
the present work has implications for spectroscopic interpretation as it gives 
information about the particle (i.e., size) and grain organization which have significant 
influence on spectral properties \citep{Poulet2004}.

In future work, both challenges of map creation and improvements for
non-flat terrains shall be taken into account, with a particular emphasis
on the complex handling of local slope effects which is not straightforward
in the case of multiple reflection between surface facets. Note that
MARS-ReCO may be used for mapping the photometric properties of the
Martian surface even when using a single CRISM targeted observation.
In this case, however, the photometric inversion should be focused
to retrieve only single scattering albedo for the best surface bidirectional reflectance sampling. 
Furthermore, use of a single-lobed Henyey-Greenstein function would reduce the number of unknown
photometric parameters and therefore could allows us to constrain
the asymmetry parameter besides the single scattering albedo. Likewise,
laboratory studies bearing on the photometric behavior of natural
surfaces under controlled conditions and under a large range of observational
geometries \citep{souchon2011} are necessary to better document and
interpret the variations of the photometric parameters in relation
with surface geological processes such as eolian, aqueous and impact
processes. 

\section*{Acknowledgments }

This work was supported by the French Space Agency CNES (Centre National
d\textquoteright{}Etudes Spatiales) and PNP (Programme National de
Plan\'{e}tologie) from INSU (Institut National des Sciences de l'Univers).
The authors would like to thank Michael Wolff for making his aerosol
optical thickness values available for this study. We would like to gratefully thank Jeffrey Johnson, Mathieu Vincendon and the anonymous reviewer for their constructive comments that substantially improved this article.

\section*{Appendix A}

\paragraph{Non-uniform PDF}

Central moments $\mu_{n}$ (such as the variance $\mu_{2}$ of order
two) are commonly used for statistical purpose while cumulants $k_{n}$
have the advantage to present unbiased statistical estimator for all
orders \citep{Fisher1930}. Also, the first three cumulants are equivalent
to the central moments. For a random variable following the PDF $f(x)$
in (0,1), the cumulant generating function is: 

\begin{equation}
\Phi(t)=\ln TF\left\{ f(x)\right\} =\ln\intop_{0}^{1}e^{itx}f(x)dx
\end{equation}

Cumulants of order $n$ are defined by :

\begin{equation}
k_{n}=\frac{\partial^{n}\Phi(t)}{\partial it^{n}}+o(t)
\end{equation}

The first four cumulants of a uniform PDF are:

\begin{equation}
k_{1}=\frac{1}{2}=\mu_{1}
\end{equation}

\begin{equation}
k_{2}=\frac{1}{12}=\mu_{2}
\end{equation}

\begin{equation}
k_{3}=0=\mu_{3}
\end{equation}

\begin{equation}
k_{4}=-\frac{1}{120}=\mu_{4}-3\left(\mu_{2}\right)^{2}
\end{equation}

Thus we propose to estimate the non-uniformity of the results of the
inversion with:

\begin{equation}
k=max\mid\frac{k_{1}-12}{12},\frac{k_{2}-112}{112},\frac{k_{3}}{160},\frac{k_{4}+1120}{1120}\mid
\end{equation}

We perform 10,000 uniform random vectors of 500 samples (identical
to the inversion procedure). Since the maximum is $k$=0.47 for the
most extreme event, we propose to have non-uniform PDF for $k$>0.5.
For the inversion purpose, since the a priori PDF's on the parameters
are uniform if the results of the inversion on one parameter has $k$<0.5,
we conclude that this parameters is not constrained by the observations.


\begin{thebibliography}{58}
\providecommand{\natexlab}[1]{#1}
\expandafter\ifx\csname urlstyle\endcsname\relax
  \providecommand{\doi}[1]{doi:\discretionary{}{}{}#1}\else
  \providecommand{\doi}{doi:\discretionary{}{}{}\begingroup
  \urlstyle{rm}\Url}\fi

\bibitem[{\textit{Arvidson et~al.}(2006)\textit{Arvidson, Poulet, Morris,
  Bibring, Bell, Squyres, Christensen, Bellucci, Gondet, Ehlmann, Farrand,
  Fergason, Golombek, Griffes, Grotzinger, Guinness, Herkenhoff, Johnson,
  Klingelhöfer, Langevin, Ming, Seelos, Sullivan, Ward, Wiseman, and
  Wolff}}]{arvidson2006b}
Arvidson, R.~E., F.~Poulet, R.~V. Morris, J.-P. Bibring, I.~Bell, J.~F., S.~W.
  Squyres, P.~R. Christensen, G.~Bellucci, B.~Gondet, B.~L. Ehlmann, W.~H.
  Farrand, R.~L. Fergason, M.~Golombek, J.~L. Griffes, J.~Grotzinger, E.~A.
  Guinness, K.~E. Herkenhoff, J.~R. Johnson, G.~Klingelhofer, Y.~Langevin,
  D.~Ming, K.~Seelos, R.~J. Sullivan, J.~G. Ward, S.~M. Wiseman, and M.~Wolff,
  Nature and origin of the hematite-bearing plains of terra meridiani based on
  analyses of orbital and mars exploration rover data sets, \textit{J. Geophys.
  Res.}, \textit{111}, E12S08, \doi{10.1029/2006JE002728}, 2006.

\bibitem[{\textit{Bell~III et~al.}(1999)\textit{Bell~III, Wolff, Daley, Crisp,
  Lee, Trauger, and Evans}}]{BellIII1999}
Bell~III, J.~F., M.~J. Wolff, T.~C. Daley, P.~B. Crisp, S.~W. Lee, J.~T.
  Trauger, and R.~W. Evans, Near-infrared imaging of Mars from HST: Surface
  reflectance, photometric properties, and implications for MOLA data,
  \textit{Icarus}, \textit{138}, 25--35, \doi{10.1006/icar.1998.6057}, 1999.

\bibitem[{\textit{{Bell et al}}(2004)}]{bell2004}
{Bell, J. F. et al}, Pancam multispectral imaging results from the Spirit rover
  at Gusev Crater, \textit{Science}, \textit{305, no. 5685}, 800--806,
  \doi{10.1126/science.1100175}, 2004.

\bibitem[{\textit{Brown and Wolff}(2009)}]{Brown2009}
Brown, A.~J., and M.~J. Wolff, Atmospheric modeling of the Martian polar
  regions: One Mars year of CRISM EPF observations of the south pole,
  \textit{Lunar Planet. Sci.}, \textit{abstract 1675}, 2009.

\bibitem[{\textit{{Calvin et al.}}(2008)}]{Calvin2008}
{Calvin, W. M. et al.}, Hematite spherules at Meridiani: Results from MI,
  Mini-TES and Pancam, \textit{J. Geophys. Res.}, \textit{113}, E12S37,
  \doi{10.1029/2007JE003048}, 2008.

\bibitem[{\textit{Ceamanos et~al.}(2013)\textit{Ceamanos, Dout\'e, Fernando,
  Schmidt, Pinet, and Lyapustin}}]{Ceamanos2012}
Ceamanos, X., S.~Dout\'e, J.~Fernando, F.~Schmidt, P.~Pinet, and A.~Lyapustin,
  Surface reflectance of mars observed by CRISM/MRO: 1. multi-angle approach
  for retrieval of surface reflectance from crism observations (MARS-ReCO),
  \textit{J. Geophys. Res. in press}, \doi{10.1029/2012JE004195}, 2013.

\bibitem[{\textit{Chandrasekhar}(1960)}]{chandrasekhar1960}
Chandrasekhar, S., \textit{Radiative Transfer}, Dover, Mineola, NY., 1960.

\bibitem[{\textit{Cheng and Domingue}(2000)}]{cheng2000}
Cheng, A., and D.~L. Domingue, Radiative transfer models for light scattering
  from planetary surfaces., \textit{J. Geophys. Res.}, \textit{105 (E4)},
  9477--9482, \doi{10.1029/1999JE001170}, 2000.

\bibitem[{\textit{Clancy and Lee}(1991)}]{Clancy1991}
Clancy, R.~T., and S.~W. Lee, A new look at dust and clouds in the Mars
  atmosphere: analysis of Emission-Phase-Function sequences from global Viking
  IRTM observations, \textit{Icarus}, \textit{93}, 135--158,
  \doi{10.1016/0019-1035(91)90169-T}, 1991.

\bibitem[{\textit{Clancy et~al.}(2003)\textit{Clancy, Wolff, and
  Christensen}}]{Clancy2003}
Clancy, R.~T., M.~J. Wolff, and P.~R. Christensen, Mars aerosol studies with
  the MGS TES emission phase function observations: Optical depths, particle
  sizes, and ice cloud types versus latitude and solar longitude, \textit{J.
  Geophys. Res.}, \textit{E9 5098}, \doi{10.1029/2003JE002058}, 2003.

\bibitem[{\textit{Cord et~al.}(2003)\textit{Cord, Pinet, Daydou, and
  Chevrel}}]{cord2003}
Cord, A.~M., P.~C. Pinet, Y.~Daydou, and S.~D. Chevrel, Planetary regolith
  surface analogs: optimized determination of Hapke parameters using
  multi-angular spectro-imaging laboratory data, \textit{Icarus}, \textit{165},
  414--427, \doi{10.1016/S0019-1035(03)00204-5}, 2003.

\bibitem[{\textit{de~Grenier and Pinet}(1995)}]{Grenier1995}
de~Grenier, M., and P.~C. Pinet, Near-opposition martian limb-darkening:
  quantification and implication for visible-near-infrared bidirectional
  reflectance studies, \textit{Icarus}, \textit{115}, 354--368,
  \doi{10.1006/icar.1995.1103}, 1995.

\bibitem[{\textit{Dout\'e and Schmitt}(1998)}]{doute1998}
Dout\'e, S., and B.~Schmitt, A multilayer bidirectional reflectance model for
  the analysis of planetary surface hyperspectral images at visible and
  near-infrared wavelengths, \textit{J. Geophys. Res.}, \textit{103},
  31,367--31,389, \doi{10.1029/98JE01894}, 1998.

\bibitem[{\textit{{Dout\'e, S.}}(2009)}]{Doute2009}
{Dout\'e, S.}, Retrieving Mars surface reflectance from OMEGA/MEx imagery,
  \textit{IEEE Workshop on Hyperspectral Image and Signal Processing :
  Evolution in Remote Sensing}, 2009.

\bibitem[{\textit{Farrand et~al.}(2006)\textit{Farrand, Bell~III, Johnson,
  Squyres, Soderblom, and Ming}}]{Farrand2006}
Farrand, W.~H., J.~F. Bell~III, J.~R. Johnson, S.~W. Squyres, J.~Soderblom, and
  D.~Ming, Spectral variability among rocks in visible and near-infrared
  multispectral Pancam data collected at Gusev Crater: Examinations using
  spectral mixture analysis and related techniques, \textit{J. Geophys. Res.},
  \textit{111}, E02S15, \doi{10.1029/2005JE002495}, 2006.

\bibitem[{\textit{Fisher}(1930)}]{Fisher1930}
Fisher, R.~A., \textit{The Genetical Theory Of Natural Selection}, Clarendon
  Press, 1930.

\bibitem[{\textit{{Golombek et al.}}(2005)}]{Golombek2005}
{Golombek, M. P. et al.}, Assessment of Mars Exploration Rover landing site
  predictions, \textit{Nature}, \textit{436}, 44--48,
  \doi{10.1038/nature03600}, 2005.

\bibitem[{\textit{Golombek et~al.}(2010)\textit{Golombek, Robinson, McEwen,
  Bridges, Tornabene, and R.}}]{Golombek2010}
Golombek, M., K.~Robinson, A.~McEwen, B.~Bridges, N.~Ivanov, L.~Tornabene, and
  S.~R., Constraints on ripple migration at Meridiani Planum from Opportunity
  and HiRISE observations of fresh craters, \textit{J. Geosphys. Res.},
  \textit{115}, E00F08, \doi{10.1029/2010JE003628}, 2010.

\bibitem[{\textit{Grynko and Shkuratov}(2007)}]{Grynko2007}
Grynko, Y., and Y.~Shkuratov, Ray tracing simulation of light scattering by
  spherical clusters consisting of particles with different shapes,
  \textit{Journal of Quantitative Spectroscopy \& Radiative Transfer},
  \textit{106}, 56--62, \doi{10.1016/j.jqsrt.2007.01.005}, 2007.

\bibitem[{\textit{Guinness et~al.}(1997)\textit{Guinness, Arvidson, Clark, and
  Shepard}}]{guinness1997}
Guinness, E.~A., R.~E. Arvidson, I.~H.~D. Clark, and M.~K. Shepard, Optical
  scattering properties of terrestrial varnished basalts compared with rocks
  and soils at the Viking Lander sites, \textit{J. Geophys. Res.},
  \textit{102}, 28,687--28,703, \doi{10.1029/97JE03018}, 1997.

\bibitem[{\textit{Hapke}(1981{\natexlab{a}})}]{hapke1981}
Hapke, B., Bidirectional reflectance spectroscopy 1. theory, \textit{J.
  Geophys. Res.}, \textit{86}, 3039--3054, \doi{10.1029/JB086iB04p03039}, 1981{\natexlab{a}}.

\bibitem[{\textit{{Hapke}}(1981{\natexlab{b}})}]{hapke11981}
Hapke, B., and E.~Wells, Bidirectional Reflectance Spectroscopy 2. Experiments
  and Observations, \textit{J. Geophys. Res.}, \textit{86}, 3055--3060, 1981{\natexlab{b}}.

\bibitem[{\textit{Hapke}(1984)}]{hapke1984}
Hapke, B., Bidirectional Reflectance Spectroscopy 3. Correction for Macroscopic
  Roughness, \textit{Icarus}, \textit{59}, 41--59,
  \doi{10.1016/0019-1035(84)90054-X}, 1984.

\bibitem[{\textit{Hapke}(1986)}]{hapke1986}
Hapke, B., Bidirectional Reflectance Spectroscopy 4. The Extinction Coefficient
  and the Opposition Effect, \textit{Icarus}, \textit{67}, 264--280,
  \doi{10.1016/0019-1035(86)90108-9}, 1986.

\bibitem[{\textit{Hapke}(1993)}]{Hapke1993}
Hapke, B., \textit{Theory of Reflectance and Emittance Spectroscopy}, Cambridge
  Univ. Press, New York, 1993.

\bibitem[{\textit{Hapke}(2002)}]{hapke2002}
Hapke, B., Bidirectional Reflectance Spectroscopy 5. The Coherent Backscatter
  Opposition Effect and Anisotropic Scattering, \textit{Icarus}, \textit{157},
  523--534, \doi{10.1006/icar.2002.6853}, 2002.

\bibitem[{\textit{Hapke}(2008)}]{hapke2008}
Hapke, B., Bidirectional reflectance spectroscopy 6. effects of porosity,
  \textit{Icarus}, \textit{195}, 918--926, \doi{10.1016/j.icarus.2008.01.003},
  2008.

\bibitem[{\textit{Hartman and Domingue}(1998)}]{hartman1998}
Hartman, B., and D.~Domingue, Scattering of light by individual particles and
  the implications for models of planetary surfaces, \textit{Icarus},
  \textit{131}, 421--448, \doi{10.1016/j.jqsrt.2011.04.006}, 1998.

\bibitem[{\textit{Helfenstein}(1988)}]{Helfenstein1988}
Helfenstein, P., The geological interpretation of photometric surface
  roughness, \textit{Icarus}, \textit{73}, 462--481,
  \doi{10.1016/0019-1035(88)90056-5}, 1988.

\bibitem[{\textit{Helfenstein et~al.}(1991)\textit{Helfenstein, Bonne, Stolovy,
  and Veverka}}]{Helfenstein1991}
Helfenstein, P., U.~A. Bonne, S.~Stolovy, and J.~Veverka, Laboratory
  photometric measurements of particulate soils out to very large phase
  angles., \textit{Reports of Planetary Geology and Geophysics Program - 1990},
  \textit{NASA-TM-4300}, 280--282, 1991.

\bibitem[{\textit{{Jakosky et al.}}(2006)}]{Jakosky2006}
{Jakosky, B. M. et al.}, Thermophysical properties of the MER and Beagle II
  landing site regions on Mars, \textit{J. Geophys. Res.}, \textit{111},
  E08,008, \doi{10.1029/2004JE002320}, 2006.

\bibitem[{\textit{Jehl et~al.}(2008)}]{Jehl2008}
Jehl, A., et~al., Gusev photometric variability as seen from orbit by
  HRSC/Mars-express, \textit{Icarus}, \textit{197}, 403--428,
  \doi{10.1016/j.icarus.2008.05.022}, 2008.

\bibitem[{\textit{{Johnson et al.}}(1999)}]{johnson1999}
{Johnson, J. R. et al.}, Preliminary results on photometric properties of
  materials at the Sagan Memorial Station, Mars, \textit{J. Geophys. Res.},
  \textit{104}, 8809--8830, \doi{doi:10.1029/98JE02247}, 1999.

\bibitem[{\textit{{Johnson et al.}}(2006{\natexlab{a}})}]{Johnson2006a}
{Johnson, J. R. et al.}, Spectrophotometric properties of materials observed by
  Pancam on the Mars Exploration Rovers: 1. Spirit, \textit{J. Geophys. Res.},
  \textit{111}, E02S14, \doi{10.1029/2005JE002494}, 2006{\natexlab{a}}.

\bibitem[{\textit{{Johnson et al.}}(2006{\natexlab{b}})}]{Johnson2006b}
{Johnson, J. R. et al.}, Spectrophotometric properties of materials observed by
  Pancam on the Mars Exploration Rovers: 2. Opportunity, \textit{J. Geophys.
  Res.}, \textit{111}, E12S16, \doi{10.1029/2006JE002762}, 2006{\natexlab{b}}.

\bibitem[{\textit{{Johnson et al.}}(2006{\natexlab{c}})}]{johnson2006c}
{Johnson, J. R. et al.}, Radiative transfer modeling of dust-coated Pancam
  calibration target materials: Laboratory visible/near-infrared
  spectrogoniometry, \textit{J. Geophys. Res.}, \textit{111}, E12S07,
  \doi{10.1029/2005JE002658}, 2006{\natexlab{c}}.

\bibitem[{\textit{Johnson et~al.}(2008)\textit{Johnson, Bell, Geissler, Grundy,
  Guinness, Pinet, and Soderblom}}]{Johnson2008}
Johnson, J.~R., J.~F. Bell, P.~Geissler, W.~M. Grundy, E.~A. Guinness, P.~C.
  Pinet, and J.~Soderblom, The martian surface, chap. physical properties of
  the martian surface from spectrophotometric observations, \textit{Cambridge
  University Press}, 2008.

\bibitem[{\textit{Kamei and Nakamura}(2002)}]{Kamei2002}
Kamei, A., and A.~M. Nakamura, Laboratory study of the bidirectional
  reflectance of powdered surfaces: On the asymmetry parameter of asteroid
  photometric data, \textit{Icarus}, \textit{156}, 551--561,
  \doi{10.1006/icar.2002.6818}, 2002.

\bibitem[{\textit{{Lichtenberg et al.}}(2007)}]{Lichtenberg2007}
{Lichtenberg, K. A. et al.}, Coordinated analyses of orbital and Spirit Rover
  data to characterize surface materials on the cratered plains of Gusev
  Crater, Mars, \textit{J. Geophys. Res.}, \textit{112}, E12S90,
  \doi{10.1029/2006JE002850}, 2007.

\bibitem[{\textit{Lucht et~al.}(2000)\textit{Lucht, Schaaf, and
  Strahler}}]{Lucht2000}
Lucht, W., C.~B. Schaaf, and A.~H. Strahler, An algorithm for the retrieval of
  albedo from space using semiempirical BRDF models, \textit{IEEE Transaction
  on Geoscience and Remote Sensing}, \textit{38(2)}, 977--998, 2000.

\bibitem[{\textit{{Lyapustin et al.}}(2010)}]{Lyapustin2010}
{Lyapustin, A., C. K. Gatebe, R. Kahn, R. Brandt, J. Redemann, et al.},
  Analysis of snow bidirectional reflectance from ARCTAS Spring-2008 Campaign,
  \textit{Atmospheric Chemistry and Physics}, \textit{10}, 4359--4375,
  \doi{10.5194/acp-10-4359-2010}, 2010.

\bibitem[{\textit{Martinez-Alonso et~al.}(2005)\textit{Martinez-Alonso,
  Jakosky, Mellon, and Putzig}}]{Martinez-Alonso2005}
Martinez-Alonso, S., B.~M. Jakosky, M.~T. Mellon, and N.~E. Putzig, A volcanic
  interpretation of Gusev Crater surface materials from thermophysical,
  spectral, and morphological evidence, \textit{J. Geophys. Res.},
  \textit{110}, E01,003, \doi{10.1029/2004JE002327}, 2005.

\bibitem[{\textit{McCord et~al.}(2007)}]{mcCord2007}
McCord, J. B.~A., T.B., et~al., The Mars Express High Resolution Stereo Camera
  spectrophotometric data: Characteristics and science analysis., \textit{J.
  Geophys. Res.}, \textit{112}, E06,004, \doi{10.1029/2006JE002769. E06004},
  2007.

\bibitem[{\textit{McGuire and Hapke}(1995)}]{mcGuire1995}
McGuire, A., and B.~Hapke, An experimental study of light scattering by large
  irregular particles, \textit{Icarus}, \textit{113}, 134--155, 1995.

\bibitem[{\textit{{McGuire et al.}}(2008)}]{McGuire2008}
{McGuire, P. C., M. J. Wolff, M. D. Smith, R. E. Arvidson, S. L. Murchie, et
  al.}, MRO/CRISM retrieval of surface lambert albedos for multispectral
  mapping of Mars with DISORT-based radiative transfer modeling: Phase 1 -
  Using historical climatology for temperatures, aerosol optical depths, and
  atmospheric pressures, \textit{IEEE Transactions on Geoscience and Remote
  Sensing}, \textit{46 (12)}, 4020--4040, 2008.

\bibitem[{\textit{Mellon et~al.}(2000)\textit{Mellon, Jakosky, Kieffer, and
  Christensen}}]{Mellon2000}
Mellon, M.~T., B.~M. Jakosky, H.~H. Kieffer, and P.~R. Christensen,
  High-Resolution Thermal Inertia Mapping from the Mars Global Surveyor
  Thermal Emission Spectrometer, \textit{Icarus}, \textit{148}, 437--455,
  \doi{10.1006/icar.2000.6503}, 2000.

\bibitem[{\textit{Mishchenko et~al.}(1999)\textit{Mishchenko, Dlugach,
  Yanovitskij, and Zakharova}}]{mishchenko1999}
Mishchenko, M., J.~M. Dlugach, E.~G. Yanovitskij, and N.~T. Zakharova,
  Bidirectional reflectance of flat, optically thick particulate layers: An
  efficient radiative transfer solution and applications to snow and soil
  surfaces., \textit{J. Quant. Spectrosc. Radiat. Trans.}, \textit{63},
  409--432, \doi{10.1016/S0022-4073(99)00028-X}, 1999.

\bibitem[{\textit{Mosegaard and Tarantola}(1995)}]{Mosegaard1995}
Mosegaard, K., and A.~Tarantola, Monte Carlo sampling of solutions to inverse
  problems, \textit{J. Geophys. Res.}, \textit{100}, 12,431--12,447, 1995.

\bibitem[{\textit{{Murchie et al.}}(2007)}]{murchie2007}
{Murchie, S. et al.}, Compact Reconnaissance Imaging Spectrometer for Mars
  (CRISM) on Mars Reconnaissance Orbiter (MRO), \textit{J. Geophys. Res.},
  \textit{112}, E05S03, \doi{10.1029/2006JE002682}, 2007.

\bibitem[{\textit{Murchie}(2012)}]{Murchie2012}
Murchie, S., CRISM on MRO - instrument and investigation overview, \textit{in
  MRO/CRISM Data Users' Workshop, The Woodlands, TX}, 2012.

\bibitem[{\textit{Pinet and Rosemberg}(2001)}]{Pinet2001}
Pinet, P., and C.~Rosemberg, Regional photometry and spectral albedo of the
  eastern hemisphere of Mars in the 0.7-1 micron domain, \textit{Lunar Planet.
  Sci.}, \textit{abstract 1640}, 2001.

\bibitem[{\textit{Pinet et~al.}(2005)}]{pinet2005}
Pinet, P.~C., et~al., Derivation of Mars surface scattering properties from
  OMEGA spot pointing, \textit{Lunar Planet. Sci.}, \textit{XXXVI}, abstract
  1694, 2005.

\bibitem[{\textit{Poulet and Erard}(2004)}]{Poulet2004}
Poulet, F., and S.~Erard, Nonlinear spectral mixing: Quantitative analysis of
  laboratory mineral mixtures, \textit{J. Geophys. Res.}, \textit{109},
  E02,009, \doi{10.1029/2003JE002179}, 2004.

\bibitem[{\textit{Seelos et~al.}(2011)\textit{Seelos, Murchie, Humm, Barnouin,
  Morgan, Taylor, Hash, and Team}}]{Seelos2011}
Seelos, F.~P., S.~Murchie, D.~Humm, O.~S. Barnouin, F.~Morgan, H.~W. Taylor,
  C.~Hash, and T.~C. Team, CRISM data processing and analysis products update -
  calibration, correction, and visualization, \textit{Lunar Planet. Sci.},
  \textit{XXXXII, abstract 1438}, 2011.

\bibitem[{\textit{Shaw et~al.}(2012)\textit{Shaw, Arvidson, Wolff, Seelos,
  Wiseman, and Cull}}]{Shaw2012}
Shaw, A., R.~E. Arvidson, M.~J. Wolff, F.~P. Seelos, S.~M. Wiseman, and
  S.~Cull, Determining surface roughness and additional terrain properties:
  unsig opportunity mars rover results to interpret orbital data for extended
  mapping, \textit{Lunar Planet. Sci.}, \textit{XXXXIII}, abstract 1644, 2012.

\bibitem[{\textit{Shepard and Helfenstein}(2007)}]{shepard2007}
Shepard, M.~K., and P.~Helfenstein, A test of the Hapke photometric model,,
  \textit{J. Geophys. Res.}, \textit{112}, E03,001,
  \doi{10.1029/2005JE002625.}, 2007.

\bibitem[{\textit{Shepard and Helfenstein}(2011)}]{Shepard2011}
Shepard, M.~K., and P.~Helfenstein, A laboratory study of the bidirectional
  reflectance from particulate samples, \textit{Icarus}, \textit{215},
  526--533, \doi{10.1016/j.icarus.2011.07.033}, 2011.

\bibitem[{\textit{Shkuratov and Starukhina}(1999)}]{Shkuratov1999a}
Shkuratov, Y., and L.~Starukhina, A model of spectral albedo of particulate
  surfaces: implications for optical properties of the Moon, \textit{Icarus},
  \textit{137}, 235--246, \doi{10.1006/icar.1998.6035}, 1999.

\bibitem[{\textit{Shkuratov et~al.}(2007)\textit{Shkuratov, Bondarenko,
  Kaydash, Videen, Munoz, and Volten}}]{Shkuratov2007}
Shkuratov, Y., S.~Bondarenko, V.~Kaydash, G.~Videen, O.~Munoz, and H.~Volten,
  Photometry and polarimetry of particulate surfaces and aerosol particles over
  a wide range of phase angles, \textit{Jounal of Quantitative Spectroscopy \&
  Radiative Transfer}, \textit{106}, 487--508,
  \doi{10.1016/j.jqsrt.2007.01.031}, 2007.

\bibitem[{\textit{Soderblom et~al.}(2004)\textit{Soderblom, Bell, Arvidson,
  Johnson, Johnson, and (2004)}}]{soderblom2004}
Soderblom, J.~M., J.~F. Bell, R.~E. Arvidson, J.~R. Johnson, M.~J. Johnson, and
  F.~P. Seelos, Mars Exploration Rover Pancam photometric data QUBs:
  Definition and example uses, \textit{AGU, 85(47) Fall Meet}, 
\textit{Abstract P21A-0198.}, 2004.

\bibitem[{\textit{Souchon et~al.}(2011)\textit{Souchon, Pinet, Chevrel, Daydou,
  Baratoux, Kurita, Shepard, and Helfenstein}}]{souchon2011}
Souchon, A.~L., P.~Pinet, S.~Chevrel, Y.~Daydou, D.~Baratoux, K.~Kurita, M.~K.
  Shepard, and P.~Helfenstein, An experimental study of Hapke's modeling of
  natural granular surface samples, \textit{Icarus}, \textit{215}, 313--331,
  \doi{10.1016/j.icarus.2011.06.023}, 2011.

\bibitem[{\textit{Souchon}(2012)}]{Souchon2012}
Souchon, A., Influence des phases amorphes dans la r\'eponse optique des
  r\'egolites plan\'etaires : caract\'erisation des propri\'et\'es physiques et
  application a l'\'etude g\'eologique de la lune, Ph.D. thesis, Univ. de
  Toulouse, France, 2012.

\bibitem[{\textit{Tarantola and Valette}(1982)}]{tarantola1982}
Tarantola, A., and B.~Valette, Inverse problems=quest for information,
  \textit{J. Geophys. Res.}, \textit{50}, 159--170, 1982.

\bibitem[{\textit{Vincendon et~al.}(2007)\textit{Vincendon, Langevin, Poulet,
  Bibring, and Gondet}}]{Vincendon2007}
Vincendon, M., Y.~Langevin, F.~Poulet, J.-P. Bibring, and B.~Gondet, Recovery
  of surface reflectance spectra and evaluation of the optical depth of
  aerosols in the near-IR using a Monte Carlo approach: Application to the
  OMEGA observations of high-latitude regions of Mars, \textit{J. Geophys.
  Res.}, \textit{112, E08S13}, \doi{10.1029/2006JE002845}, 2007.

\bibitem[{\textit{Ward et~al.}(2005)\textit{Ward, Arvidson, and
  Golombek}}]{ward2005}
Ward, J.~G., R.~E. Arvidson, and M.~Golombek, The size-frequency and areal
  distribution of rock clasts at the Spirit landing site, Gusev Crater, Mars,
  \textit{Geophys. Res. Lett.}, \textit{32}, L11,203,
  \doi{10.1029/2005GL022705}, 2005.

\bibitem[{\textit{Wiseman et~al.}(2012)\textit{Wiseman, Arvidson, Wolff,
  Morris, Seelos, Smith, Humm, Murchie, and Mustard}}]{Wiseman2012}
Wiseman, S.~M., R.~E. Arvidson, M.~J. Wolff, R.~V. Morris, F.~P. Seelos, M.~D.
  Smith, D.~Humm, S.~L. Murchie, and J.~F. Mustard, Retrieval of
  atmospherically corrected CRISM spectra using radiative transfer modeling,
  \textit{Lunar Planet. Sci.}, \textit{XXXXIII}, abstract 2146, 2012.

\bibitem[{\textit{Wolff et~al.}(2006)\textit{Wolff, Smith, Clancy, Spanovich,
  Whitney, Lemmon, Bandfiel, Bandfiel, Ghosh, Landis, Christensen, Bell~III,
  and Squyres}}]{Wolff2006}
Wolff, M.~J., M.~D. Smith, R.~T. Clancy, N.~Spanovich, B.~A. Whitney, M.~T.
  Lemmon, J.~L. Bandfiel, D.~Bandfiel, A.~Ghosh, G.~Landis, P.~R. Christensen,
  J.~F. Bell~III, and S.~W. Squyres, Constraints on dust aerosols from the Mars
  Exploration Rovers using MGS overflights and Mini-TES, \textit{J. Geophys.
  Res.}, \textit{111}, E12S17, \doi{10.1029/2006JE002786}, 2006.

\bibitem[{\textit{{Wolff et al.}}(2009)}]{wolff2009}
{Wolff, M. J. et al.}, Wavelength dependence of dust aerosol single scaterring
  albedo as observed by the Compact Reconnaissance Imaging Spectrometer,
  \textit{J. Geophys. Res.}, \textit{114}, E00D64,
  \doi{doi:10.1029/2009JE003350}, 2009.

\bibitem[{\textit{Wolff et~al.}(2010)\textit{Wolff, Clancy, Goguen, Malin, and
  Cantor}}]{Wolff2010}
Wolff, M.~J., R.~T. Clancy, J.~D. Goguen, M.~C. Malin, and B.~A. Cantor,
  Ultraviolet dust aerosol properties as observed by MARCI, \textit{Icarus},
  \textit{208}, 143--155, \doi{10.1016/j.icarus.2010.01.010}, 2010.

\end{thebibliography}

\end{document}